\documentclass[twocolumn,showpacs,aps,prd,nobibnotes,notitlepage,floatfix]{revtex4-1} 

\usepackage{graphicx,color}
\usepackage{amsmath,amssymb}
\usepackage{verbatim}

\usepackage{float}
\usepackage{wasysym}
\usepackage{ulem} 
\usepackage{hyperref,url}
\usepackage{multimedia}
\usepackage{enumerate}

\newcommand{\bra}[1]{\left( #1 \right)}
\newcommand{\av}[1]{\langle #1 \rangle}

\newcommand{\mnras}{Mon.\ Not.\ R.\ Astron.\ Soc.\ }

\newcommand{\aap}{Astron.\ and\ Astrophys.\ }

\newcommand{\beq}{\begin{equation}}
\newcommand{\eeq}{\end{equation}}
\newcommand{\beqn}{\begin{eqnarray}}
\newcommand{\eeqn}{\end{eqnarray}}

\newcommand{\llabel}[1]{\label{#1}}              
\newcommand{\labeq}[2]{ \begin{equation} \llabel{#1}
{#2}
\end{equation}}

\begin{document}

\title{Accretion disks around binary black holes of unequal mass: \\ GRMHD simulations near decoupling}

\author{Roman Gold${}^1$, Vasileios Paschalidis${}^1$, Zachariah B. Etienne${}^{1,2,3,4}$, Stuart L. Shapiro${}^{1,5}$}
\affiliation{
${}^1$Department of Physics, University of Illinois at Urbana-Champaign, Urbana, IL~61801\\
${}^2$Physics Department \& Joint Space-Science Institute, University of Maryland, College Park, MD 20742 \\
${}^3$Gravitational Astrophysics Laboratory, NASA Goddard Space Flight Center, Greenbelt, MD 20771 \\
${}^4$Department of Mathematics, West Virginia University, Morgantown, WV 26506 \\
${}^5$Department of Astronomy \& NCSA, University of Illinois at Urbana-Champaign, Urbana, IL 61801
\vspace{-10pt}
}
\author{and \\ Harald P. Pfeiffer${}^{6,7}$}
\affiliation{
${}^6$Canadian Institute for Theoretical Astrophysics, University of Toronto, ON M5S 3H8, Canada\\
${}^7$Canadian Institute for Advanced Research, Toronto, ON M5G 1Z8, Canada
}

\begin{abstract}

We report on simulations in general relativity of magnetized disks
onto black hole binaries. We vary the binary mass ratio from 1:1 to
1:10 and evolve the systems when they orbit near the binary-disk
decoupling radius. We compare (surface) density profiles, accretion
rates (relative to a single, non-spinning black hole), variability, effective
$\alpha$-stress levels and luminosities as functions of the mass
ratio. We treat the disks in two limiting regimes: rapid radiative cooling and
no radiative cooling.  The magnetic field lines clearly reveal jets emerging
from both black hole horizons and merging into one common jet at large
distances. The magnetic fields give rise to much stronger shock
heating than the pure hydrodynamic flows, completely alter the disk
structure, and boost accretion rates and luminosities.  Accretion
streams near the horizons are among the densest structures; in fact,
the 1:10 no-cooling evolution results in a refilling of the cavity.
The typical effective temperature in the bulk of the disk is $\sim
10^5 (M/10^8 M_\odot)^{-1/4} (L/L_{\rm edd})^{1/4} {\rm K}$ yielding
characteristic thermal frequencies $\sim 10^{15} (M/10^8
M_\odot)^{-1/4} (L/L_{\rm edd})^{1/4}(1+z)^{-1}{\rm Hz} $. These systems are
thus promising targets for many extragalactic optical surveys, such as
LSST, WFIRST, and PanSTARRS.

\end{abstract}

\maketitle

\section{Introduction and Motivation}

Supermassive black hole (SMBH) binaries can form in magnetized plasma
following galaxy mergers, via bar-mode instability in rapidly rotating
supermassive stars, or by other dynamical processes
\cite{Begelman:1980vb}. After formation, a combination of dynamical
friction and gas-driven migration is likely to catalyze the binary
inspiral into the gravitational radiation-driven regime
\cite{Ivanov:1998qk,Haiman:2009te,Rafikov:2012hd,Begelman:1980vb,Armitage:2002uu,Armitage:2005xq,Milosavljevic:2001vi}. The
exact details of these processes, including the ``last-parsec
problem'', remain active areas of research (see
e.g.~\cite{Milosavljevic:2002ht,Merritt:2004gc,Khan:2013wbx} and
references therein). As a result, event rates and population synthesis
studies at this stage are highly uncertain \cite{Dotti:2011um}.

The exciting prospect of a simultaneous observation of both
electromagnetic (EM) and gravitational waves (GWs) arising from
accreting binary BHBHs makes these systems prime targets in the era
of multi-messenger astronomy. Such observations will enable us to
determine the binary masses, BH spins, redshift and even determine the
Hubble constant to better than 1\% \cite{Schutz:1986gp,Holz:2005df,Nissanke13}.

Gravitational waves (GWs) from SMBHs are expected to be detected by planned GW
interferometers such as eLISA/NGO detectors
\cite{AmaroSeoane:2012km}, sensitive to GW frequencies $10^{-5}-1 \rm
Hz$, and the currently operating Pulsar Timing Arrays
\cite{Hobbs:2009yy,Tanaka:2011af,Sesana:2012ak}, sensitive to
frequencies $10^{-9} - 10^{-6}\rm Hz$.  As SMBHs are typically
believed to have masses in the range $10^6-10^9M_\odot$, the GW strain
at orbital separation $d=10M$ - the value adopted in this work - is
$h\sim 10^{-16} (d/10M)^{-1} (M/10^8M_\odot)(D/16 \rm Gpc)^{-1}$. Here
$M$ is the total mass of the binary, and we normalized to a luminosity
distance $D$ corresponding to redshift $z\simeq 2$ in a standard
$\Lambda$CDM Universe. The corresponding gravitational-wave frequency
is $f_{\rm GW} \sim 10^{-4} \bra{M/10^8M_\odot}^{-1}
\bra{d/10M}^{-3/2} [(1+z)/3]^{-1}$Hz. The expected
GW strain is above the eLISA strain sensitivity at these frequencies
\cite{AmaroSeoane:2012km}, hence these systems will be detectable by
eLISA. In particular, for $M\sim 10^6M_\odot$, the value of $f_{\rm
  GW}$ at $d=10M$ falls well within the eLISA sensitivity band even
for larger redshifts, while for $M\sim 10^8M_\odot$ and large
redshifts ($z \sim 10$), the value of $f_{\rm GW}$ at $d=10M$ is
marginally within the eLISA sensitivity band. However, the inspiral
time from these separations is $t_{\rm GW} \sim 20(M/10^8
M_\odot)\rm\ days$ assuming equal masses.  Hence, these
systems will quickly enter the eLISA sweet spot, and EM precursor
signals can trigger targeted GW searches with a convenient lead time
of several days.

While awaiting the first detection of GWs, currently operating and
future electromagnetic (EM) detectors such as LSST
\cite{Abell:2009aa}, WFIRST \cite{Green:2012mj} and PanSTARRS
\cite{Kaiser:2002zz} are promising instruments to identify accreting
BHBH systems in the EM spectrum. Important steps have already been
made toward realizing this goal.

Currently, we know of one spatially resolved SMBH binary candidate at
an orbital separation $d\sim 7$pc: 0402+379 \cite{Rodriguez:2006th}.
Other spatially unresolved, SMBH binary candidates have
been found, including OJ287
\cite{Sillanpaa88,lehto96,Valtonen:2011ny,Tanaka:2013cva} and SDSS
J1536+0441 \cite{Boroson:2009va,Chornock09}.  Binary AGN candidates
have been singled out based on offsets in the broad line and narrow
line regions, emission line profiles, and time variability \cite{Decarli:2013xwa,Eracleous:2011ua}. Recently,
very-long baseline interferometric observations interpreted ejection
components from AGN cores as undulations caused by the precession of
the accretion disks around a SMBH binary \cite{Roland:2013jda}. A
simplified model was applied to two AGN sources; for 1823+568 their
analysis yields $d \sim 0.42$pc, and a mass ratio $q$ in the range $0.095<q<0.25$, while for 3C 279 $d \sim 2.7$pc, $q\sim
0.36$. However, given the lack of a robust circumbinary accretion disk
theory these results are at best preliminary. Nevertheless, they still
motivate a study of accretion flows onto supermassive BH
binaries with different mass ratios, which we initiate in this work in general relativity (GR).

Other features identified as ``smoking guns'' for binary black holes include BH
recoil/kicks \cite{Komossa:2012cy}, used to explain the large
velocity offsets between emission lines in AGN spectra, as well as
observed kinks in jets probably due to changes in BH spin (X-shaped
radio sources), a past merger event, or precession effects
\cite{Romero2000,Roos1993,Sudou:2003hv}. Modifications to the line
profiles have also been proposed as promising characteristic features
to distinguish binary BH AGN from classical, single BH AGN sources
\cite{McKernan:2013cha}.

To assist and solidify all these detection efforts, it is crucial to
identify and model possible electromagnetic ``precursor'' and
``afterglow'' signatures
\cite{Milosavljevic:2004cg,Kocsis:2007yu,Haiman:2008zy,Tanaka:2009iy,Shapiro:2009uy,Liu:2010mh,Shapiro:2013qsa}.

Depending on the physical regime the properties characterizing the gas
can differ considerably, and different accretion models are
applicable. For example, if the gas has little angular momentum, the
accretion flow resembles the binary analog of a Bondi-Hoyle-Lyttleton
solution \cite{bondi44,bondi52,1989ApJ...336..313P} (see
\cite{Farris:2009mt,Zanotti:2011mb,Giacomazzo:2012iv} for GR studies).
If the gas has significant angular momentum, then the gas can become
rotationally supported and form a disk.

For a BHBH embedded in a disk, one can identify several different
regimes based on the time scales for migration of the binary. For a
SMBH binary engulfed by a (thin) disk at large separation, the
migration of the binary is initially governed by binary-disk angular
momentum exchange mediated by (effective) viscosity
\cite{Haiman:2009te}. At large enough separations, the reduced mass
$\mu$ of the binary is less than the local interior disk mass ($4\pi
d^2\Sigma$ where $\Sigma$ is the surface density of the disk). This
leads to the so-called disk-dominated type II migration occurring on
the viscous time scale $t_{\rm vis}$. As the migration proceeds, the
reduced mass of the system becomes larger than the local disk mass and
the migration enters the secondary-dominated type II regime, which
occurs on a longer time scale $t_{\rm sd} \equiv (4\pi
d^2\Sigma/\mu)^{-k} \times t_{\rm vis} \geq t_{\rm vis}$, where $k$
a constant of order $\sim 0.4$ if $4\pi d^2\Sigma/\mu < 1$ and $0$
otherwise.  Ultimately, the binary enters a regime at smaller orbital
separations where angular momentum losses due to GWs dominate, and the
binary migrates on the GW time scale $t_{\rm GW}$.  In all regimes,
the disk moves inwards on the viscous time scale. When $t_{\rm
  vis}\leq t_{\rm sd}<t_{\rm GW}$ or $t_{\rm vis}<t_{\rm GW} < t_{\rm
  sd}$ the disk can follow the inspiral of the binary and settle in a
quasi-steady state - this is called the \textit{predecoupling}
regime. When $t_{\rm GW}<t_{\rm vis}$, the binary decouples from the
disk, i.e.~the inward migration of the binary out-paces the inward
drift of the disk.

In this paper we focus on the phase near \textit{decoupling},
while the postdecoupling regime will be the subject of a future
paper. We note that unlike eLISA, Pulsar Timing Arrays are sensitive
only to SMBH binaries well within the predecoupling
regime.

Accretion onto a single BH has been studied in great detail for
decades, and magnetohydrodynamic studies in GR have drastically
improved our understanding of these flows (see \cite{lrr-2013-1} for a recent review).
Many different disk models have been proposed in
the literature. These models range from geometrically-thin, optically
thick disks \cite{Shakura73,Novikov73} and slim disks
\cite{Abramowicz1988}, to geometrically thick, optically thin,
radiatively inefficient accretion flows (RIAF)
\cite{Ichimaru1977,Blandford:1998qn,Igumenshchev:1997vi,Esin:1997he,Narayan:1996gp}.
However, our understanding of accretion flows onto BHBHs remains poor
and studies of these systems are still in their infancy.

The first analytic Newtonian model and smooth particle hydrodynamic simulation of a
circumbinary accretion disk was given in
\cite{Artymowicz:1994bw}. Since then, other Newtonian (semi-)analytic
studies \cite{Tanaka:2009iy,Shapiro:2009uy,Liu:2010mh,Kocsis:2012ui,Kocsis:2012cs}
and hydrodynamic simulations in 2D
\cite{Artymowicz:1996zz,MacFadyen:2006jx,D'Orazio:2012nz,Farris:2013uqa}, and 3D
\cite{Dotti:2006ef,Cuadra:2008xn,Rodig:2011jz,Roedig:2012nc,Hayasaki:2012wu}
have followed. Newtonian magnetohydrodynamic (MHD) simulations were presented in
\cite{Shi:2011us}, and Post-Newtonian MHD simulations in
\cite{Noble:2012xz}. Many of these earlier studies
excluded the binary and most of the inner cavity from the
computational domain, introducing an artificial inner boundary
condition. The importance of treating the inner regions
self-consistently has been discussed in \cite{Roedig:2012nc}, and only
full GR calculations can achieve this goal reliably.

A ``GR-hybrid'' orbit-averaged model for thin disks, in which the viscous part is handled in GR and
the tidal torques in Newtonian gravity was introduced in
\cite{Shapiro:2013qsa}, and GR hydrodynamical simulations of accretion
onto BHBH binaries - taking into account the dynamical spacetime
- have been performed in
\cite{Bode:2009mt,Farris:2009mt,Bogdanovic:2010he,Bode:2011tq}. To
date the only GR magnetohydrodynamic (GRMHD) simulations of disk accretion onto BHBHs that
account both for the dynamical spacetime and the BH horizons were
presented in \cite{Farris:2012ux}.

Using a different approach by assuming that the B-field is anchored to
a circumbinary disk \textit{outside} the computational domain,
\cite{Mosta:2009rr,Neilsen:2010ax,Palenzuela:2010xn,Palenzuela:2010nf}
modeled EM signatures by solving the GR force-free electrodynamic
equations.

Close to merger a single BH remnant is formed on a time scale much shorter than the
dynamical time scale in the disk. The mass and angular momentum of the
remnant BH is different from the total mass and angular momentum prior
to merger due to GW emission, causing a quasi-instantaneous perturbation to the disk.
This effect has been modeled using hydrodynamical and MHD simulations in
Newtonian gravity and in GR
\cite{Corrales:2009nv,Rossi:2009nk,Anderson:2009fa,Megevand:2009yx,Ponce:2011kv}.

A realistic and ideal 3D global model for a \textit{circumbinary} disk
around a SMBH binary near the decoupling radius requires: a) a fully
relativistic treatment of gravitation in a \textit{dynamical}
spacetime, b) GRMHD for the plasma flow, c) realistic cooling
processes, and d) radiative transfer in curved spacetime. Simulations
incorporating these effects must also have high resolution and long
integration times (several viscous time scales). However, including all
these ingredients in one simulation would make these computations
prohibitive, because the wall-clock times required to integrate for
even $\sim 5$ viscous time scales at the inner disk radius at high
resolution are far too long with current supercomputer resources. Thus,
some of these ingredients must be relaxed in order to obtain a
qualitative understanding of these systems. For this reason, the
models in this work feature a) and b). High resolution is only adopted
for a few models. In addition, we model radiative cooling by a simple
cooling function and consider the extreme opposite limits of ``rapid''
cooling and ``no cooling'' to bracket the possibilities.

In this paper we study the effects of the binary mass ratio on the
disk near decoupling. We vary the BHBH binary mass ratio $q \equiv
M_1/M_2 \leq 1$, considering 1:1, 1:2, 1:4, 1:8 and 1:10 mass ratios. The mass
ratio regime $0.1\lesssim q\lesssim 1$ is shown to be of high
astrophysical relevance in \cite{Gergely:2007ny,Sesana:2007sh}. Also,
it has been argued that BHBHs forming in major galaxy mergers will
typically have mass ratios in the range $(0.01, 1)$
\cite{Sesana:2011zv}. These results motivate our choice of mass
ratios. Note that Newtonian simulations studying the effects of mass
ratio in the context of 2D thin-disk models were performed
in~\cite{D'Orazio:2012nz,Farris:2013uqa}.

The new aspects of our work are the inclusion of relativistic
gravitation to resolve the crucial physics near the BH horizons,
effective viscosity arising from the magnetorotational
instability \cite{Balbus:1991ay,Balbus:1998ja}, geometrically thick
disks, three spatial dimensions (3D), a $\Gamma$-law equation of state (EOS),
and effective cooling.  We model the decoupling epoch by fixing
the binary separation $d$, evolving the spacetime by rotating our
conformal-thin-sandwich initial data~\cite{Pfeiffer:2002iy,Cook:2004kt,Caudill:2006hw,BSBook} as we have done in
\cite{Farris:2011vx,Farris:2012ux}. The dependence of the flow
variability, EM signatures, the magnetic field structure and the
matter dynamics inside the low-density ``hollow'' on the mass ratio
$q$ are investigated. We estimate thermodynamic properties of the gas
and scale our results with the binary total mass and the
luminosity in units of the Eddington luminosity where feasible. From
this, emission characteristics including typical thermal frequencies and
luminosities are given and relevant detectors are discussed.

In the absence of any observational constraints on the thermodynamic
state of accreting BHBHs, the models we consider in this work adopt an
adiabatic EOS governed by an adiabatic index
$\Gamma=4/3$, appropriate for radiation pressure-dominated, optically
thick disks. This choice is motivated in part both by theory and
observations. First, accretion disk theory \cite{lrr-2013-1} predicts
that the inner part of circumbinary disks around SMBH binaries are
optically thick, radiation pressure dominated for a large set of
possible disk parameters (see e.g. \cite{Novikov73,Rafikov:2012hd} and the
next section). Furthermore, as discussed in \cite{Haiman:2009te},
steady-state disk models predict that radiation pressure-dominated
circumbinary disks will channel more material into the cavity for a
given central mass. This finding makes radiation pressure-dominated
circumbinary disks promising sources for electromagnetic counterparts.

Second, AGN data from the Sloan Digital Sky Survey (SDSS)
\cite{Heckman:2004zf,Shen:2011ge,Kelly:2012vz}, the AGN and Galaxy
Evolution Survey (AGES) \cite{Kollmeier:2005cw}, XMM-COSMOS
\cite{Lusso:2012yv} and others \cite{Schulze:2010wc} covering mostly
type I AGNs and the local Universe to redshift of $z\lesssim 5$,
reveal Eddington ratio distributions in the range $0.01 \lesssim
L/L_{\rm Edd} \lesssim 1$ with the tendency that higher $L/L_{\rm Edd}$ values
occur at higher $z$ (and possibly smaller central mass $M$). A
comparison of accretion time scales with the age of the Universe
suggests that earlier accretion episodes are closer to the Eddington limit,
similar to the recently discovered quasar at $z\sim 7.1$ with a
central mass of $M \sim 2\times 10^9M_\odot$ \cite{Mortlock:2011va}
accreting at $L/L_{\rm Edd}\sim 1.2 \pm 0.5$. These surveys therefore
motivate studies of disks accreting near the Eddington
luminosity, for which radiation pressure is important. Moreover,
radiation pressure-dominated disks accreting near the Eddington limit
are more likely to be detectable, even at large cosmological redshifts.

This paper is structured as follows: In Sec.~\ref{sec:qualitative} we
present a qualitative discussion of the range of parameters and the
associated physical regimes for which our simulations are
appropriate. In Sec.~\ref{sec:methods} we describe our methods and
techniques for evolving the spacetime, fluid, and magnetic fields, as
well as our simple cooling prescription. We also present the different
cases we consider and list our diagnostics for characterizing the
accretion flow and EM signatures. In Sec.~\ref{sec:tests} we show
several tests we performed to motivate our numerical setup. In
Sec.~\ref{sec:results} we present the results from our simulations. We
conclude in Sec.~\ref{sec:conclusions} with a summary of the main
results and a discussion of future work. 

Here and throughout we adopt geometrized units, where $G=c=1$, 
unless otherwise stated.

\section{Qualitative considerations}
\label{sec:qualitative}

In this section we use the Shakura-Sunyaev/Novikov-Thorne thin-disk
model \cite{Shakura73,Novikov73} to make rough analytic estimates
regarding the physical regime our disk models probe.  While the
Shakura-Sunyaev/Novikov-Thorne model strictly applies to a viscous
flow onto a single BH neglecting the binary tidal torques, we expect
that our qualitative analysis will apply roughly to our circumbinary
disk models, which have $H/R = \mathcal{O}(0.1)$, where $H$ is the disk scale
height. This expectation should be best realized outside the binary
orbit because the ratio of the tidal to viscous torques decays quickly
far away from the binary orbit (see e.g.~Fig.~6 in
\cite{Shapiro:2013qsa}).

\subsection{Radiation pressure dominance}

\subsubsection{Shakura-Sunyaev/Novikov-Thorne model}
The simulations reported here apply to {\it any} total (ADM)
binary black hole mass $M$ and, since we neglect the self-gravity
of the gas, to {\it any} rest-mass density $\rho_0$, provided it has the 
same initial disk profile adopted here. The quantities that are fixed, 
in addition to the initial disk profile, are the adiabatic index
$\Gamma$ appearing in the adopted ideal gas EOS, the ratio of the initial
magnetic-to-total disk pressure, the initial magnetic field profile, the initial 
binary separation and the cooling law. In this section we use the steady-state
thin-disk solution for a Shakura-Sunyaev/Novikov-Thorne disk about a single BH 
of mass $M$ to estimate the physical values of some of the gas dynamical quantities
as functions of $M$ and luminosity $L$. We show below that for a range of astrophysically
relevant choices of these parameters the disk is thermal radiation pressure-dominated, 
and this fact motivates our setting $\Gamma=4/3$.  

Neglecting the perturbative role of the secondary tidal torque, the
steady-state accretion flow in a geometrically thin disk is uniquely
specified by the Shakura-Sunyaev viscosity parameter $\alpha$, the
central mass $M$ and the accretion rate $\dot{M}$. The quantity $\dot M$
is specified in turn by the disk luminosity $L$ and efficiency $\varepsilon$.
The Shakura-Sunyaev/Novikov-Thorne model \cite{Shakura73,Novikov73}
describes a disk that is radiation pressure-dominated inside a radius
$r_{\rm inner}$. In this region the radiation to gas pressure ratio
is~\cite{lrr-2013-1}
\beqn
\frac{P_{\rm rad}}{P_{\rm gas}} & \sim & 5.4\times 10^5 
                   \bra{\frac{r}{20M}}^{-21/8}
                   \bra{\frac{\alpha}{0.1}}^{1/4} \nonumber\\
          &      & \bra{\frac{M}{10^8 M_\odot}}^{1/4} 
                   \bra{\frac{L}{L_{\rm edd}}}^{2} 
                   \bra{\frac{\varepsilon}{0.1}}^{2},
                   \label{eq:Prad_over_Pgas}
\eeqn
where $\varepsilon\equiv L/\dot M c^2$ is the radiative efficiency, 
$L_{\rm Edd} \simeq 1.3\times 10^{46} (M/10^8M_\odot) \rm erg/s$,
and we chose the normalization for the $\alpha$ parameter based
on typical values found in our simulations.  For a thin-disk model the
efficiency ranges from $0.057$ for a non-spinning BH to $0.42$ for a
maximally spinning BH, so we scale to a value residing between these
limits. 
The size of the inner region is determined by the condition
$P_{\rm rad}/P_{\rm gas}=1$, for which Eq. \eqref{eq:Prad_over_Pgas}
yields
\beqn
r_{\rm inner}  / M
        &\sim & 3000
        \bra{\frac{\alpha}{0.1}}^{2/21}
        \bra{\frac{M}{10^8 M_\odot}}^{2/21} \nonumber\\
        & & \bra{\frac{L}{L_{\rm edd}}}^{16/21} 
        \bra{\frac{\varepsilon}{0.1}}^{-16/21}.
\label{eq:Prad}
\eeqn

 The geometrically thick disks we evolve in this work extend radially
 out to $r_{\rm out}\sim 100M-200M$. Hence, when scaling the accretion
 rate such that our models accrete near the Eddington limit, our
 models are fully immersed in this inner radiation pressure-dominated
 region.

In this region the typical rest mass densities are~\cite{lrr-2013-1}
\beqn
\rho_0 & \sim & 5.5 \cdot 10^{-12}
                   \bra{\frac{r}{20M}}^{3/2}
                   \bra{\frac{\alpha}{0.1}}^{-1} \nonumber\\
          &      & \bra{\frac{M}{10^8 M_\odot}} 
                   \bra{\frac{\dot M}{2.3M_\odot/\rm yr}}^{-2}\frac{\rm g}{\rm cm^3}\\
      & \sim & 5.5 \cdot 10^{-12}
                   \bra{\frac{r}{20M}}^{3/2}
                   \bra{\frac{\alpha}{0.1}}^{-1} \nonumber\\
          &      & \bra{\frac{M}{10^8 M_\odot}}^{-1} 
                   \bra{\frac{L}{L_{\rm edd}}}^{-2} 
                   \bra{\frac{\varepsilon}{0.1}}^{2} \frac{\rm g}{\rm cm^3}.
\eeqn

As we will show later, these characteristic values for $P_{\rm
  rad}/P_{\rm gas}$ and $\rho_0$ are comparable to the
values found in our simulations. 

The Shakura-Sunyaev/Novikov-Thorne model predicts that the effective optical depth to
absorption is $\tau^* \simeq 0.02$ adopting the same normalizations as
in the equation above.
This well-known inconsistency near the Eddington limit of the
Novikov-Thorne model is removed with the generalization to the slim
disk model \cite{Abramowicz1988}. This model differs from a thin disk
in that it allows for cooling to occur via advection, which dominates
radiative cooling at high accretion rates ($L \gtrsim 0.3 L_{\rm
  Edd}$), thereby puffing up the disk.  When scaling our models to
accrete near the Eddington limit, they are closer to slim-disk models,
which remain optically thick in this high luminosity limit
\cite{Abramowicz1988,lrr-2013-1}.

As we discuss later, near the Eddington limit and $M=10^8M_\odot$,
the effective optical depth satisfies $\tau^\star\gtrsim 1$ in the
bulk of our disk models, implying that the gas is optically thick to
absorption and the photons eventually \textit{thermalize}. Thus, these
qualitative considerations motivate the adoption of an adiabatic index
$\Gamma=4/3$, appropriate for thermal radiation pressure dominance.

Note also that alternative disk models have been proposed,
e.g.~\cite{Rafikov:2012hd}. However, they largely share the prediction
that the inner regions of the disk are radiation pressure dominated.

\subsubsection{Decoupling radius}
\label{sec:decoupling}
We estimate the decoupling radius $a_d$ by equating $t_{\rm GW}=t_{\rm
  vis}$ and solve for the separation to find \cite{Farris:2012ux}
\labeq{dec_rad}{
\frac{a_d}{M} \approx  13.3\bigg(\frac{\alpha}{0.1}\bigg)^{\hspace{-4pt}-2/5} \bigg(\frac{H/R}{0.3}\bigg)^{\hspace{-4pt}-4/5} \bigg(\frac{\tilde{\eta}}{1}\bigg)^{\hspace{-4pt}2/5} 
}
where we assumed that the inner disk edge radius settles to $r_{\rm in}\approx 1.5d$ as
typically found in our simulations, $\alpha$ is the
Shakura-Sunyaev turbulent viscosity parameter, and $\tilde{\eta}\equiv
4q/(1+q)^2$ (see also \cite{Armitage:2002uu}). Notice that in contrast to 
geometrically thick disks, where the decoupling radius is a few tens of $M$, geometrically 
thin disks have $a_d/M \simeq 100$.
The decoupling radius estimate \eqref{dec_rad} for the mass ratios considered in this work
ranges from $a_d(q=0.1) \approx 8.5M$ to $a_d(q=1)\approx 13.3M$. The
normalizations in Eq. \eqref{dec_rad} are based on typical values
obtained from our simulations.  We choose the binary separation $d\sim 10M$ for all cases
studied in this work, a value which is consistent with the crude estimate of Eq.~\eqref{dec_rad}.
In the future we intend to start our evolutions at larger
separations in order to dynamically determine the decoupling radius and evolve through it
as in \cite{Shapiro:2013qsa}.

\section{Methods}
\label{sec:methods}

The models we adopt here assume: a) circular binary orbits,
neglecting the binary inspiral (justified for large separations; see
Sec.~\ref{sec:decoupling}), b) non self-gravitating disks, which
likely is a good assumption (see, e.g. \cite{Goodman:2002gv}), c)
ideal MHD, d) no radiative feedback, e) an
effective cooling scheme that brackets no cooling and rapid cooling.

Some of these assumptions may not be obeyed strictly, e.g. the binary
may become eccentric in the predecoupling
regime~\cite{Armitage:2005xq,Dotti:2006ef,Cuadra:2008xn,Roedig:2011rn,Rodig:2011jz,Roedig:2012nc},
or radiative feedback may become important near Eddington accretion
rates. However, our simulations still provide a qualitative
understanding of the physics that will be useful in designing the next
set of more realistic models of binary black holes immersed in
circumbinary disks. In this section, we describe the initial data and
computational methods we adopt to account for a)-e).

\subsection{Initial data and AMR hierarchy}
\label{sec:initialdata}

\subsubsection{Metric initial data}
\label{sec:tests1}
At large separation the binary inspiral time scale is much longer than
the binary orbital period and the viscous time scale at the inner edge
of the disk just beyond the binary orbit. Accordingly, the inspiral can be
neglected over many orbital periods. To model this epoch in GR, we
adopt quasi-equilibrium conformal-thin-sandwich (CTS) solutions for
the black hole
binary~\cite{Pfeiffer:2002iy,Cook:2004kt,Caudill:2006hw,BSBook}.  The
spacetime initial data satisfying the CTS equations correspond to a
circular orbit and possess a helical Killing vector. The CTS initial
data have been generated using the spectral techniques described in
\cite{Pfeiffer:2002wt} as implemented in the Spectral Einstein Code
(SpEC) \cite{SpECwebsite} (see also \cite{Mroue:2013xna}). We list the
initial data parameters describing our spacetimes in
Table~\ref{tab:cts_evolve}.

\begin{center}
 \begin{table}[th]
  \caption{CTS initial data parameters for the BHBH vacuum spacetime.
    Columns show mass ratio ($q$), ADM mass ($M_{\rm ADM}$) and
    angular momentum ($J_{\rm ADM}$), and irreducible masses
    ($M^i_{irr}$), and apparent horizon radii ($r^i_{hor}$) for the two
    black holes. Diagnostics generating these quantities, but computed
    from vacuum, test simulations agree with these values to within
    one part in $10^4$.
\label{tab:cts_evolve}}
  \begin{tabular}{ccccccc} \hline\hline

 $q$   & $M_{\rm ADM}$ & $J_{\rm ADM}$ &  $M^1_{\rm irr}$ &  $M^2_{\rm irr}$ & $r^1_{\rm hor}$ & $r^2_{\rm hor}$ \\ \hline 
$1:1$    & 0.98989 & 0.96865 & 0.50000 & 0.50000 & 0.42958 & 0.42958 \\ 
$1:2$  & 0.99097 & 0.85933 & 0.66667 & 0.33333 & 0.60192 & 0.27312 \\ 
$1:4$  & 0.99342 & 0.61603 & 0.80000 & 0.20000 & 0.75140 & 0.15832 \\ 
$1:8$  & 0.99589 & 0.37868 & 0.88889 & 0.11111 & 0.85640 & 0.08618 \\ 
$1:10$ & 0.99656 & 0.31652 & 0.90909 & 0.09091 & 0.88081 & 0.07022 \\ \hline\hline 
  \end{tabular}
 \end{table}
\end{center}

\subsubsection{Matter and B-field initial data}
\label{sec:tests2}
For the fluid we use the same family of equilibrium disk
models around single BHs as in \cite{Chakrabarti:1985,DeVilliers:2003gr,Farris:2011vx}.
We choose the initial inner disk edge radius $r_{in,0}=18M$ and
specific angular momentum $l(r_{in,0})=5.15 M^2$ around a non-spinning
BH as in \cite{Farris:2012ux}. However, the disk model
is not identical to \cite{Farris:2012ux} due to the different
polytropic index ($\Gamma=4/3$ here versus $\Gamma=5/3$ in
\cite{Farris:2012ux}).

\begin{figure}[h!]
    \includegraphics[trim =0cm 0cm 1.15cm 0cm,clip=True,width=0.45\textwidth]{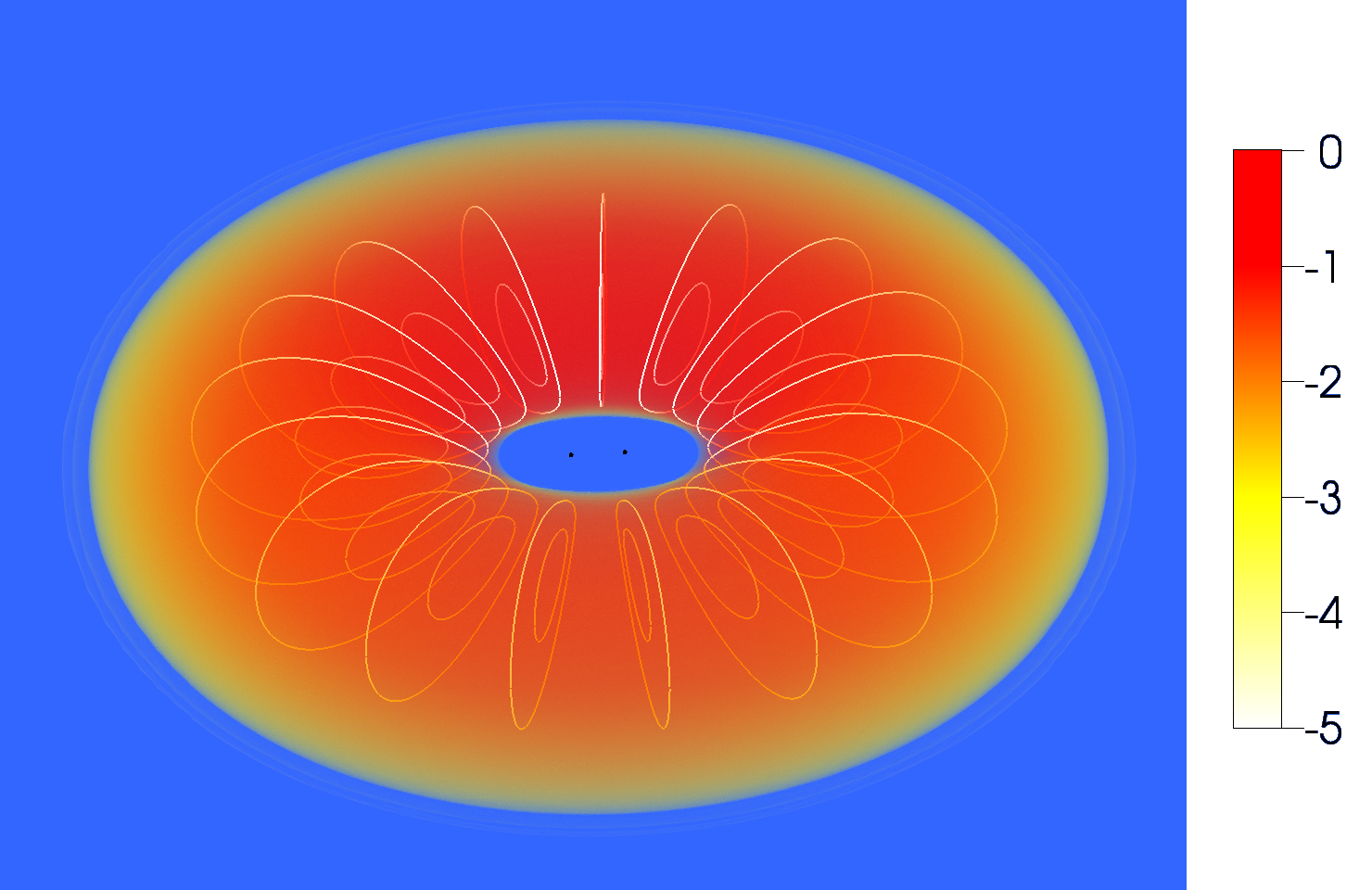}
      \caption{B-field lines (white curves) and volume rendering of
        rest-mass density normalized to its maximum value at
        $t=0$. The two black dots at the center indicate the BH
        apparent horizons for the $q=1$ case.
            \label{fig:visit-bfield-id}}
\end{figure}

We seed the initial disk with a small, purely poloidal B-field using
the same procedure as in \cite{Farris:2012ux,Etienne:2011ea}; see
Fig.~\ref{fig:visit-bfield-id}. The field is dynamically unimportant
initially: the initial maximum value for the ratio of magnetic $P_M$
to total pressure $P$ is 0.025. All cases we consider in this work are
initialized with the same disk and magnetic field.

Although we will qualitatively discuss how the evolution of these matter
initial data depends on the binary mass ratio, it is important to stress
that the goal of our work is to assess how the {\it final relaxed state} 
of the disk depends on the binary mass ratio. The initial data, which
governs the early evolution, has no physical significance.

\subsection{Evolution equations and techniques}

\subsubsection{GRMHD evolution}

We use the Illinois GRMHD adaptive-mesh-refinement (AMR) code
\cite{Duez:2005sf,Etienne:2010ui,Etienne:2011re}, which adopts the
Cactus/Carpet infrastructure
\cite{Goodale2002a,Carpet-cactusweb,Carpet-carpetweb}. This code has
been extensively tested and used in the past to study numerous systems
involving compact objects and/or magnetic fields (see
e.g.~\cite{Paschalidis:2011ez,Etienne:2011ea,Etienne:2012te,Paschalidis:2012ff,Etienne:2013qia,Paschalidis:2013jsa}),
including black hole binaries in gaseous media
\cite{Farris:2009mt,Farris:2011vx,Farris:2012ux}.

The code solves the equations of ideal GRMHD in a flux-conservative
formulation [see Eqs. (27)-(29) in \cite{Etienne:2010ui}] employing
a high-resolution-shock-capturing scheme (see
\cite{Duez:2005sf,Etienne:2010ui} for details), and including
effective cooling source terms [see Eqs.~(65) and (66) in
\cite{Paschalidis:2011ez}]. To enforce the zero-divergence constraint
of the magnetic field, we solve the magnetic induction equation
using a vector potential formulation 
[see Eq.~(9) in
\cite{Etienne:2011re}]. As our EM gauge choice we use the generalized
Lorenz gauge condition 
we developed in \cite{Farris:2012ux} and used in
\cite{Etienne:2012te,Paschalidis:2013jsa}. We choose the damping
parameter $\xi=8/M$. The advantage of this gauge condition is that it
leads to \textit{damped}, propagating EM gauge waves preventing
spurious magnetic fields from arising near AMR boundaries even more
effectively than the standard Lorenz gauge choice ($\xi=0$)
\cite{Etienne:2011re}. The damping properties of the generalized
Lorenz gauge are crucial for stable and accurate long-term GRMHD evolutions.
The ``algebraic'' gauge condition used in the first GRMHD simulations
adopting A-field evolution (see
e.g.~\cite{Etienne:2010ui,Giacomazzo:2012iv,Rezzolla:2011da}) was
shown in \cite{Etienne:2011re} to suffer from spurious conversion of
EM gauge modes into physical modes and vice-versa, due to
interpolation at AMR boundaries. These spurious magnetic fields
contaminate the evolution and the effect is exacerbated when matter
crosses refinement boundaries.

To close the system of equations we use a $\Gamma$-law EOS
\labeq{}{
P=\rho_0\epsilon (\Gamma-1),
}
where $P$ is the total pressure, $\rho_0$ the rest-mass density, and
$\epsilon$ the specific internal energy. This EOS allows
for shock heating. We choose $\Gamma=4/3$ appropriate for
radiation pressure-dominated, optically thick disks.

\subsubsection{Metric evolution}

The metric evolution is treated under the approximation that
the inspiral time scale due to GW emission is long compared to both
the binary orbital period and the viscous time scale of the
disk. Hence, we can neglect the inspiral for multiple binary orbits. The CTS initial data we
adopt possess a helical Killing vector, which implies that the
gravitational fields are stationary in a frame corotating with the
binary. As a result, we can perform the metric evolution in the
center-of-mass frame of the binary by simply rotating the initial
spacetime fields as was done in
\cite{Farris:2009mt,Farris:2012ux}. This technique simplifies our
computations substantially. In addition, the rotating metric method
facilitates our evolving dynamically to relaxed disk/magnetic field
initial data for the inspiral.

To implement this method, we map the CTS solution from the spectral
grid onto three grids corresponding to three partially overlapping
spherical coordinate systems: One spherical coordinate system covers
the entire evolution domain and is centered on the binary center of
mass, and two smaller ones are centered on each BH. These new grids
employ a logarithmic radial coordinate. We use the CTS solution stored on
these spherical grids to interpolate the data onto the Cartesian evolution grids
whenever we perform the rotation transformation.

We have checked that the mapping from the spectral grid to the
spherical grids is implemented correctly by performing vacuum
simulations that use the CTS solution stored in the spherical grids as
initial data. More specifically, we have computed several diagnostic
quantities which characterize the BHs and the global spacetime and
compared them with the values known from the spectral CTS initial data
(see Table~\ref{tab:cts_evolve}). These agree to within 1 part in
$10^4$. Moreover, we have verified that a crude estimate for the
orbital frequency of the binary (orbital trajectory traverses a full
phase of $2\pi$) as determined by a dynamical vacuum evolution agrees
with the value given by the initial data ($M\Omega=0.028$) to within
$\sim 10\%$. We have computed the normalized $L_{2,N}$ norm of the
Hamiltonian and momentum constraint violations as introduced in
Eqs.~(59) and (60) in \cite{Duez:2002bn}, with the modification that
we split up the Laplacian operator into its individual components when
computing normalized norms. We find the normalized norm of the
Hamiltonian constraint to be dominant, with $L_{2,N}(\mathcal{H})\sim
2\%$. We conclude that the CTS solutions are mapped correctly and
accurately.

\subsubsection{Cooling}
Without cooling, the binary tidal and the viscous torques act to
gradually heat and puff up the disk.  ``Advective'' cooling, which 
is crucial in slim-disk and ADAF models \cite{lrr-2013-1}, 
is self-consistently accounted for in our simulations. However, adding
radiative cooling may be necessary to achieve a steady state. Realistic radiative cooling
based on actual physical mechanisms depends on complicated
microphysics, which we do not model here, but intend to incorporate in
future studies.  To model steady-state solutions in this work, we
introduce ``radiative'' cooling via an effective cooling leakage scheme. This
scheme is, strictly speaking, valid in the optically thin
regime. While this is a very crude approach to
radiative cooling, treating both extremely rapid radiative cooling as well as
no radiative cooling can help bracket the possibilities. Different formulae for the
cooling emissivity $\Lambda$ have been proposed in the literature:
\begin{enumerate}[I.]
    \item Non-exponential cooling \cite{Shafee:2008mm,Noble:2012xz}:
      \labeq{}{ \Lambda = \frac{\rho_0 \epsilon}{\tau_{\rm cool}}
        \big( \frac{\Delta S}{S_0} + \big| \frac{\Delta S}{S_0} \big|
        \big), 
      }
      where $S\equiv K=\frac{P}{\rho_0^\Gamma}$ is an entropy
      parameter, and $S_0$ the initial target value. We call this
      emissivity ``non-exponential'', because the effective cooling
      time scale for this scheme is not just $\tau_{\rm cool}$, but
      depends on the internal energy $\epsilon$ (see
      Appendix~\ref{Appendix:Courant-stability}).
      \label{k-driver}
    \item Exponential cooling
      \cite{Paschalidis:2011ez,Paschalidis:2012ff} 
      \labeq{}{
      \Lambda =
      \frac{\rho_0 \epsilon_{th}}{\tau_{\rm cool}} = \frac{\rho_0
        \epsilon_0}{2\tau_{\rm cool}} \big( \frac{\Delta S}{S_0} +
      \big| \frac{\Delta S}{S_0} \big| \big),
      }
      where $\epsilon_0$ is the internal energy calculated using
      $S_0$, and $\epsilon_{th} = \epsilon -\epsilon_0$.  In this
      scheme the effective cooling time scale is $\tau_{\rm cool}$.
       \label{exp-cooling}
\end{enumerate}
Both emissivities dissipate all shock-induced thermal energy, driving
the entropy of the gas to its initial value.  We use prescription
\ref{exp-cooling}, instead of \ref{k-driver}, because we have found
that prescription \ref{k-driver} is prone to the development of a
Courant instability, as the \textit{effective} cooling time scale of
this scheme depends on the amount of shock heating, which can be very
strong in low-density regions (see e.g. Appendix B in
\cite{Etienne:2008re}). Thus, to stabilize the simulations with
prescription \ref{k-driver}, one typically excludes cooling of the
low-density regions or unbound matter
\cite{Noble:2012xz,Farris:2012ux}. As both the BH horizons and the
low-density cavity is included in our computational domain (unlike
earlier studies), we find that the strong shock heating inside
the cavity in conjunction with emissivity \ref{k-driver} leads to a
numerical instability. In order to bracket the effect of cooling, this
inner cavity needs to be cooled when cooling is enabled.  In
Appendix \ref{Appendix:Courant-stability}, we present an analytic
calculation illustrating the above considerations. We demonstrate
that shock heating of matter in the cavity is important in Sec. \ref{subsec:cooling}.

To model ``rapid'' radiative cooling we set the cooling time scale equal to 10\% of the
local, Keplerian time scale $\tau_{\rm cool}/M = 0.1\tau_{\rm Kep}/M=
0.1 \cdot 2\pi (r/M)^{3/2}$, where $r$ is the cylindrical radial
coordinate measured from the center of mass of the binary. In order to
prevent the cooling time from becoming prohibitively small as $r
\rightarrow 0$ we floor the cooling time at $\tau_{\rm cool} \geq
10M$. Throughout this paper we refer to cases with $\Lambda \neq 0$ as the cooling
cases and $\Lambda =0$ as the no-cooling cases.

\begin{center}
 \begin{table*}[th]
  \caption{List of grid parameters for all models. Equatorial symmetry
    is imposed in all cases. The computational mesh consists of two
    sets of nested AMR grids, one centered on each BH, with the outer
    boundary at $240$M in all cases. From left to right the columns
    indicate the case label, mass ratio $q$, whether cooling is
    included or not, the coarsest grid spacing $\Delta x_{\rm max}$,
    number of AMR levels around the primary (BH) and the secondary
    (bh), and the half length of each AMR box centered on each BH. The
    grid spacing of all other levels is $\Delta x_{\rm
      max}/2^{n-1},\ n = 1, 2,\ldots, n_{\rm max}$, where $n$ is the
    level number such that $n = 1$ corresponds to the coarsest
    level. A dash ``--''~indicates ``no information available''. 
    \label{tab:models}}
  \begin{tabular}{cccccccc} \hline\hline
Case    & $q$ & cooling? & $\Delta x_{\rm max}$ &\ levels(BH)\ &\ levels(bh)\ & Grid hierarchy \\ \hline
1:1nc-hr \hspace{-0.01cm} & 1:1  & No & $4.8M$ & 7 & 7 & $240M/2^{n-1},\ n =2,\ldots 5$, $240M/2^{n},\ n =6,7$ \hspace{0.5cm} \\
1:1nc-mr                 & 1:1   & No & $6.0M$ & 7  & 7  & $240M/2^{n-1},\ n =2,\ldots 5$, $240M/2^{n},\ n =6,7$ \hspace{0.5cm} \\
1:1nc-lr \hspace{0.08cm} & 1:1   & No & $8.0M$ & 7  & 7  & $240M/2^{n-1},\ n =2,\ldots 5$, $240M/2^{n},\ n =6,7$ \hspace{0.5cm} \\
1:2nc-lr \hspace{0.08cm} & 1:2  & No & $8.0M$ & 7 & 8 & $240M/2^{n-1},\ n =2,\ldots 5$, $240M/2^{n},\ n =6,7,8$ \hspace{0.2cm} \\
1:4nc-lr \hspace{0.08cm} & 1:4  & No & $8.0M$ & 7 & 9 & $240M/2^{n-1},\ n =2,\ldots 5$, $240M/2^{n},\ n =6,\ldots, 9$ \hspace{-0.1cm} \\
1:8nc-lr \hspace{0.08cm} & 1:8  & No & $8.0M$ & 6 & 10 & $240M/2^{n-1},\ n =2,\ldots 5$, $240M/2^{n},\ n =6,\ldots, 10$\hspace{-0.15cm}  \\
1:10nc-lr & 1:10 & No & $8.0M$    & 6  &  10  & \hspace{0.03cm} $240M/2^{n-1},\ n =2,\ldots 5$, $240M/2^{n},\ n =6,\ldots, 10$ \\
0nc-hr \hspace{0.22cm} & 0 & No & $4.8M$    & 6  &  --  & $240M/2^{n-1},\ n =2,\ldots 5$, $240M/2^{n},\ n =6$ \hspace{0.8cm} \\
0nc-mr \hspace{0.13cm} & 0 & No & $6.0M$    & 6  & --   & $240M/2^{n-1},\ n =2,\ldots 5$, $240M/2^{n},\ n =6$  \hspace{0.8cm} \\
0nc-lr \hspace{0.30cm} & 0 & No & $8.0M$    & 6  & --   & $240M/2^{n-1},\ n =2,\ldots 5$, $240M/2^{n},\ n =6$ \hspace{0.8cm} \\ \hline
1:1c-mr \hspace{0.08cm}  & 1:1   & Yes & $6.0M$ & 7  & 7  & $240M/2^{n-1},\ n =2,\ldots 5$, $240M/2^{n},\ n =6,7$ \hspace{0.5cm} \\
1:2c-lr \hspace{0.25cm}  & 1:2   & Yes & $8.0M$    &  7 & 8  & $240M/2^{n-1},\ n =2,\ldots 5$, $240M/2^{n},\ n =6,7,8$ \hspace{0.2cm} \\ 
1:4c-lr \hspace{0.25cm}  & 1:4     & Yes & $8.0M$     & 7  & 9  &  $240M/2^{n-1},\ n =2,\ldots 5$, $240M/2^{n},\ n =6,\ldots, 9$\hspace{0.05cm} \\ 
1:8c-lr \hspace{0.25cm}  & 1:8     & Yes & $8.0M$     & 6  &  10  & \hspace{0.03cm} $240M/2^{n-1},\ n =2,\ldots 5$, $240M/2^{n},\ n =6,\ldots, 10$ \\ 
1:10c-lr \hspace{0.08cm}  & 1:10     & Yes &  $8.0M$    & 6  & 10   &\hspace{0.03cm} $240M/2^{n-1},\ n =2,\ldots 5$, $240M/2^{n},\ n =6,\ldots, 10$\\ 
0c-lr \hspace{0.49cm} & 0 & No & $8.0M$    & 6  &  --  & $240M/2^{n-1},\ n =2,\ldots 5$, $240M/2^{n},\ n =6$ \hspace{0.8cm} \\ \hline\hline
  \end{tabular}
 \end{table*}
\end{center}

\subsubsection{Evolution grids \& models}
We use a hierarchical, box-in-box adaptive mesh provided by the
Cactus/Carpet infrastructure \cite{Carpet-cactusweb,Carpet-carpetweb}.
We constructed two sets of nested boxes, with one set centered on each
BH, on which we discretize the GRMHD evolution equations. The coarsest
level has an outer boundary at $r=240M$. Due to a range of resolution
requirements related to the different sizes of the BHs, different
models use different number of refinement levels, which in turn yields
different finest grid spacings (see Table~\ref{tab:models}). We set up
the locations of our AMR boundaries such that the computational grids
resolve both the BHs and the inner disk region for the given
resources.  The grid spacing is also motivated by both MRI resolution
requirements and the results gleaned from test runs involving
hydrodynamic disk evolutions around a single, non-spinning BH, which
we report on in Sec.~\ref{sec:tests}.

In Table~\ref{tab:models} we also list the distinguishing characteristics of the
different cases we consider in this work. The labels are chosen to designate
the mass ratio, whether cooling is applied or not, and the resolution. For example, 
the label 1:1nc-hr means mass ratio $q=1$, no-cooling, and high resolution. 

We point out that the disparity in length scales (horizon vs.~disk
size) and time scales (the Courant condition vs.~viscous time scale)
intrinsic to the circumbinary BHs disk problem introduces a large
computational cost. Most of our simulations were run continuously (excluding
queue waiting times) for $\sim 2$ months on {\tt Blue Waters}, {\tt Kraken}, {\tt
  Lonestar}, as well as the Illinois Relativity group 36-node Beowulf
cluster. The CPU hours used depended on the computer cluster, but we
estimate that the simulations presented here required $\sim 2\times
10^6$ CPU hours.

\subsection{Diagnostics}
\label{sec:diagnostics}

We distinguish two types of diagnostics. The first type consists of
probes of the MHD flow, including the density and velocity profiles,
accretion rates, luminosity estimates, magnetic field profiles, the establishment of MRI
turbulence and jets, etc.  The second type concerns
properties of the plasma such as local temperatures, optical depths, 
characteristic frequencies of emitted radiation, etc.  The first type are
straightforward to calculate from our simulation data, as they depend
on the overall MHD behavior of the disk, and once we have chosen an
EOS and cooling prescription, are independent of
detailed microphysics. We are quite careful and confident in reporting
these diagnostic quantities. The second type depends crucially on the
specific physical values we assign to our nondimensional input
parameters (e.g. BH mass and disk density) and to the microphysics
that is not incorporated in our calculation (e.g. realistic radiative
cooling and radiative transport). Nevertheless, we can make crude
estimates for the latter quantities, and do so below. Although
considerable caution must be applied to these estimates, they may
serve as useful guides to subsequent, more detailed investigations and
to astronomical instruments that may be able to observe the scenarios
we are simulating.

The first type of diagnostics includes: 
1) Accretion rate $\dot{M}$ as defined in \cite{Farris:2009mt}. We compute the total
accretion rate onto the binary, and also the accretion rate onto each
individual BH. 
2) Fourier analysis of the accretion rate
$FFT(\dot{M})$, targeted to identify possible quasiperiodic signatures in
the accretion flow. 
3) Surface density profile $\Sigma(r)$ as defined
in \cite{Farris:2011vx}. This diagnostic is also useful to compare
with studies of 1D orbit-averaged disk models.
4) Shakura-Sunyaev $\alpha$-stress parameter computed as $\alpha \sim
\alpha_{\rm EM} \equiv \av{\frac{\av{T_{\hat r\hat \phi}^{\rm
        EM}}}{\av{P}}}_t$ where $T^{\rm EM}_{\hat r \hat\phi}$ is the
dominant orthonormal component to the Maxwell stress-energy tensor
evaluated using the tetrad defined in \cite{Penna:2010hu} (the
brackets denote an orbit averaged quantity). More
specifically, we report an azimuthally- and $z$- averaged
$\alpha=\alpha(r)$ profile, which can be used in 1D orbit-averaged
disk models.  
5) Disk scale height $H = \Sigma/\rho_0(z=0)$.  
6) Inner disk edge radius $r_{\rm in}$: In all $\Sigma(r)$-profiles we
observe that inside the cavity $\Sigma(r)$ declines rapidly with decreasing $r$ and
becomes convex.  We fit a fifth order polynomial to the orbit-averaged
$\Sigma(r)$ in the convex region at small $r$, and define $r_{\rm in}$
as the radius where the curvature of the $\Sigma(r)$ fitting function
is maximized, $[\Sigma''(r)/(1+\Sigma'(r)^2)^{3/2} ]'=0$ where
$'\equiv d/dr$.
7) EM Poynting luminosity ($L_{\rm EM}$) as defined in Eq. (1) of \cite{Paschalidis:2013jsa}.
8) Energy loss rate carried off by the outflowing gas $L_{\rm gas} = \oint_s T_{0,}{}_{\rm
  (gas)}^r dS$.  This surface integral must be performed in the
asymptotically flat regime. Given that we do not perform the integration
at an infinite radius, as a crude approximation to $L_{\rm gas}$ 
we include in the integration only matter that is
unbound, i.e., matter for which at large radii $E=-u_0>1$. 
We compute 7) and 8) at several radii including $90M,140M,210M$.  
9) For cases where our cooling scheme is enabled, we compute the cooling luminosity $L_{\rm cool} =
\oint_s T_{0,}{}_{\rm (rad)}^r dS$, which we estimate via the volume
integral $L_{\rm cool} = \int \Lambda u_0 \alpha\sqrt{\gamma}d^3x$. The volume
integration is exactly equal to the surface integration at steady-state and in
spacetimes possessing a timelike Killing  vector and when we ignore any 
radiation captured by the BHs.  We also compute the bolometric radiative luminosity 
$L_{b} = L_{\rm EM} + L_{\rm cool}$.

The second type of diagnostics includes:
10) Optical depth to {\it true} absorption $\tau^* = \sqrt{3\tau_{\rm
    es}\tau_{\rm ff}}$ (Eddington approximation), where we assume pure hydrogen 
and where $\tau_{\rm es}$ ($\tau_{\rm ff}$) is the optical depth to electron scattering
(free-free absorption), calculated using the Thompson scattering
opacity $\kappa_{\rm es}=0.4\rm cm^2/g$ (Rosseland mean opacity
$\bar{\kappa}_{\rm ff}=6.45 \cdot 10^{23}\rho_0 T^{-3.5}\rm cm^2/g$)
as $\tau_{\rm es} = \kappa_{\rm es} \Sigma$ ($\tau_{\rm ff} =
\kappa_{\rm ff} \Sigma$)~\cite{Shapiro:1983book}. $\tau^*>1$ implies
the matter is optically thick to absorption.  
11) Local temperature of the matter,
calculated by solving $\epsilon = a T^4/\rho_0 + 2 k_B T \rho_0 /m_p$, where
$m_p$ is the proton mass and $k_B$ the Boltzmann constant.  
12) In the cases with cooling, the effective disk temperature (in cooling cases),
estimated by assuming that all cooling luminosity is emitted as
black body radiation: 
\labeq{Teffective}{
T_{\rm eff}=[L_{\rm cool}/4\pi\sigma(r_{\rm
    out}^2-r_{\rm in}^2)]^{1/4},
}
 where $\sigma$ is the
Stefan-Boltzmann constant and $r_{\rm out}$ is the disk outer radius.
13) Characteristic observed thermal radiation frequencies $\nu_{\rm
  bb} = k_B T_{\rm eff}/h(1+z)$, where h is the Planck
constant and $z$ the cosmological redshift. This is calculated only when $\Lambda \neq 0$.
14) Cyclotron emission. While we find the bulk
of the disk to be optically thick near Eddington accretion rates, the low-density ``cavity'' is optically thin. 
From these regions cyclotron lines may be detectable.  
15) In cases where $\Lambda \neq 0$ we compute the characteristic
cyclotron frequencies $\nu_{\rm cy} = eB/mc(1+z)$, where $e$
is the electron charge, $m$ the electron mass and $B$ the magnitude
of the magnetic field.

\section{Tests and resolution requirements}
\label{sec:tests}

In this section we describe tests we performed that motivate the
choices for the grid resolution and cooling time scale.

\subsection{Hydrodynamical evolutions with $B=0$}

To set a lower limit on the necessary resolution to perform our GRMHD
simulations, we found the minimum resolution required so that our code
maintains stable equilibrium of an unmagnetized disk around a single
non-spinning BH for several thousands of $M$ of evolution. The
equilibrium disk solution we use is described at the beginning of
Sec.~\ref{sec:tests2}. Our study indicates that for the low resolution
($\Delta x_{\rm max}=8.0M$) listed in Table~\ref{tab:models} the surface
density profile of the initial disk is maintained to within $2\%$
throughout the disk for at least $t\sim 5000M$. 


\subsection{MRI resolution requirements}
Here we check the conditions for MRI to operate in our disk models. For this to be
the case three conditions have to be satisfied: (I) A magnetic field
configuration must be present that violates the stability condition
for MRI $d\Omega/dr \geq 0$, (II) The wavelength of the
fastest-growing mode $\lambda_{\rm MRI}$ has to be resolved by
$\gtrsim 10$ gridpoints
\cite{Shibata:2006hr,Sano:2003bf,Shiokawa:2012}, and (III) the B-field
must be sufficiently weak for $\lambda_{\rm MRI}\lesssim 2H$. In other
words the wavelength of the fastest growing mode should fit in the
disk
\footnote{Condition (III) is satisfied everywhere shortly after our
  evolutions begin because magnetic winding converts poloidal magnetic
  field into toroidal, thereby decreasing $\lambda_{\rm MRI}$. Our
  high-resolution grids satisfy condition (II) in the innermost parts
  of the disk even after the evolution begins.}.

\begin{figure}[h]
    \includegraphics[width=0.49\textwidth]{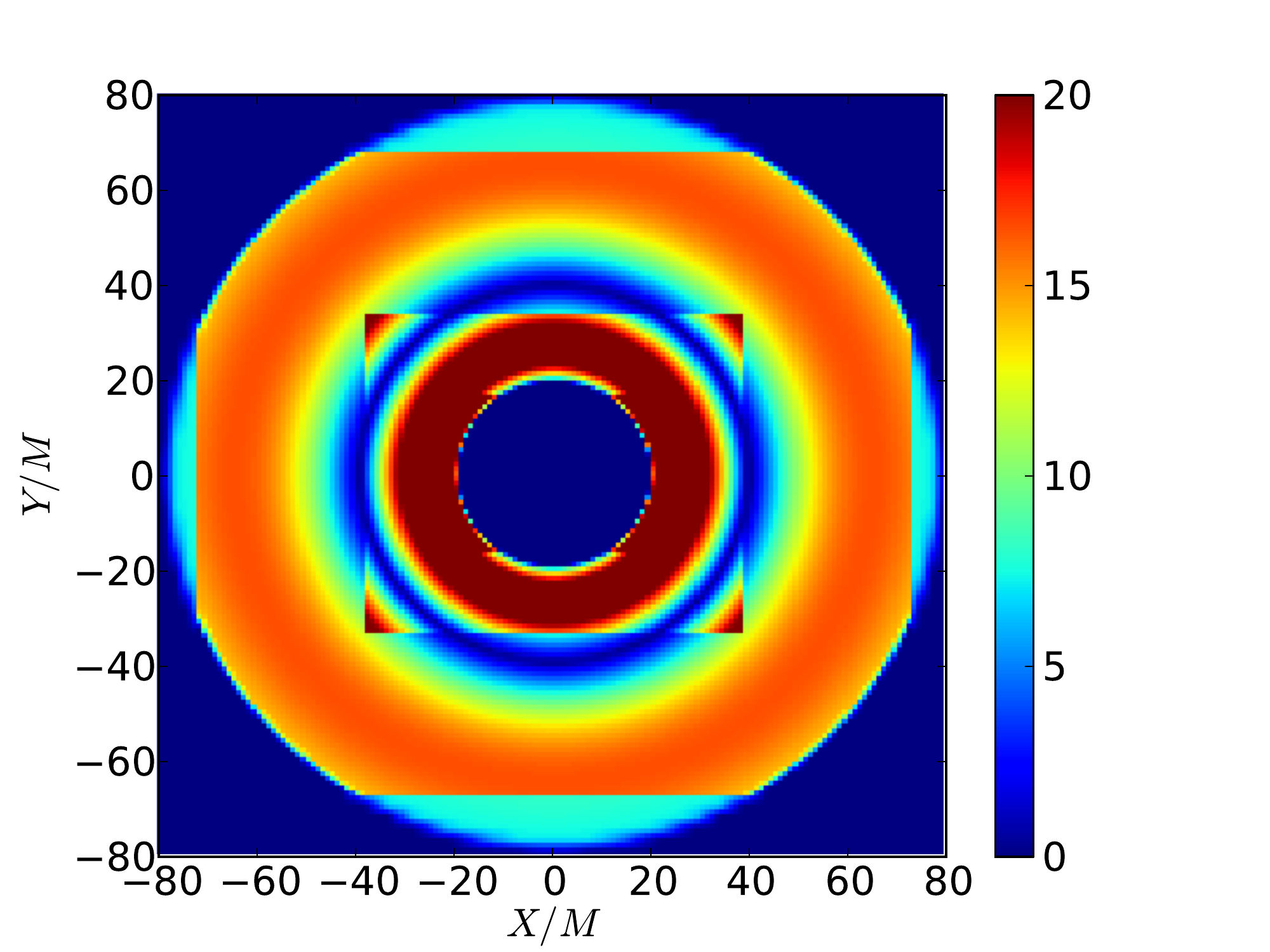} \caption{Contours
      of the $\lambda_{\rm MRI}$-quality factor $Q=\lambda_{\rm
        MRI}/dx$ in the equatorial plane, corresponding to the equal
      mass ($q=1$) medium resolution case [divide (multiply) by $1.33$
        ($1.25$) for the low (high) resolution case] at $t=0M$. We
      resolve the fastest growing MRI mode by $\gtrsim 10$ gridpoints
      over a large part of the disk (the blue ring stems from extremely
      small values of $\lambda_{\rm MRI}$, when the vertical component
      of the B-field changes sign).
        \label{fig:q01nc-hr_mri_xy}}
\end{figure}

Regarding (I) our initial disk model is unstable to the MRI because of
the outwardly decreasing angular velocity and the presence of an
initially small poloidal magnetic field. Regarding (II) we plot the
quality factor $Q\equiv \lambda_{\rm MRI}/dx$ where $dx$ is the local
grid spacing (which jumps by a factor of two at AMR boundaries); see
Fig.~\ref{fig:q01nc-hr_mri_xy}. One can see that we satisfy the
criterion $Q \gtrsim 10$ over a rather large portion of the disk
initially except for the region near the radius where the poloidal
field changes sign and becomes very small. We chose our low-resolution
grids such that condition (II) is satisfied at $t=0$ in the innermost
parts of the disk.

\begin{figure}[h]
      \includegraphics[width=0.49\textwidth]{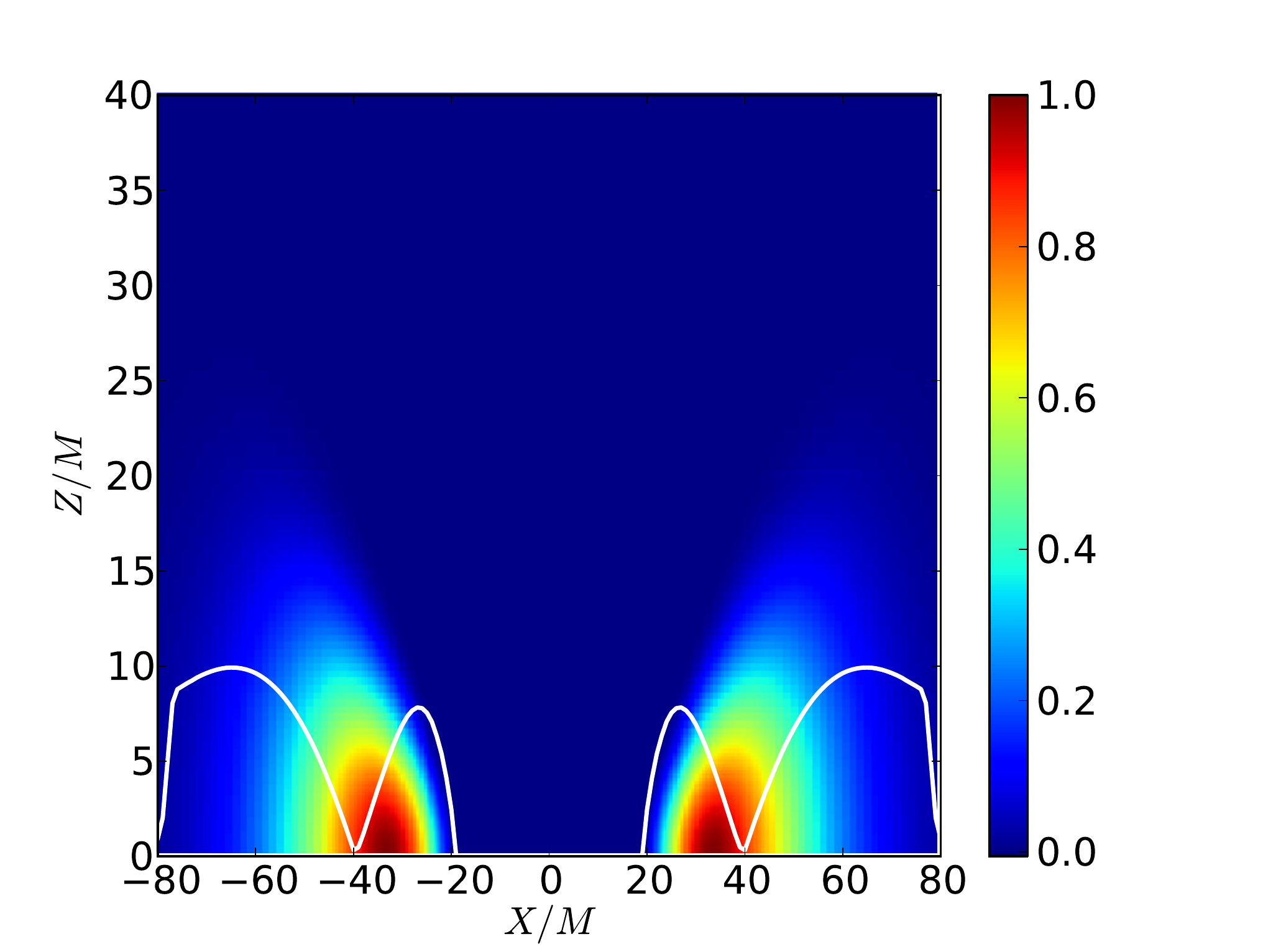}
      \caption{ Rest-mass density contours (color coded) on a
        meridional slice, and $\lambda_{\rm MRI}/2$ (black solid line)
        at $t=0M$. The plot corresponds to equal mass ($q=1$) but is
        the same among all cases considered in this work. For the most
        part $\lambda_{\rm MRI}/2$ fits within
        disk. \label{fig:q01nc-hr_mri_xz}}
\end{figure}

Regarding (III) we plot a meridional $(x,z)$- slice of the rest-mass
density overlayed by a line plot showing $\lambda_{\rm MRI}$ as a
function of $x$ in Fig.~\ref{fig:q01nc-hr_mri_xz}. The plot 
shows that for the most part $\lambda_{\rm MRI}$ fits within the 
disk. At the inner disk edge the MRI
is likely to be suppressed initially, but as the evolution proceeds
magnetic winding converts poloidal fields into toroidal ones lowering
$\lambda_{\rm MRI}$ and eventually triggering the MRI near the inner
disk-edge.

\subsection{Cooling}
\label{subsec:cooling}
We seek to compare two opposite limiting cases for each mass ratio:
(I) No-cooling, $\tau_{\rm cool} \gg \tau_{\rm Kep}$ for which
$\Lambda =0$.  Here $\tau_{\rm Kep}$ is the local Keplerian time scale
which is comparable to the  (shock) heating
time scale; (II) extremely rapid cooling, $\tau_{\rm cool} \ll \tau_{\rm Kep}$
which we model with the effective emissivity $\Lambda = \frac{\rho_0
  \epsilon_{th}}{\tau_{\rm cool}}$.

\begin{figure*}[t]
  \includegraphics[width=0.99\textwidth]{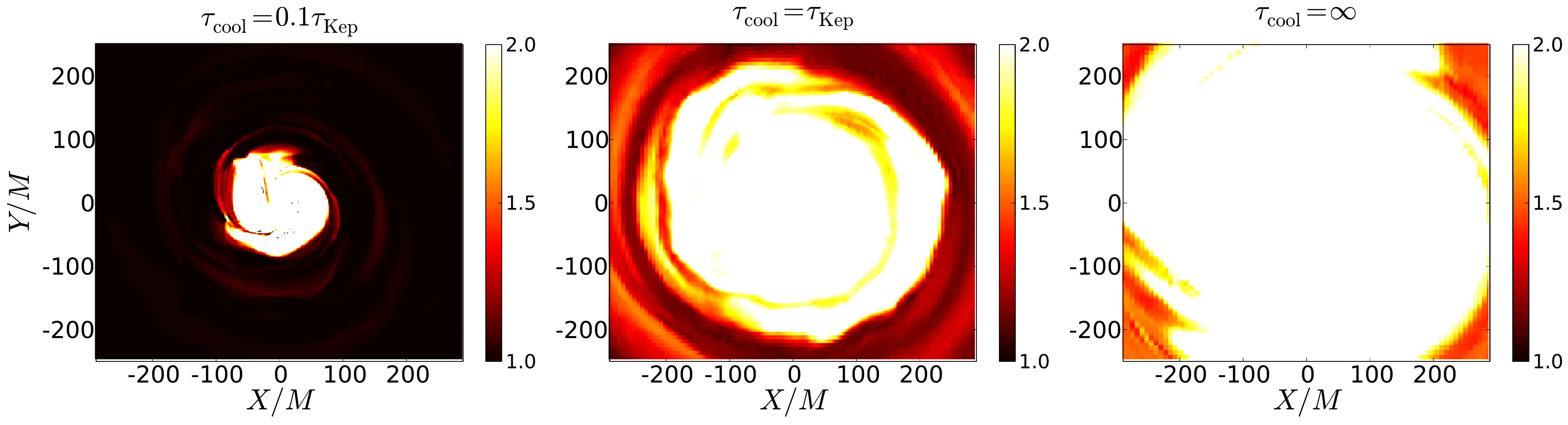}
      \caption{$K/K_0$ for three $q=1$ runs with different $\tau_{\rm
          cool}$ at $t=7700M$. Left panel: $\tau_{\rm
          cool}=0.1\tau_{\rm Kep}$. Middle panel: $\tau_{\rm
          cool}=\tau_{\rm Kep}$. Right panel: No-cooling ($\tau_{\rm
          cool}=\infty$).  Cooling in the $\tau_{\rm cool}=\tau_{\rm
          Kep}$ case cannot keep up with heating, causing $K$ to drift
        away from its initial value. Values much smaller than
        $\tau_{\rm Kep}$ are required for rapid cooling. Steady state
        is not achieved in the $\tau_{\rm cool}\geq \tau_{\rm Kepler}$
        cases.
        \label{fig:dK}}
\end{figure*}

To determine the value for $\tau_{\rm cool}$ at which cooling becomes
rapid (at least in the bulk of the disk), we performed the $q=1$
cooling case using different cooling time scales. We concluded that
rapid cooling requires a cooling time scale significantly shorter than
the local Keplerian time scale. In Fig.~\ref{fig:dK}, we plot $K/K_0$,
where the entropy parameter $S\equiv K= P/(\rho_0^\Gamma)$ and
$S_0=K_0=K(t=0)$ for a run with $\tau_{\rm cool} = 0.1\tau_{\rm Kep}$
(left panel), a run with $\tau_{\rm cool} = \tau_{\rm Kep}$ (middle
panel) and a run without cooling $\tau_{\rm cool} = \infty$ (right
panel). It becomes apparent that when $\tau_{\rm cool} = \tau_{\rm
  Kep}$, $K$ is not driven back to its initial value. Physically, this
means that not all shock generated entropy is radiated away, hence
$\tau_{\rm cool} = \tau_{\rm Kep}$ does not correspond to rapid
cooling and steady state is not achieved. For $\tau_{\rm
  cool}=0.1\tau_{\rm Kep}$ we find $K/K_0 \sim 1$ in the bulk of the
disk. The values depart from unity only inside the cavity where low
density gas exists and can be shock heated to very high $K/K_0$,
demonstrating that shock heating is extremely strong in the
low-density cavity.  We adopt $\tau_{\rm cool}=0.1\tau_{\rm Kep}$ in
all our cooling simulations because it leads to rapid cooling, at
least throughout the bulk of the disk.

\section{Results}
\label{sec:results}

In this section we present the results of our numerical simulations.
First, in Sec.~\ref{sec:singleBH} we show results from our resolution
study. In Sec.~\ref{sec:to-B-or-not-to-B}, we directly compare the
$q=1$ binary case with $B\neq 0$ to the $B=0$ case.  Lastly, we report on the variation of our diagnostics 
with mass ratio for all magnetized cases
in Sec.~\ref{sec:q-trend}.

Our simulations have two parameters that scale out of the problem: the
total mass of the BHBH binary $M$ and the rest-mass density of the
disk. Alternatively, we can exchange one of these parameters with
another parameter that depends on these two free parameters. So,
instead of the rest-mass density, in the results we quote we choose
the Eddington ratio $L_{\rm b}/L_{\rm Edd}$, where $L_{\rm b}$ is the
bolometric EM luminosity described in Sec.~\ref{sec:diagnostics}. The
relation between these parameters is determined as follows: the
accretion rate through the horizon must scale like $\dot M \propto
\rho_{0,\rm ref} M^2$, where $\rho_{0,\rm ref}$ is a reference
rest-mass density in the disk. We choose the maximum rest-mass density
at $t=0$ as the reference density, and our simulations determine the
proportionality constant. For example, in the single, nonspinning BH case with
cooling we find
\beq 
\rho_{\rm ref} = 9 \times 10^{-12}  \nonumber 
\bra{\frac{L_{\rm b}}{L_{\rm Edd}}}\bra{\frac{M}{10^8
    M_\odot}}^{-1} \rm g\ cm^{-3},
\eeq
where we have replaced the accretion rate via the following expression
\labeq{dotMref}{
\begin{split}
\dot M = &\ L_{\rm b} \varepsilon^{-1} = \bigg(\frac{L_{\rm b}}{L_{\rm Edd}}\bigg) L_{\rm Edd} \varepsilon^{-1} \\
\approx  &\ 2.75 \bigg(\frac{L_{\rm b}}{L_{\rm Edd}}\bigg)\bra{\frac{M}{10^8M_\odot}} M_\odot\ \rm  yr^{-1},
\end{split}
}
and where $\varepsilon \equiv L_{\rm b}/\dot M = 0.08$ is the radiative
efficiency as computed via our simulations for our adopted cooling law
in the single, nonspinning BH case. In the no-cooling cases the
only radiation luminosity estimate we have is the Poynting luminosity
which is expected to be a small fraction of the total radiative luminosity. Hence, in the
no-cooling cases we do not scale our results with the Eddington
ratio. Instead, we choose a fiducial accretion rate similar to the one
given in Eq. \eqref{dotMref}.

\begin{figure*}[t]
      \includegraphics[width=0.49\textwidth]{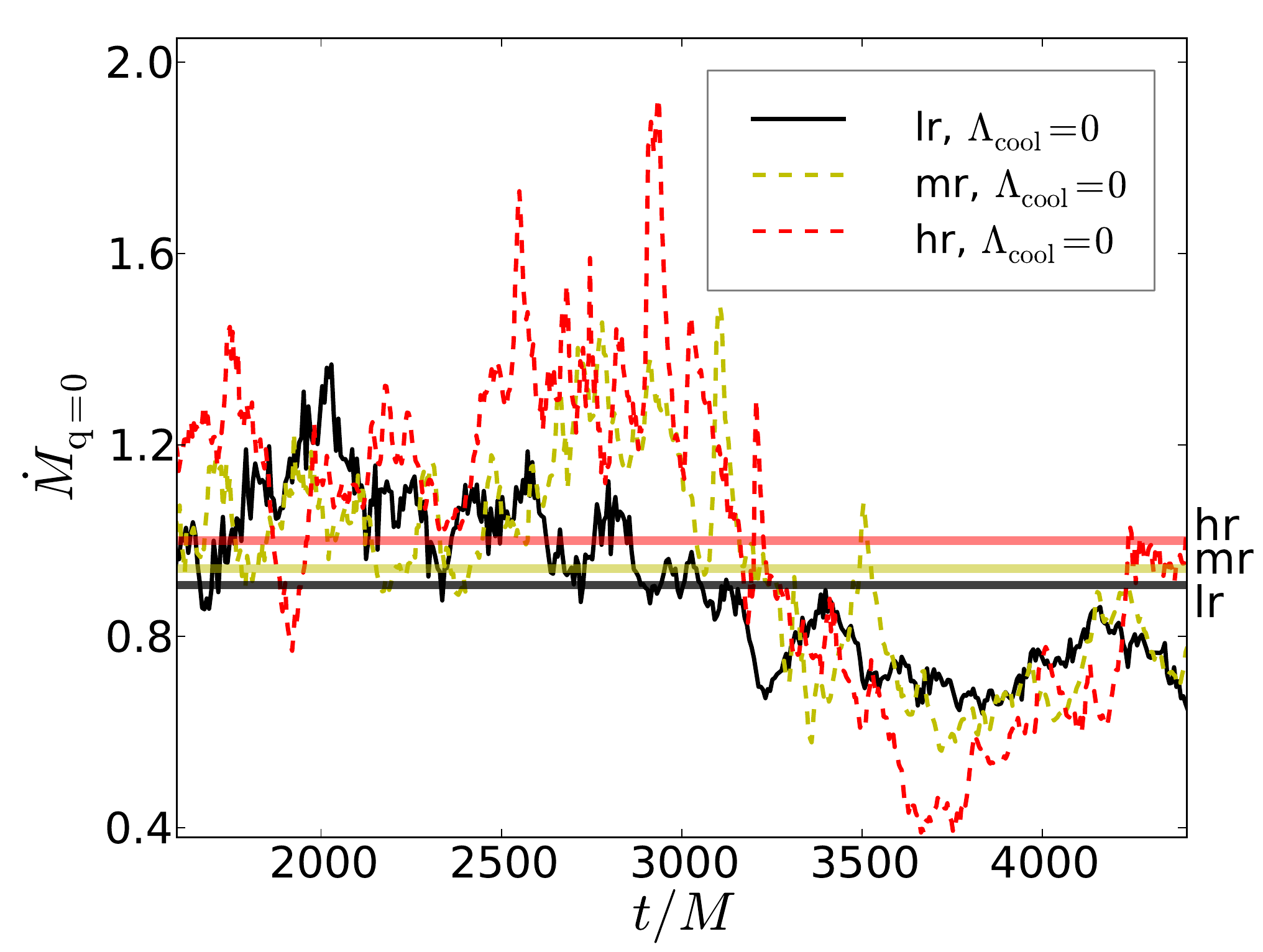}
      \includegraphics[width=0.49\textwidth]{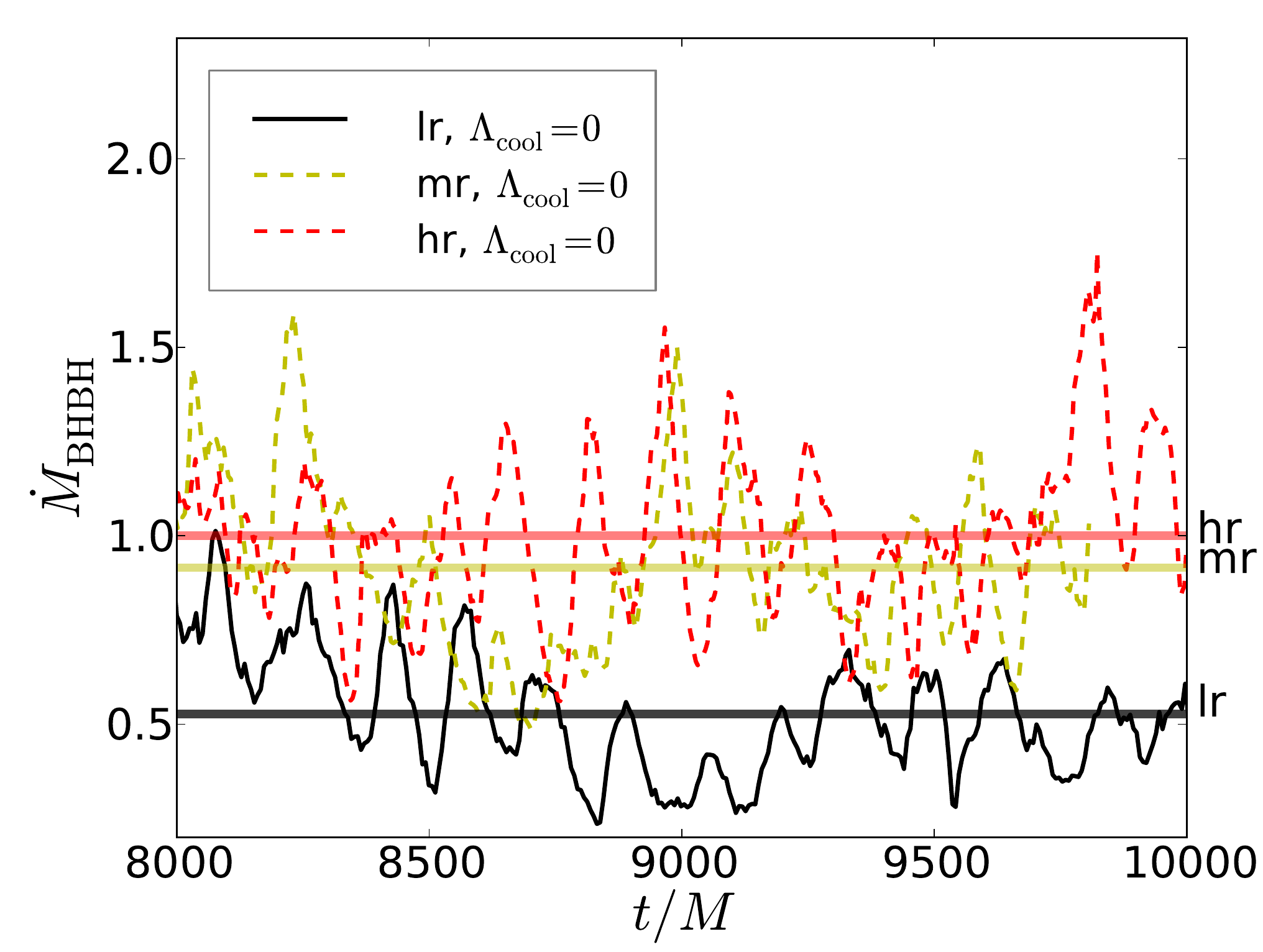}
        \caption{Left panel: rest-mass accretion rate vs time
          for the $q=0$ no-cooling cases at low (lr), medium (mr), and high (hr)
          resolutions.  Right panel: rest-mass accretion rate vs time
          for the $q=1$ no-cooling cases at low, medium, and high
          resolutions. The accretion rate in the $q=0$ ($q=1$) case is
          normalized by the mean accretion of the high resolution
          $q=0$ ($q=1$) run. The horizontal lines indicate the mean
          accretion rate at low, medium and high resolution. 
          \label{fig:q00nc-lr-mr-hr_mdot} }
\end{figure*}

\subsection{Resolution study}
\label{sec:singleBH}

Here we present the results of our resolution study.  For the single BH,
no-cooling and BHBH equal mass, no-cooling cases we use the
three resolutions (see Table~\ref{tab:models}). 

In the single BH-case the average accretion rate varies little with
resolution (see left panel in Fig.~\ref{fig:q00nc-lr-mr-hr_mdot}). The
maximum fractional difference of the mean accretion rate between
different resolutions is $15\%$. Other quantities show a similar
behavior.
These results indicate that the resolutions used in this case are high
enough for the main MRI effects to be captured and the results to be
qualitatively independent of resolution. However, the resolution is
not sufficiently high to perform a formal convergence test.

For the equal-mass case we observe a different behavior. The mean
accretion rates appear to converge (see right panel in
Fig.~\ref{fig:q00nc-lr-mr-hr_mdot}), but the low resolution run
underestimates the accretion rate by almost a factor of 2, while the
medium and high resolution runs are in good agreement. For the latter resolutions
mean accretion rates agree to within $9\%$. We conclude that for the $q=1$ case our adopted medium and
high resolutions are sufficient for drawing qualitative conclusions
and that higher resolutions are necessary for accurate quantitative
results that reside in the convergent regime.

The results of the resolution study in the $q=0$ and $q=1$ cases
differ because of the distribution of matter in both cases and our
grid setup. In the $q=0$ case there is more matter close to the BH,
where very high resolution refinement levels reside, whereas in the
$q=1$ case the bulk of the matter remains outside the inner edge of
the disk, where the grid resolution is not as high. As we show later,
this is not the case in our $q < 1$ models. There, more matter
resides closer to the BHs, and hence closer to the high resolution
levels.

Based on our resolution study we conclude that the low resolution used
in the equal-mass case is not sufficiently high to yield reliable
results, but for the other mass ratios we consider in this work, our
adopted low resolution suffices for a qualitative discussion. Thus, in
the equal-mass cases results will be reported mainly from our medium
resolution runs.

We stress here that these simulations, which include all relativistic
effects and resolve the BHs, are computationally very demanding (see
Sec.~\ref{sec:methods}) and increasing the resolution even further
will incur a very large computational cost.  With increasing computer
power and larger computer allocations we plan to improve our results in
the near future.


\subsection{Significance of B-fields}
\label{sec:to-B-or-not-to-B}


Previous hydrodynamic simulations and (semi)analytic models of
circumbinary accretion disks using the simplified $\alpha$-disk model
(e.g.~\cite{Artymowicz:1994bw,MacFadyen:2006jx}) showed that the main
feature in the equal-mass \textit{binary} case is that the density
inside and near the binary's orbit remains substantially lower than in
the single BH case (see, e.g., Fig.~\ref{fig:q01nc-lr-hydro_rho_xy}).  Such
an inner cavity or ``hollow'' can have important consequences for the
emergent radiation, such as line emission due to small optical depth
and small bolometric luminosity from the hollow. Any such difference
between single BH vs.~circumbinary accretion disks can provide a path
to distinguishing a binary AGN versus a classical, standard (single
BH) AGN \cite{Tanaka:2013oju}.
The explanation for the existence of a hollow is that the binary tidal
torques for $q\gtrsim 0.01$, are strong enough to push most matter away
from the binary orbit~\cite{Haiman:2009te}. The effect is most prominent
in geometrically thin disks, which arise when radiative cooling is highly
efficient. 

However, even in the absence of viscosity or magnetic fields, the
time-dependent tidal field strips off matter from the inner disk
edge, giving rise to an accretion pattern consisting of two streams
which penetrate the inner cavity and extend to the horizons of the BHs
\cite{Haiman:2009te,Farris:2009mt,Bode:2009mt,Bode:2011tq,Bogdanovic:2010he,Farris:2011vx}.

Furthermore, recent MHD simulations
\cite{Shi:2011us,Noble:2012xz,Farris:2012ux} universally revealed that
the reduction of density inside the cavity in the binary case is not
as substantial as previously thought. Such simulations explore regimes
in which the disk is geometrically thick, which partially accounts
for the difference. 

Here we present a comparison between magnetized and
unmagnetized circumbinary accretion disks onto an equal mass BHBH,
while all other physical and numerical parameters remain identical, 
to illustrate the importance of magnetic fields in filling the 
hollow with dense material.

\subsubsection{Midplane-density}
\begin{figure*}[t]
      \includegraphics[width=0.85\textwidth]{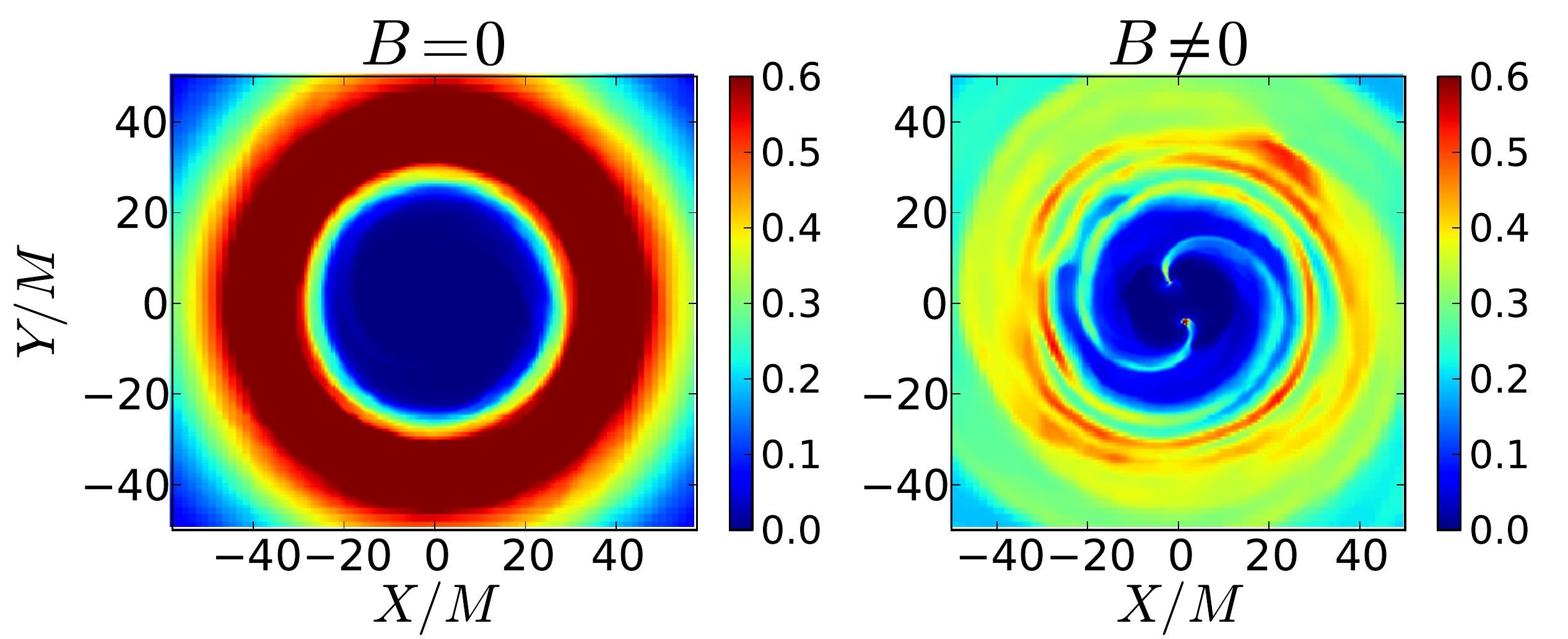}
      \caption{Contours of rest-mass density normalized to the initial
        maximum rest-mass density in the equatorial plane at $t\sim
        5000M$ for two $q=1$ no-cooling cases. Left panel: $B=0$.
        Right panel: $B \neq 0$. Notice the higher densities in spiral
        arms in the inner regions in the $B\neq 0$
        case. \label{fig:q01nc-lr-hydro_rho_xy}}
\end{figure*}

Figure~\ref{fig:q01nc-lr-hydro_rho_xy} demonstrates the striking
differences between no-cooling evolutions with and without magnetic fields at the
same integration time. Magnetic-free hydrodynamic evolutions severely
underestimate both the density in the inner regions and the
overdensity due to the spiral arms. This indicates that the amount of
matter stripped off by tidal torques is small compared to the
amount of matter flowing into the hollow due to MHD turbulence.
  \subsubsection{$\Sigma$-profiles, $q=1$}
Next we compare the surface density profiles of magnetized vs
unmagnetized disks. We find different profiles between the two cases as
shown in Fig.~\ref{fig:q01_sigma}. The $B=0$ model remains relatively
close to the initial data, apart from a slow, mild expansion due to
tidal heating and shocks. When magnetic fields are present, the final
disk profile is completely different from the initial data even though
the binary torques are identical. This implies that the MRI-driven
viscous torques have a much larger impact on the global disk structure
than the binary tidal torques, except perhaps near the inner disk edge.

\subsubsection{Sensitivity to cooling}
\begin{figure}[b]
            \includegraphics[width=0.48\textwidth]{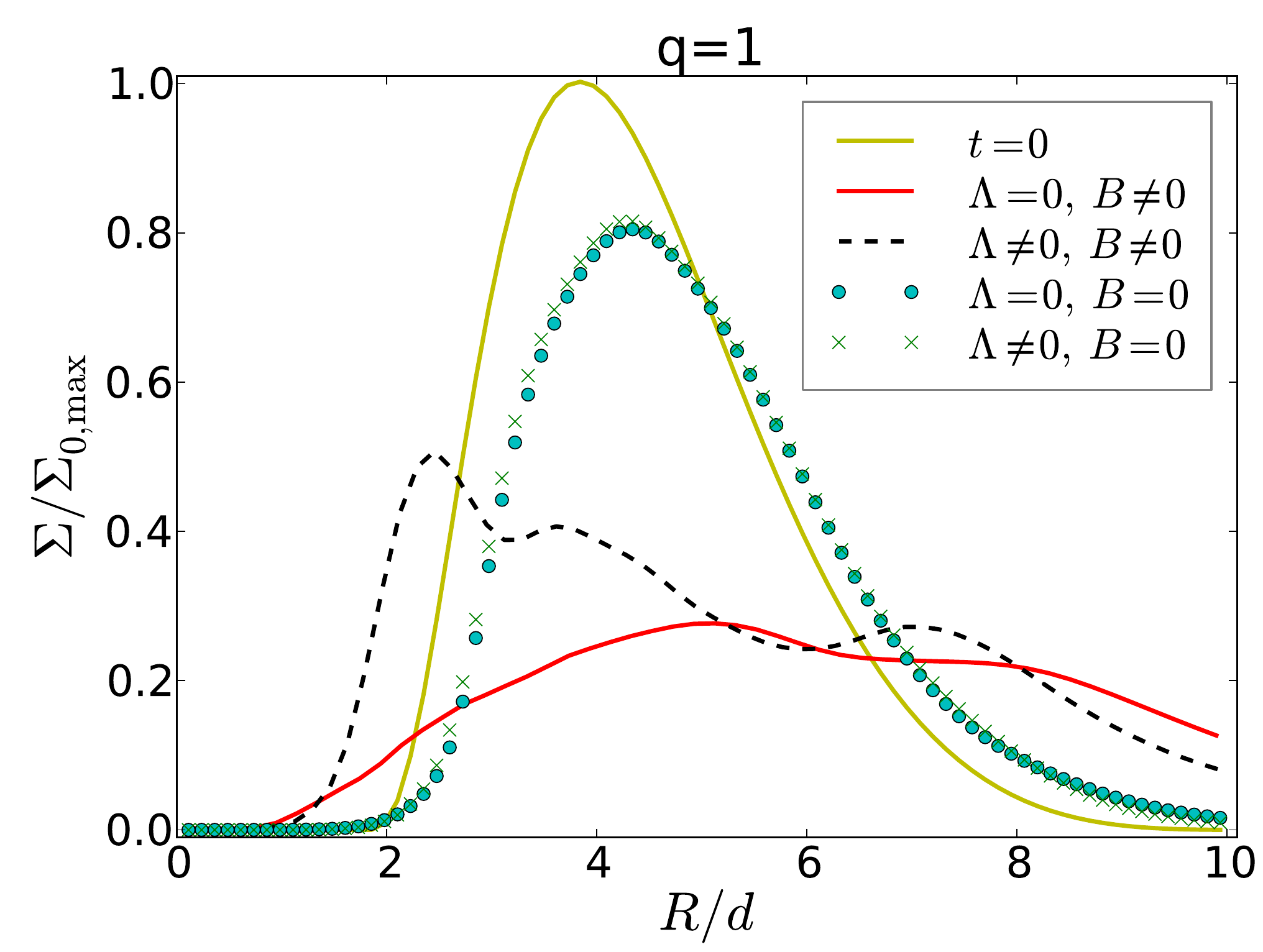}
            \caption{
              Surface density profiles for several $q=1$ cases:
              Initial profile (yellow, solid line), $B=0$ no-cooling
              (green circles) \& cooling (green crosses), $B\neq 0$
              no-cooling (red, solid line) \& cooling (black, dashed
              line) cases. Apart from the initial profile all other
              profiles are orbit-averaged over the last 10 orbits of
              evolution, beyond which the profile does not change
              appreciably.  Notice the clear emergence of a pile-up
              near the inner disk edge in the cooling $B\neq 0$ case, as
              well as the relatively small change relative to the initial data in
              the $B=0$ cases, and the similarity between $\Lambda =0$ and $\Lambda \neq 0$
              unmagnetized cases. 
              \label{fig:q01_sigma} 
              }
\end{figure}

We find a fundamental difference between $B=0$ 
and $B\neq 0$ evolutions regarding their sensitivity to
cooling. In the $B=0$ case both $\Lambda\neq 0$ and $\Lambda=0$
evolutions lead to essentially the same $\Sigma$-profile (see
Fig.~\ref{fig:q01_sigma}).  This is in stark contrast to the $B\neq 0$
(magnetized) cases shown in Fig.~\ref{fig:q01_sigma}, for which
cooling has a strong impact, leading to a matter pile-up near the
inner disk edge. As our particular choice of $\Lambda$ serves to dissipate 
heat from shocks, we conclude that magnetic fields lead
to far stronger shock heating in the disk than the binary tidal torques.

\begin{figure*}[t]
  \centering
  \includegraphics[width=0.99\textwidth]{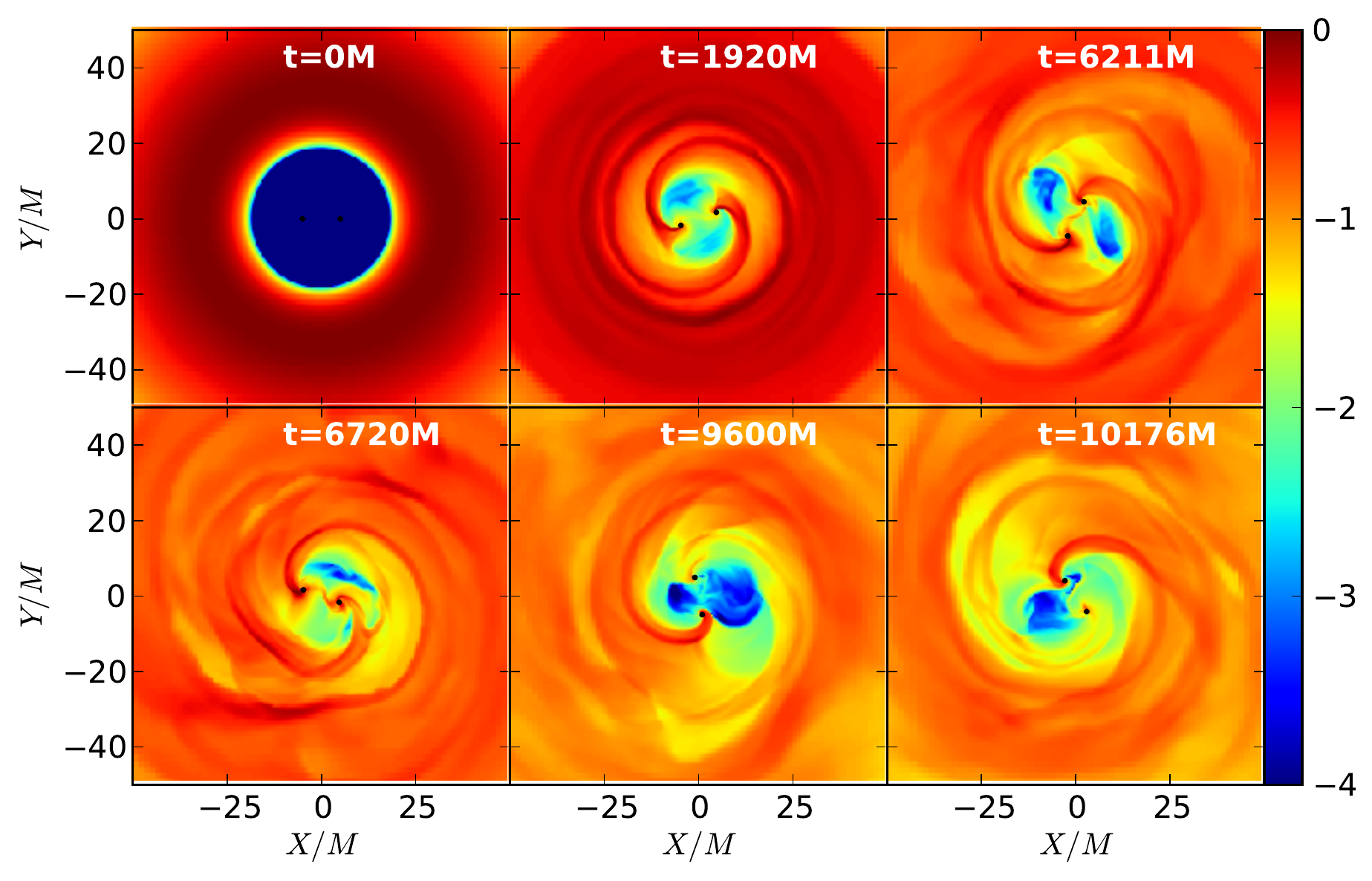}
            \caption{Contours at select times of rest-mass density normalized to the
              initial maximum $\rho_{\rm max}$ (log scale) in the
              equatorial plane. The plot corresponds
              to the $q=1$ no-cooling case.  Here $\rho_{\rm max}
              \simeq 5.6 \times 10^{-11} \bra{\frac{\dot{M}_{\rm
                    BHBH}}{10M_\odot\ \rm{ yr}^{-1}}} \bra{\frac{M}{10^8
                  M_\odot}}^{-2} \rm g\ cm^{-3}$.
            \label{fig:q01nc-hr_snapshots_rho_xy}}
\end{figure*}

\begin{figure*}[h]
  \centering
  \includegraphics[width=0.99\textwidth]{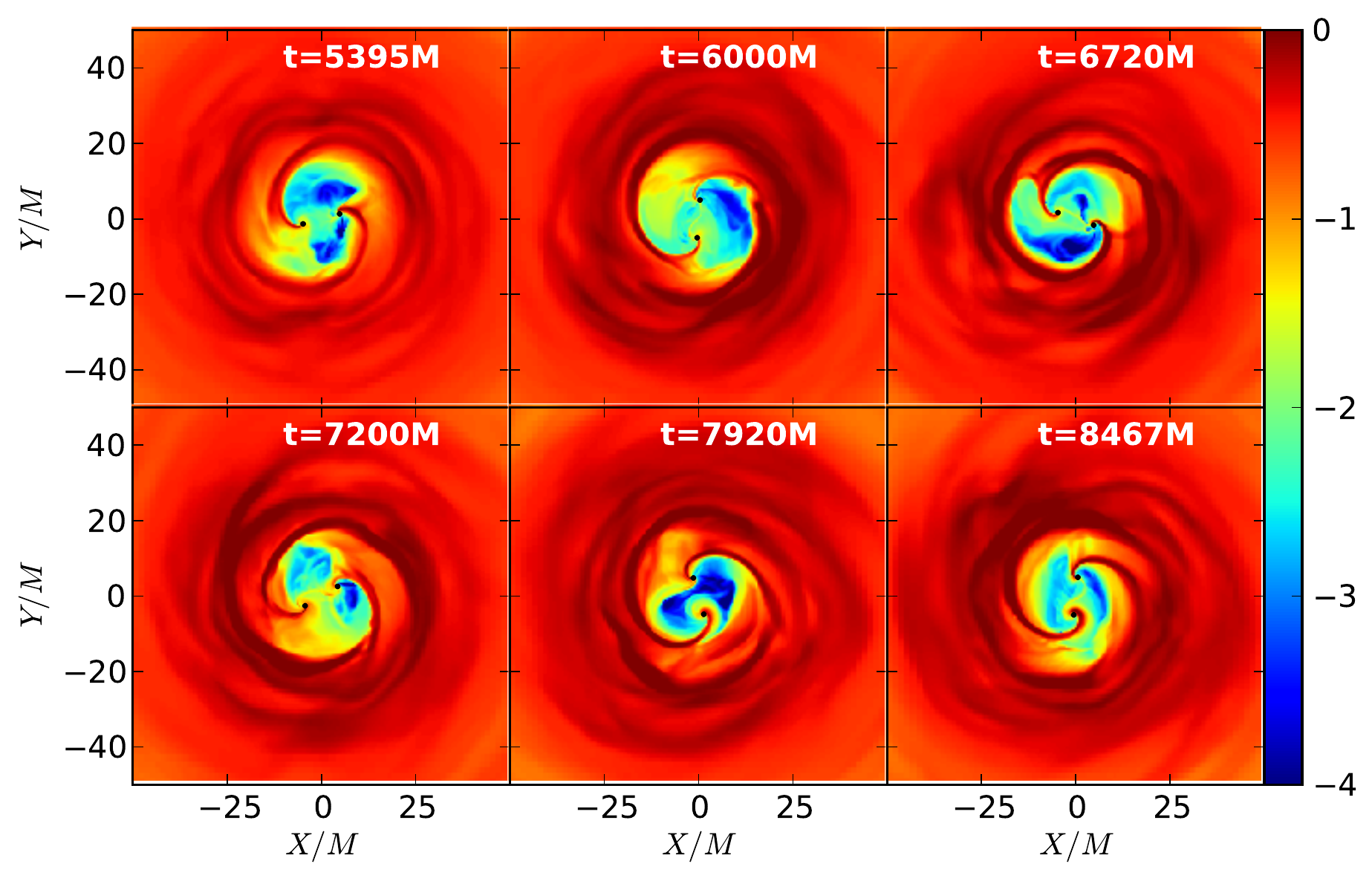}
            \caption{Contours at select times of rest-mass density normalized to the
              initial maximum $\rho_{\rm max}$ (log scale) in the
              equatorial plane. The plot corresponds
              to the $q=1$ cooling case. Here $\rho_{\rm max} \simeq 2.1 \times 10^{-11} 
              \bra{\frac{\dot M}{1.75 M_\odot/\rm yr}}\bra{\frac{M}{10^8
                  M_\odot}}^{-2} {\rm g\ cm}^{-3}
              \simeq 2.1 \times 10^{-11} 
              \bra{\frac{L_{{\rm b}}}{L_{{\rm Edd}}}}\bra{\frac{M}{10^8
                  M_\odot}}^{-1} \rm g\ cm^{-3}$. 
            \label{fig:q01c-hr_snapshots_rho_xy}}
\end{figure*}

\begin{figure*}[t]
  \centering
  \includegraphics[width=0.99\textwidth]{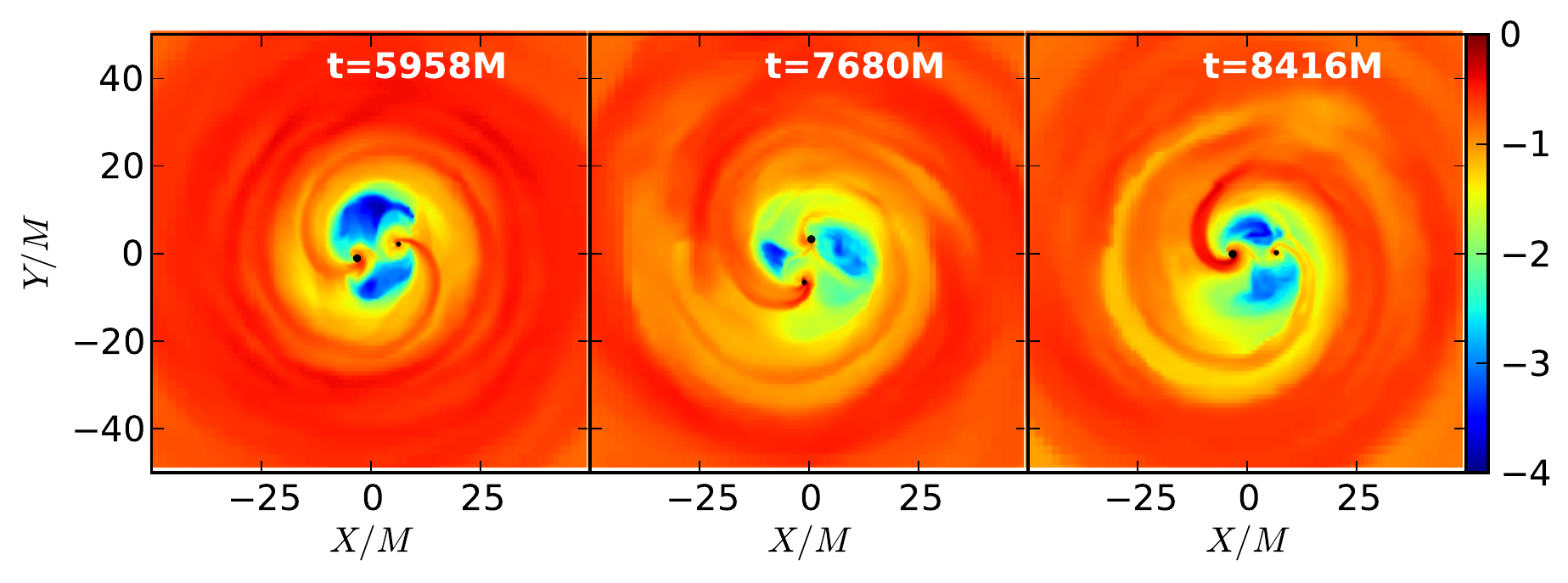}
            \caption{Contours at select times of rest-mass density normalized to the
              initial maximum $\rho_{\rm max}$ (log scale) in the
              equatorial plane. The plot corresponds
              to the $q=0.5$ no-cooling case. Here $\rho_{\rm max}
              \simeq 8.3 \times 10^{-11} \bra{\frac{\dot{M}_{\rm
                    BHBH}}{15.8M_\odot\ \rm{ yr}^{-1}}} \bra{\frac{M}{10^8
                  M_\odot}}^{-2} \rm g\ cm^{-3}$.
            \label{fig:q02nc-lr_snapshots_rho_xy}}
\end{figure*}

\begin{figure*}[t]
  \centering
  \includegraphics[width=0.97\textwidth]{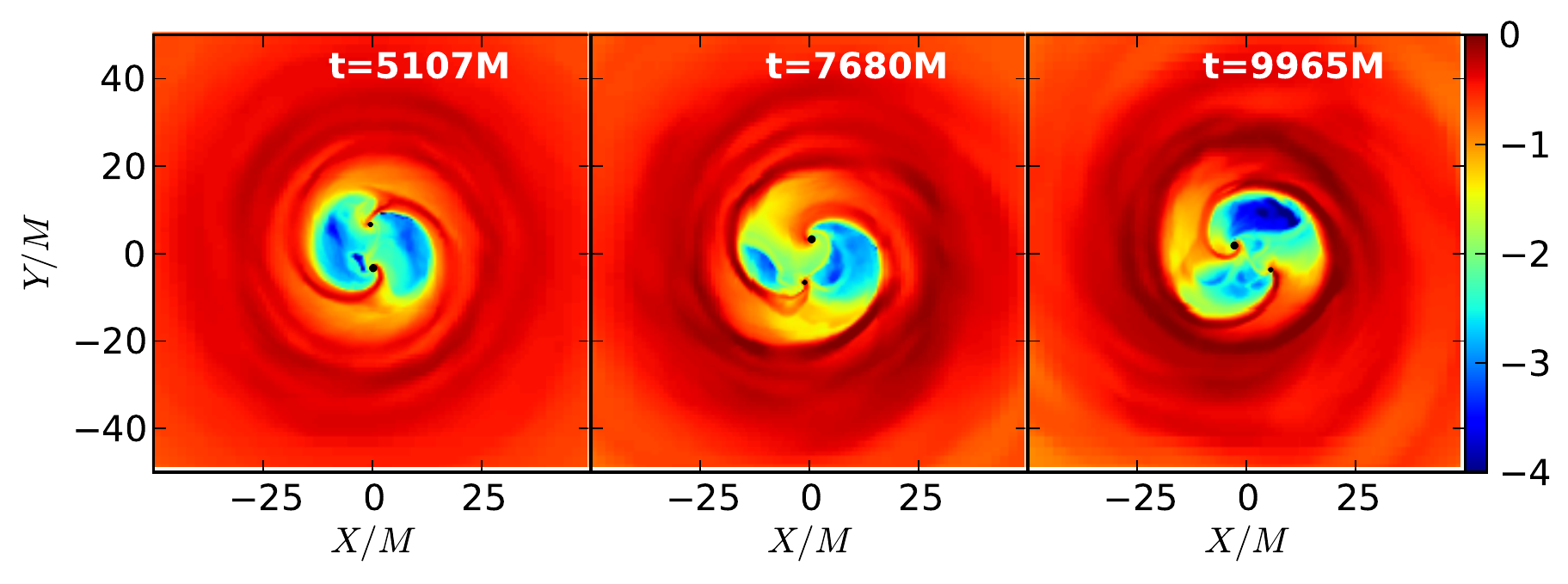}
            \caption{Contours at select times of rest-mass density normalized to the
              initial maximum $\rho_{\rm max}$ (log scale) in the
              equatorial plane. The plot corresponds
              to the $q=0.5$ cooling case. Here $\rho_{\rm max} \simeq 4.2
              \times 10^{-11} \bra{\frac{\dot M}{2.85 M_\odot/\rm yr}}\bra{\frac{M}{10^8
                  M_\odot}}^{-2} {\rm g\ cm}^{-3}
              \simeq 4.2 \times 10^{-11}  \bra{\frac{L_{\rm b}}{L_{\rm
                    Edd}}}\bra{\frac{M}{10^8 M_\odot}}^{-1}
              \rm g\ cm^{-3}$. 
            \label{fig:q02c-lr_snapshots_rho_xy}} 
\end{figure*}

\begin{figure*}[h]
  \centering
  \includegraphics[width=0.89\textwidth]{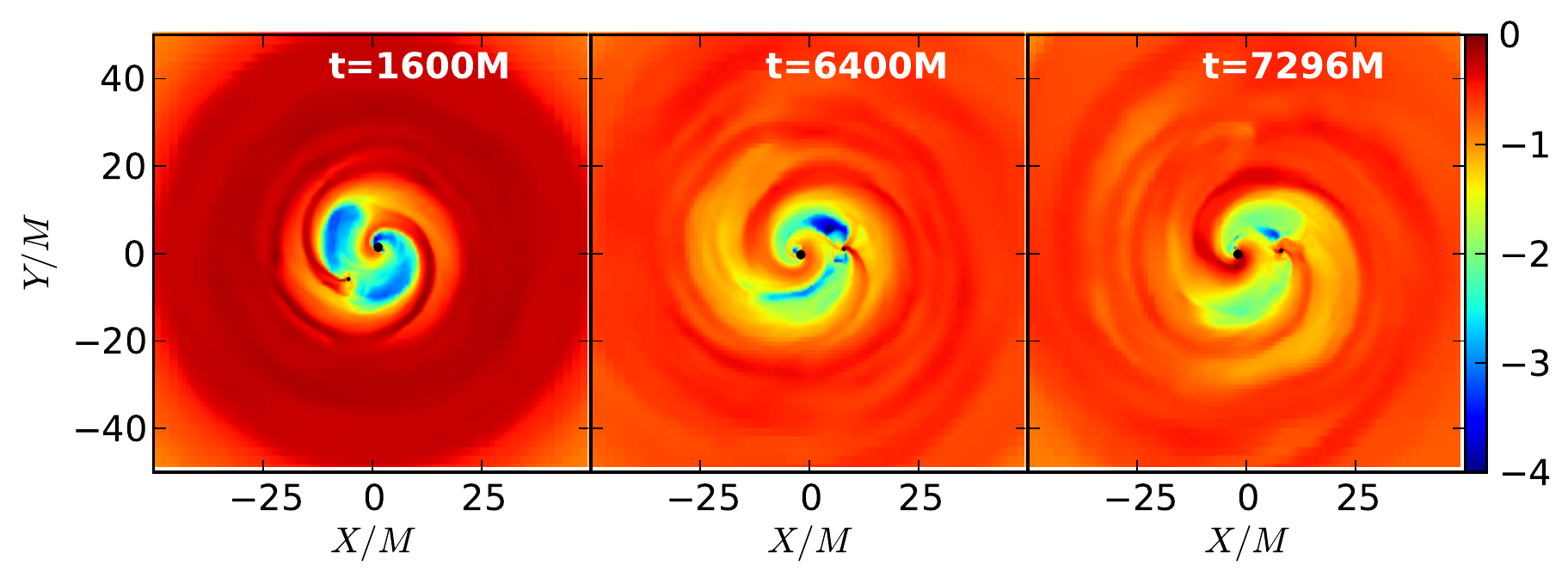}
            \caption{Contours at select times of rest-mass density normalized to the
              initial maximum $\rho_{\rm max}$ (log scale) in the
              equatorial plane. The plot corresponds
              to the $q=0.25$ no-cooling case. Here $\rho_{\rm max}
              \simeq 6.7 \times 10^{-11} \bra{\frac{\dot{M}_{\rm
                    BHBH}}{13M_\odot\ \rm{ yr}^{-1}}}
              \bra{\frac{M}{10^8 M_\odot}}^{-2} \rm
              g\ cm^{-3}$. \label{fig:q04nc-lr_snapshots_rho_xy}}
\end{figure*}

\begin{figure*}[t]
  \centering
  \includegraphics[width=0.89\textwidth]{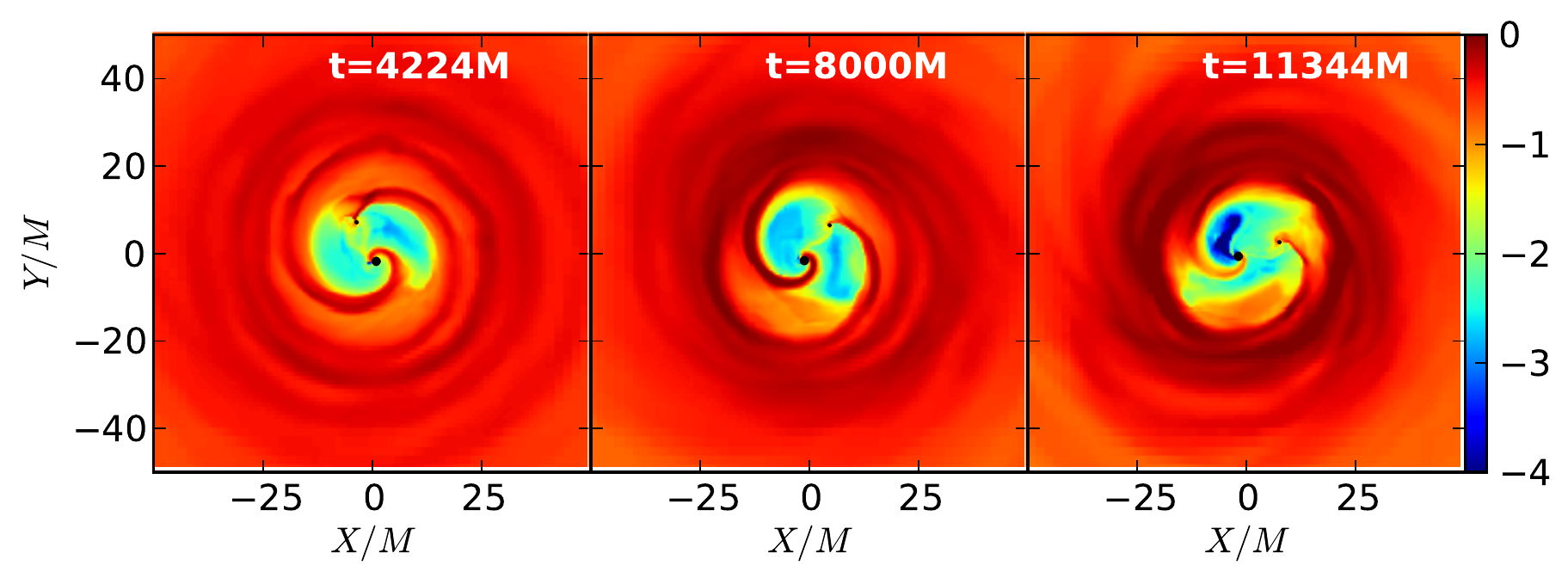}
            \caption{Contours at select times of rest-mass density normalized to the
              initial maximum $\rho_{\rm max}$ (log scale) in the
              equatorial plane. The plot corresponds
              to the $q=0.25$ cooling case.  
              Here $\rho_{\rm max} \simeq 3.75
              \times 10^{-11} \bra{\frac{\dot M}{2.27 M_\odot/\rm yr}}\bra{\frac{M}{10^8
                  M_\odot}}^{-2} {\rm g\ cm}^{-3}
              \simeq 3.75
              \times 10^{-11}  \bra{\frac{L_{\rm b}}{L_{\rm
                    Edd}}}\bra{\frac{M}{10^8 M_\odot}}^{-1}
              \rm g\ cm^{-3}$. 
            \label{fig:q04c-lr_snapshots_rho_xy}}
\end{figure*}

\begin{figure*}[t]
  \centering
  \includegraphics[width=0.89\textwidth]{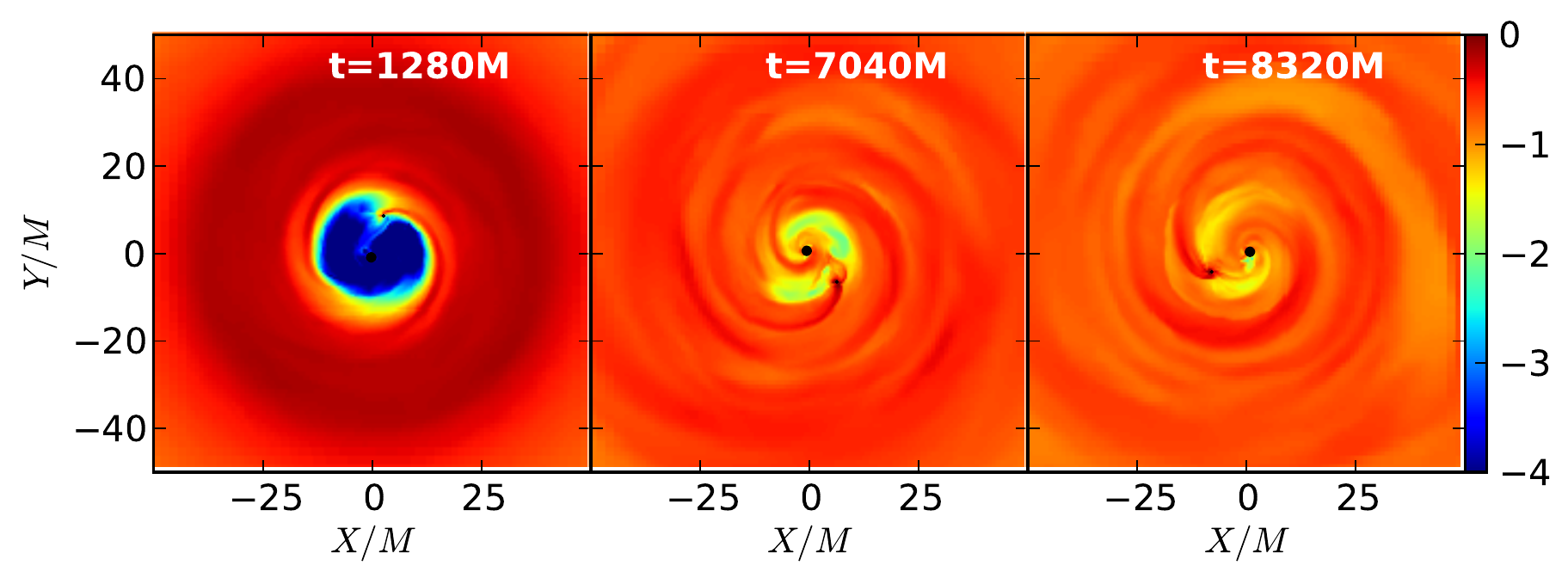}
            \caption{Contours at select times of rest-mass density normalized to the
              initial maximum $\rho_{\rm max}$ (log scale) in the
              equatorial plane. The plot corresponds
              to the $q=0.1$ no-cooling case. Here $\rho_{\rm max}
              \simeq 6 \times 10^{-11} \bra{\frac{\dot{M}_{\rm
                    BHBH}}{11M_\odot\ \rm{ yr}^{-1}}}
              \bra{\frac{M}{10^8 M_\odot}}^{-2} \rm g\ cm^{-3}$.
            \label{fig:q10nc-lr_snapshots_rho_xy}}
\end{figure*}

\begin{figure*}[t]
  \centering
            \includegraphics[width=0.89\textwidth]{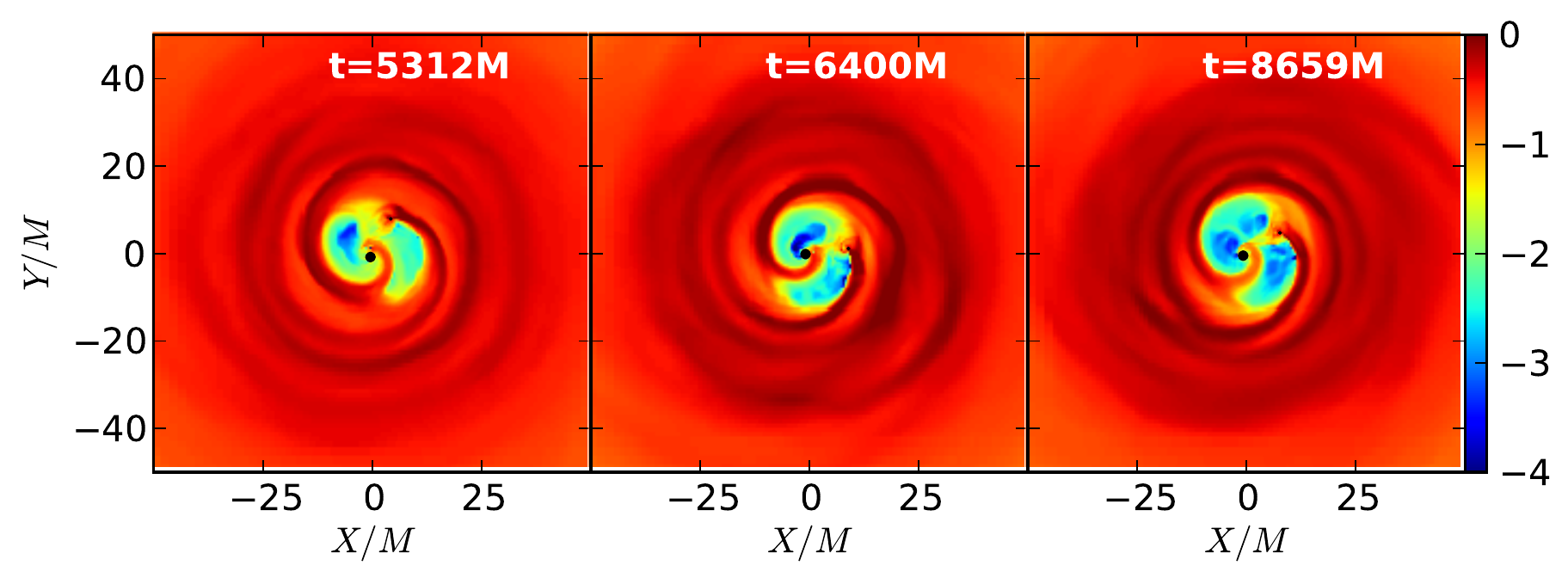}
            \caption{Contours at select times of rest-mass density
              normalized to the initial maximum $\rho_{\rm max}$ (log
              scale) in the equatorial plane. The plot corresponds to
              the $q=0.1$ cooling case. Here $\rho_{\rm max} \simeq 3.5
              \times 10^{-11} \bra{\frac{\dot M}{2.85 M_\odot/\rm
                  yr}}\bra{\frac{M}{10^8 M_\odot}}^{-2} {\rm g\ cm}^{-3}
              \simeq 3.5 \times 10^{-11} \bra{\frac{L_{\rm b}}{L_{\rm
                    Edd}}}\bra{\frac{M}{10^8 M_\odot}}^{-1} \rm
              g\ cm^{-3}$.  The gas is denser everywhere compared to
              mass ratios closer to unity; compare to
              Figs.~\ref{fig:q01nc-hr_snapshots_rho_xy},
              \ref{fig:q01c-hr_snapshots_rho_xy},
              \ref{fig:q02nc-lr_snapshots_rho_xy},
              \ref{fig:q02c-lr_snapshots_rho_xy},
              \ref{fig:q04nc-lr_snapshots_rho_xy},
              \ref{fig:q04c-lr_snapshots_rho_xy}.
            \label{fig:q10c-lr_snapshots_rho_xy}} 
\end{figure*}

  \subsubsection{Accretion rate}

The ratio of the time-averaged accretion rate without magnetic fields
to that with magnetic fields is $\langle\dot{M}_{BHBH, B =
  0}\rangle/\langle\dot{M}_{BHBH, B\neq 0}\rangle \lesssim 1\%$ (see
also Sec.~\ref{sec:q-trend}). This result applies to both cooling and 
no-cooling cases.

In summary, $B = 0$ evolutions underestimate the material inside the
cavity and accretion rates \textit{by orders of magnitude}. Hence,
incorporating magnetic fields is paramount for a proper treatment of
circumbinary accretion disks.

\subsection{Trend with mass ratio, $B\neq 0$}
\label{sec:q-trend}

In this section we discuss the dependence of our multiple diagnostics
on the binary mass ratio for our $B\neq 0$ cases. We use results from
the single non-spinning BH case ($q=0$) to normalize and compare our results for
the binary cases.

\begin{figure*}[t]
      \includegraphics[width=0.85\textwidth]{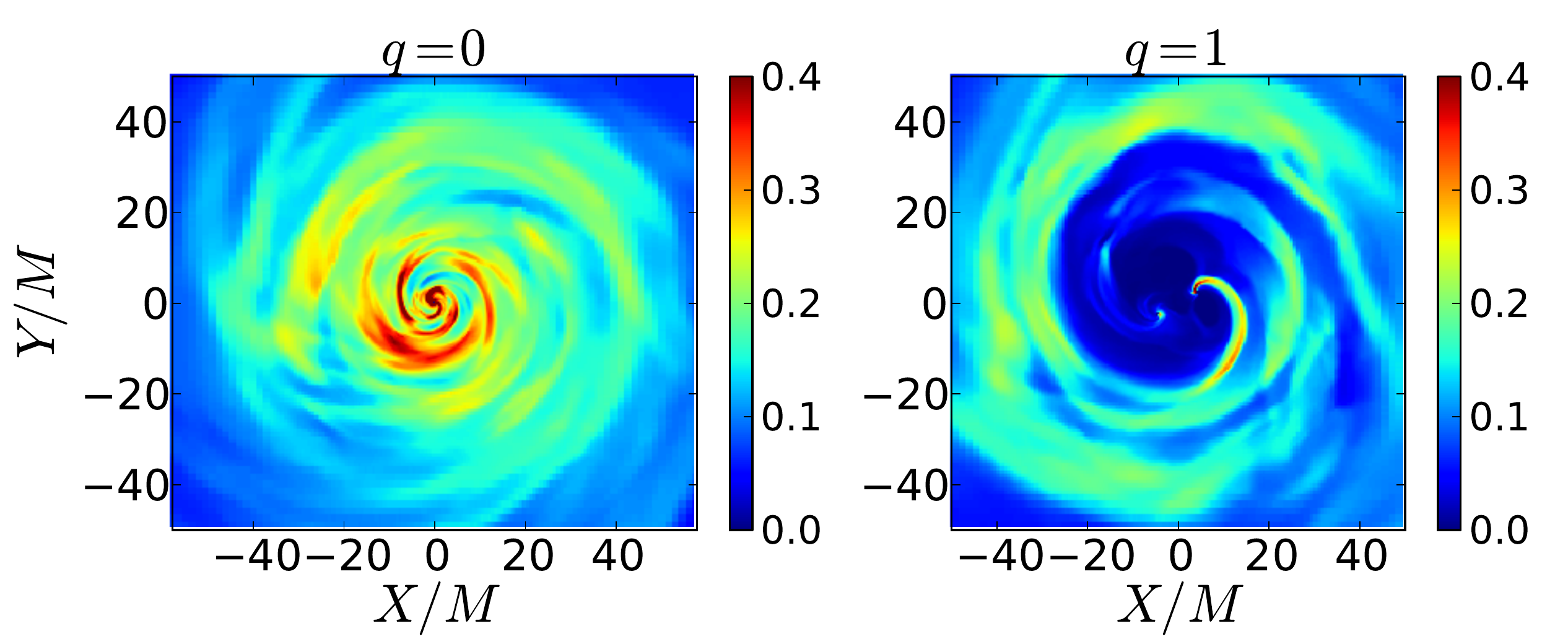}
      \caption{Contours of rest-mass density (linear color scale) normalized to the initial maximum
        in the equatorial plane at $t=10000M$. Left panel:
        $q=0$ no-cooling. Right panel: $q=1$ no-cooling case.
        \label{fig:q00-q01_xy_late}}
\end{figure*}

\subsubsection{Disk structure in the bulk and inside the cavity}
\label{sec:matter-cavity}

We begin this section by discussing the qualitative evolution of the
disk rest-mass density $\rho_0$. For $q\neq 0$, the early evolution is
similar for the different mass ratios, but departs strongly in the
subsequent evolution depending on $q$. The onset of accretion occurs
through two spiral streams, which remain attached to the horizons
throughout the evolution, as shown in
Figs.~\ref{fig:q01nc-hr_snapshots_rho_xy}--\ref{fig:q10c-lr_snapshots_rho_xy}.
Here we plot the rest-mass density contours in the equatorial
plane. These streams are among the densest structures of the accretion
flow, especially for the lowest non-zero mass ratio case $q=0.1$ (see
Figs.~\ref{fig:q10nc-lr_snapshots_rho_xy} and
\ref{fig:q10c-lr_snapshots_rho_xy}).  Spiral density waves are
launched near the inner edge of the disk, which propagate and
dissipate into the outer disk. This feature can also be seen in
Figs.~\ref{fig:q01nc-hr_snapshots_rho_xy}--\ref{fig:q10c-lr_snapshots_rho_xy}
for all mass ratios.

Late in the evolution we find that when $q\neq 0$, the density of the
matter inside the ``cavity'' in $\Lambda=0$ cases is larger than that
in $\Lambda \neq 0$ cases (see
Figs.~\ref{fig:q01nc-hr_snapshots_rho_xy} and
\ref{fig:q01c-hr_snapshots_rho_xy}). Hence, the amount of matter in the
hollow is smaller when we allow rapid cooling. This dependence on
cooling arises because of the larger disk thickness in the no-cooling
cases, which leads to a reduction in tidal torques
\cite{Lin:1993,Armitage:2002uu} near the binary orbit, allowing matter
to overflow more easily.

We also find that the smaller the mass ratio the more matter pours
into the cavity. This is anticipated from the Newtonian expression for
the binary tidal torque \cite{Lin:1993,Armitage:2002uu}, which
decreases with decreasing $q$.  As expected, the largest contrast
arises between the $q=1$ and $q=0$ cases, which becomes obvious by
simply inspecting the rest-mass density contours in the equatorial
plane as shown in Fig.~\ref{fig:q00-q01_xy_late}. One can see the main
difference: The presence of a region of lowered density with two
accretion streams near the BHs in the binary $q=1$ case -- the
``hollow'' (or ``cavity'') -- and the
absence of these features in the $q=0$ case (left panel).

The relaxed disk structure in the predecoupling regime is not
axisymmetric. The back-sloshing of material towards the inner disk
edge occurs mainly in two opposing directions and leads to a gradual
overdense feature in the disk, which has been referred to as a
``lump'' \cite{Noble:2012xz,Shi:2011us}, and its presence has been
linked with a growth in disk eccentricity (see also
\cite{MacFadyen:2006jx}). 
The nonaxisymmetric feature is stronger for models
with cooling. Hence, a more realistic calculation with radiative
transfer is necessary to assess the strength of nonaxisymmetric
structure in a circumbinary disk.

As expected, the rest-mass density contrast between the two accretion
streams becomes larger with smaller mass ratios. This effect is
easily seen when comparing the rest mass density contours in the equatorial
plane for $q=0.25$, $q=0.125$ and $q=0.1$ cases (see
Figs. \ref{fig:q04nc-lr_snapshots_rho_xy} and
\ref{fig:q10c-lr_snapshots_rho_xy}).

In the $q=1$ and $0.5$ no-cooling cases we observe time variations in
the density of the streams relative to each other: for about half an
orbit one stream is stronger than the other.

We find that the supply of material channeled onto the BHs is
sufficient to keep the BHs immersed in a persistent gaseous
environment with $b^2/\rho_0 \sim 10^{-3}$. This means that the
force-free electrodynamics approximation may be inadequate to
globally describe the systems considered here.

For $q=0.1$ there is hardly a low-density hollow (see
Fig.~\ref{fig:q10nc-lr_snapshots_rho_xy}). This is also revealed by
the inner disk edge being close to or inside the orbit of the secondary,
especially in the no-cooling case.  In the $q=0$ limit no hollow
appears, as expected. However, we observe a region of 
lowered surface density near and inside the ISCO of the primary BH.

\subsubsection{Inner disk edge}

In Newtonian 2D studies of geometrically thin disks and large binary separations,
the location of the inner disk edge is found to be roughly twice the
binary separation, independent of $q$; see, e.g., Table I in
\cite{Artymowicz:1994bw}. For the geometrically thick disks and binary
separations we are considering, we find the inner disk edge in the
relaxed state to be dependent on $q$ and whether cooling is enabled.

\begin{figure}[h]
    \includegraphics[width=0.48\textwidth]{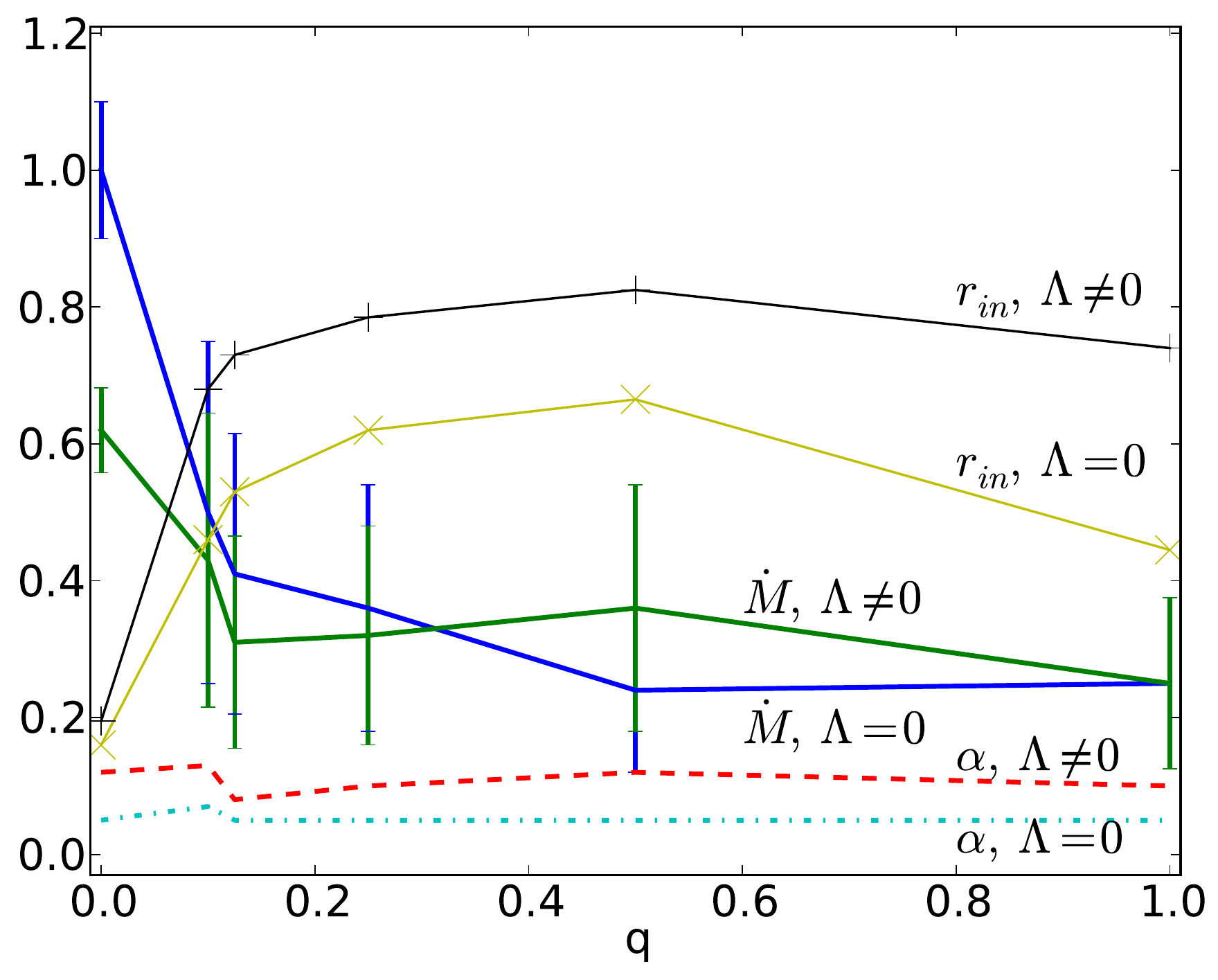}
        \caption{ Mean accretion rate $\dot M$ (normalized to the
          no-cooling single BH accretion rate), $\alpha$, and $r_{\rm in}$
          (normalized to twice the binary separation) as functions of
          $q$ for cooling and no-cooling cases (all at low
          resolution).  In the absence of medium and high resolution
          runs for $q\neq 1$ we place error bars in $\dot M$
          based on the $q=1$ resolution study. These error bars are
          chosen to be 50\%, corresponding to the fractional
          difference between the high and low resolutions runs in the
          $q=1$ case (see Sec.~\ref{sec:singleBH}). The error bar in
          the $q=0$ case is 15\% corresponding to the fractional
          difference between the high and low resolutions runs in the
          $q=0$ case. \label{fig:diag-vs-q}}
\end{figure}

In all cases (see the snapshots in
Figs.~\ref{fig:q01nc-hr_snapshots_rho_xy}--\ref{fig:q10c-lr_snapshots_rho_xy}),
the inner disk edge is closer than predicted by Newtonian thin-disk
calculations \cite{Artymowicz:1994bw,MacFadyen:2006jx,Shi:2011us}.  In
the equal-mass cooling case the inner edge remains closer to the
initial one (see Fig.~\ref{fig:q01_sigma}). The trend is such that the
smaller the mass ratio the closer the disk edge is to the binary
orbit. In the $q=0.1$ no-cooling case the inner disk edge effectively coincides
with the orbit of the secondary. We report the value for $r_{\rm in}$
found in each case in Table~\ref{tab:results}, and plot $r_{\rm in}$ vs $q$
in Fig.~\ref{fig:diag-vs-q}.

\begin{center}
 \begin{table*}
  \caption{Table summarizing our main results. Columns show case
    label, inner disk edge normalized to the binary separation
    $r_{in}/d$, the mean accretion rate and its approximate standard
    deviation normalized to the no-cooling single
    BH mean accretion rate $(\langle\dot{M}_{\rm
      BHBH}\rangle \pm \delta \dot{M}_{\rm
      BHBH})/\langle\dot{M}_{q=0}\rangle$, main Fourier
    frequencies normalized by the binary orbital period in the Fourier
    analysis of the accretion rate, Shakura-Sunyaev
    $\alpha$-parameter\footnote{In some cases we quote a range of
      values because a single radially averaged value would
      overestimate $\alpha$.}, the mean Poynting luminosity $\langle L_{\rm EM}\rangle$,
    and the mean cooling luminosity $\langle L_{\rm cool}\rangle$ both normalized to the binary mean
    accretion rate.  A dash ``--''
    ~indicates ``no information available''.
      \label{tab:results}}
  \begin{tabular}{cccccccc} \hline\hline
  Case    & $r_{\rm in}/d$& $(\langle\dot{M}_{\rm
      BHBH}\rangle \pm \delta \dot{M}_{\rm
      BHBH})/\langle\dot{M}_{q=0}\rangle$ & $f/f_{\rm orb}$    & $\alpha$   & $\langle L_{\rm EM}\rangle/\langle\dot M_{\rm BHBH}\rangle c^2$ &$\langle L_{\rm cool}\rangle/\langle\dot M_{\rm BHBH}\rangle c^2$ \\ \hline
 1:1nc-mr \hspace{-0.2cm}  &  $0.89$        & $0.43 \pm 50\%$  & $(1.0,1.5)$   & $ 0.04$    & $0.013$      & --          \\
 1:2nc-lr \hspace{-0.01cm} &  $1.33$        & $0.24 \pm 60\%$  & $(0.7,1.5)$   & $ 0.05$    & $0.012$      & --          \\
 1:4nc-lr \hspace{-0.01cm} &  $1.24$        & $0.36\pm 20\%$   & $(0.7,1.5)$   & $0.03-0.1$ & $0.010$      & --          \\
 1:8nc-lr \hspace{-0.01cm} &  $1.06$        & $0.41\pm 40\%$   & $0.7$         &$0.03-0.06$ & $0.010$      & --          \\
1:10nc-lr\hspace{-0.05cm} &  $0.92$        & $0.50\pm 50\%$    & --            & $ 0.07$    & $0.017$      & --          \\
 0nc-hr \hspace{0.22cm}   &  \hspace{0.27cm}$0.32$
\footnote{For ease of comparison, in the single BH case we normalize
  $r_{\rm in}$ to $10M$ even though it does not correspond to an
  orbital separation.} 
& $1.00 \pm 24\%$ & -- & $ 0.05$& $0.011$ & -- \\ \hline
 1:1c-mr  \  &  $1.48$        & $0.43 \pm 60\%$  & $(0.5,1.5)$  & $ 0.2$     & $0.004$      & $0.127$      \\
 1:2c-lr \hspace{0.18cm} &  $1.65$        & $0.36 \pm 30\%$  & $1.0$         & $ 0.12$    & $0.003$             & $0.110$      \\
 1:4c-lr \hspace{0.18cm} &  $1.57$        & $0.32\pm 60\%$   & $1.0$        & $ 0.1$     & $ 0.002$     & $0.107$       \\
 1:8c-lr \hspace{0.18cm} &  $1.46$        & $0.31\pm 30\%$   & $0.6$         & $ 0.08$    & $ 0.006$     & $0.096$       \\
 1:10c-lr \hspace{0.01cm} &  $1.36$        & $0.43\pm 30\%$   & --            & $ 0.013$    & $0.006$      & $0.081$      \\
 0c-lr \hspace{0.4cm}   & $0.39$         & $0.62\pm 10\%$              & --            & $ 0.012$    & $0.002$      & $0.115$
\\ \hline\hline
  \end{tabular}
 \end{table*}
\end{center}


\subsubsection{Surface density}
\begin{figure*}[t]
      \includegraphics[width=0.49\textwidth]{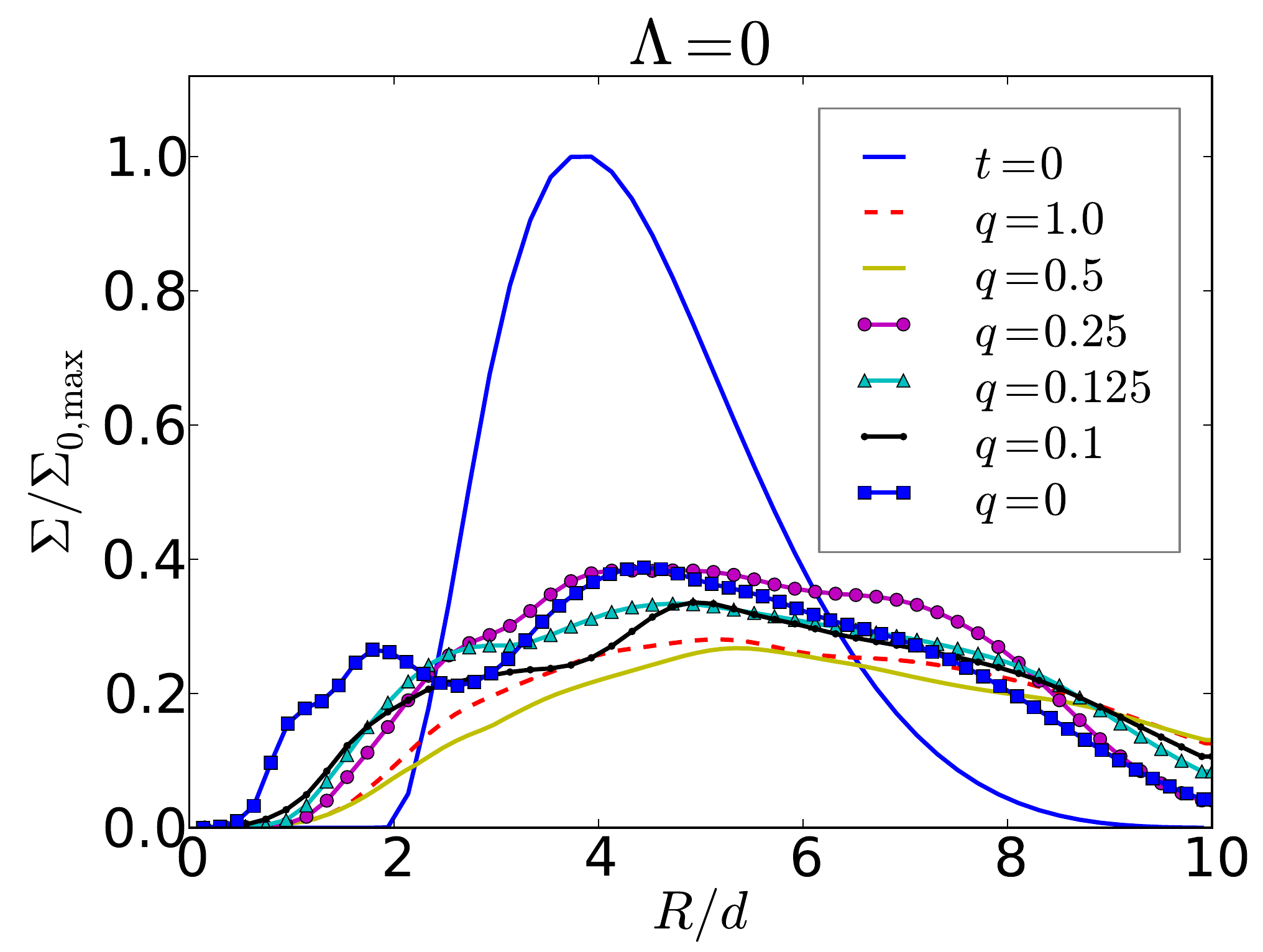}
      \includegraphics[width=0.49\textwidth]{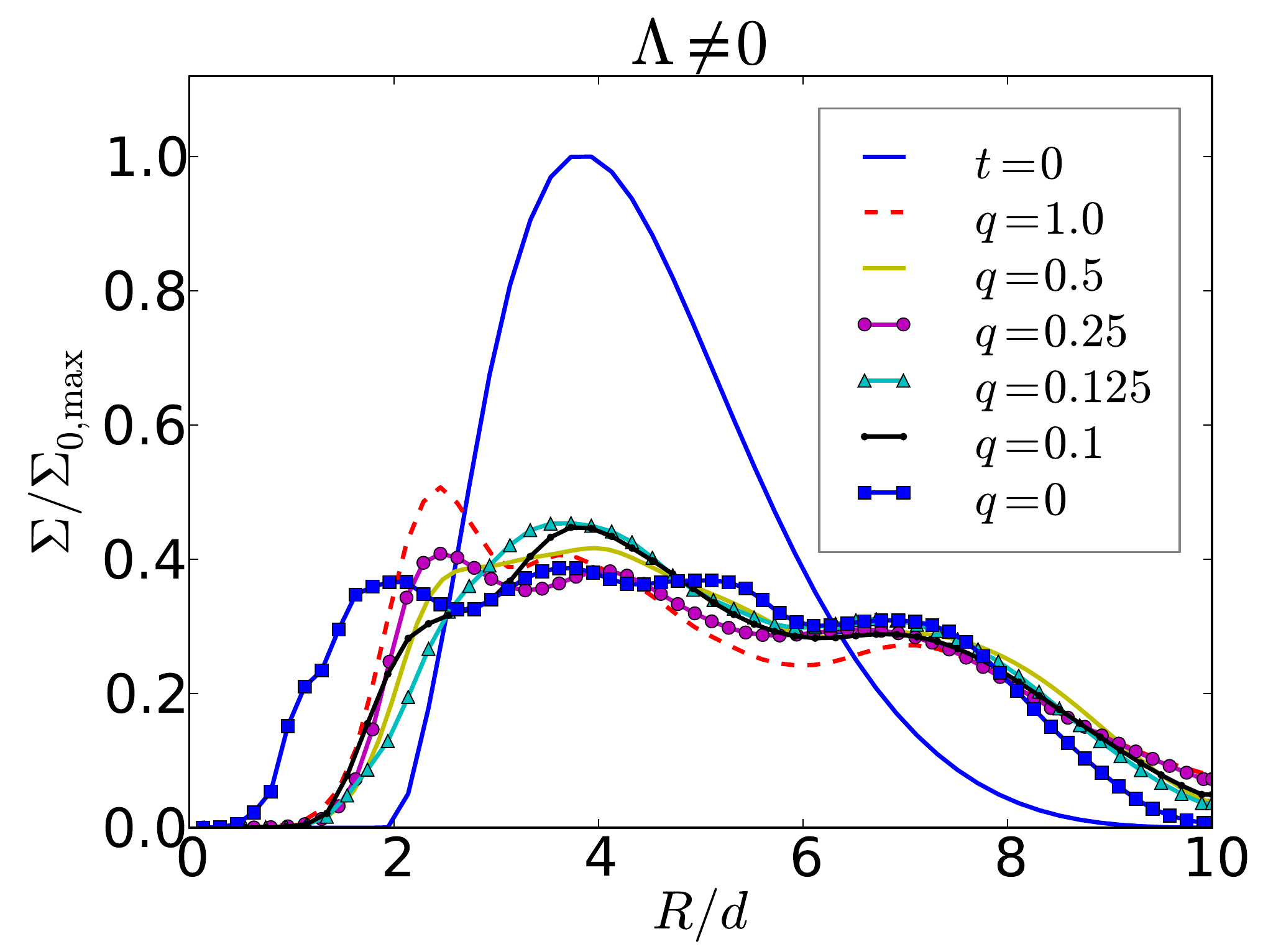}
         \caption{Disk surface density $\Sigma$-profile for different
           mass ratios as a function of cylindrical radius (normalized
           to the binary separation $d \sim 10M$). Left panel:
           no-cooling cases. Right panel: cooling cases.  Except for
           the initial profile all other curves correspond to the
           relaxed state averaged over the last 10 binary orbital
           periods.  In all these $B\neq 0$ cases the relaxed profile
           deviates strongly from the initial one.  In the $\Lambda=0$
           sequence the density reaches further toward small radii
           with decreasing $q$.  In the $\Lambda \neq 0$ sequence, one
           can see the pile-up near the inner disk edge, which is
           strongest for large $q$. \label{fig:sigma-qall-nc-vs-c}}
\end{figure*}

In Fig.~\ref{fig:sigma-qall-nc-vs-c} we show the surface density
($\Sigma$) profiles of the relaxed disks, averaged over the last 10 binary
orbital periods (for all mass ratios, cooling- and
no-cooling). For all cases the relaxed state deviates strongly from
the initial profile, highlighting the importance of evolving the system
for at least a viscous time scale $t_{\rm vis}$ at the radii of interest, where
\labeq{}{
\frac{t_{\rm vis}}{M} \equiv \frac{2R^2}{3\nu M} = 8485 \bra{\frac{\alpha}{0.1}}^{-1} \bra{\frac{H/R}{0.3}}^{-2}\bra{\frac{R}{18M}}^{3/2},
}
and where $\nu \equiv 2\alpha P/3\rho_0\Omega_{\rm Kep}$ is the shear viscosity, with $\Omega_{\rm Kep} = (M/R^3)^{1/2}$.

No-cooling: The evolutions for $q=1$ and $0.5$ are similar in terms of
their $\Sigma$ profiles. The other cases ($q=0.25-0$) yield similar
$\Sigma$ which extend further in than for $q=1$ and $0.5$. The surface
density diagnostic clearly demonstrates that the Newtonian thin-disk
feature that the cavity edge is at twice the binary separation and
independent of the mass ratio does not hold in this class of runs.

Cooling: For non-zero mass ratios, cooling yields a pile up of
dense gas near the inner disk edge, which is absent in the no-cooling
runs and is strongest for the equal-mass case.  Apart from the pile-up
at small radii, all evolutions have a rather similar profile at larger
radii. In the cooling cases the cavity edge is farther out than for
no-cooling, but still closer than twice the binary separation.
The dependence on mass ratio is weaker than in the no-cooling cases,
but still in some disagreement with the Newtonian thin-disk calculations.

\subsubsection{Effective $\alpha$-stress}

\begin{figure*}[t]
            \includegraphics[width=0.495\textwidth]{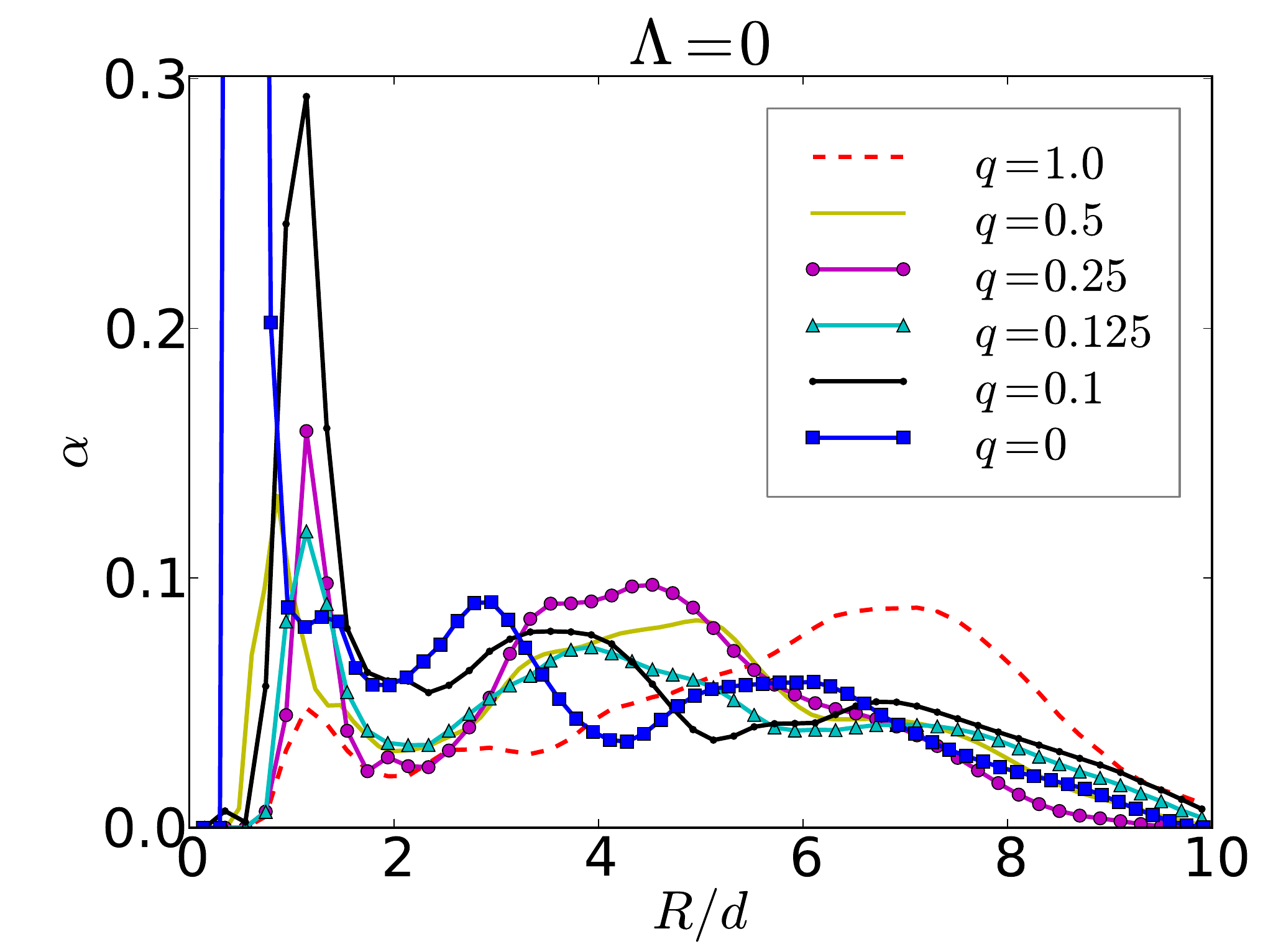}
            \includegraphics[width=0.495\textwidth]{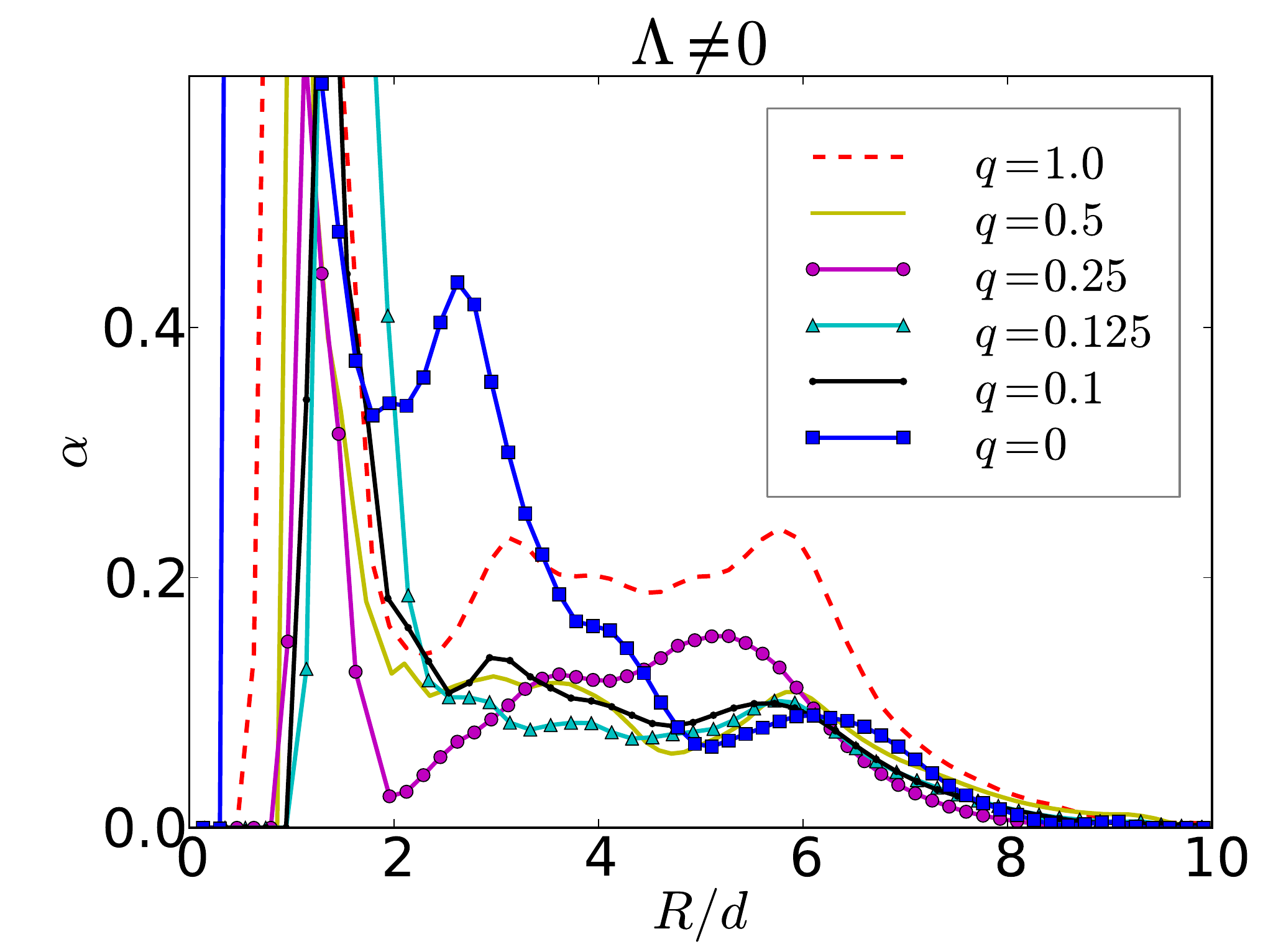}
            \caption{Shakura-Sunyaev $\alpha$ profiles for all mass
              ratios. Left panel: no-cooling cases. Right panel:
              cooling cases. Profiles have been averaged over the
              last 10 binary orbital periods.  Cooling cases tend to
              yield larger values. In all cases the stress increases
              inside the cavity.
           \label{fig:alpha-qall-c}}
\end{figure*}

Figure~\ref{fig:alpha-qall-c} shows the averaged effective
Shakura-Sunyaev $\alpha$ parameter profiles for all mass ratios, both
for cooling and no-cooling models in the relaxed state. The average is
taken over the last 10 binary orbits. In Table~\ref{tab:results} we
also quote characteristic $\alpha$ values obtained by additionally
averaging over radii from the inner disk edge out to the location of
the density maximum. We plot $\alpha$ vs $q$ in
Fig.~\ref{fig:diag-vs-q}.

In all cases we observe larger values for $\alpha$ in the cooling
cases than for the no-cooling cases, and there is always a steep
increase in $\alpha$-stress at smaller radii. For the $q=1$ cooling
case we find $\alpha(r_{\rm in}<r<r_{\rm max}) \sim 0.2$ and
$\alpha(r_{\rm in}<r<r_{\rm max}) \sim 0.1$ for all other cooling cases. 
A typical value for all no-cooling cases is $\alpha(r_{\rm
  in}<r<r_{\rm max}) \sim 0.05$. The higher stress for cooling cases
results from the additional gas compression and associated amplification
of magnetic fields when cooling is allowed. 

  \begin{figure*}[t]
      \includegraphics[width=0.85\textwidth]{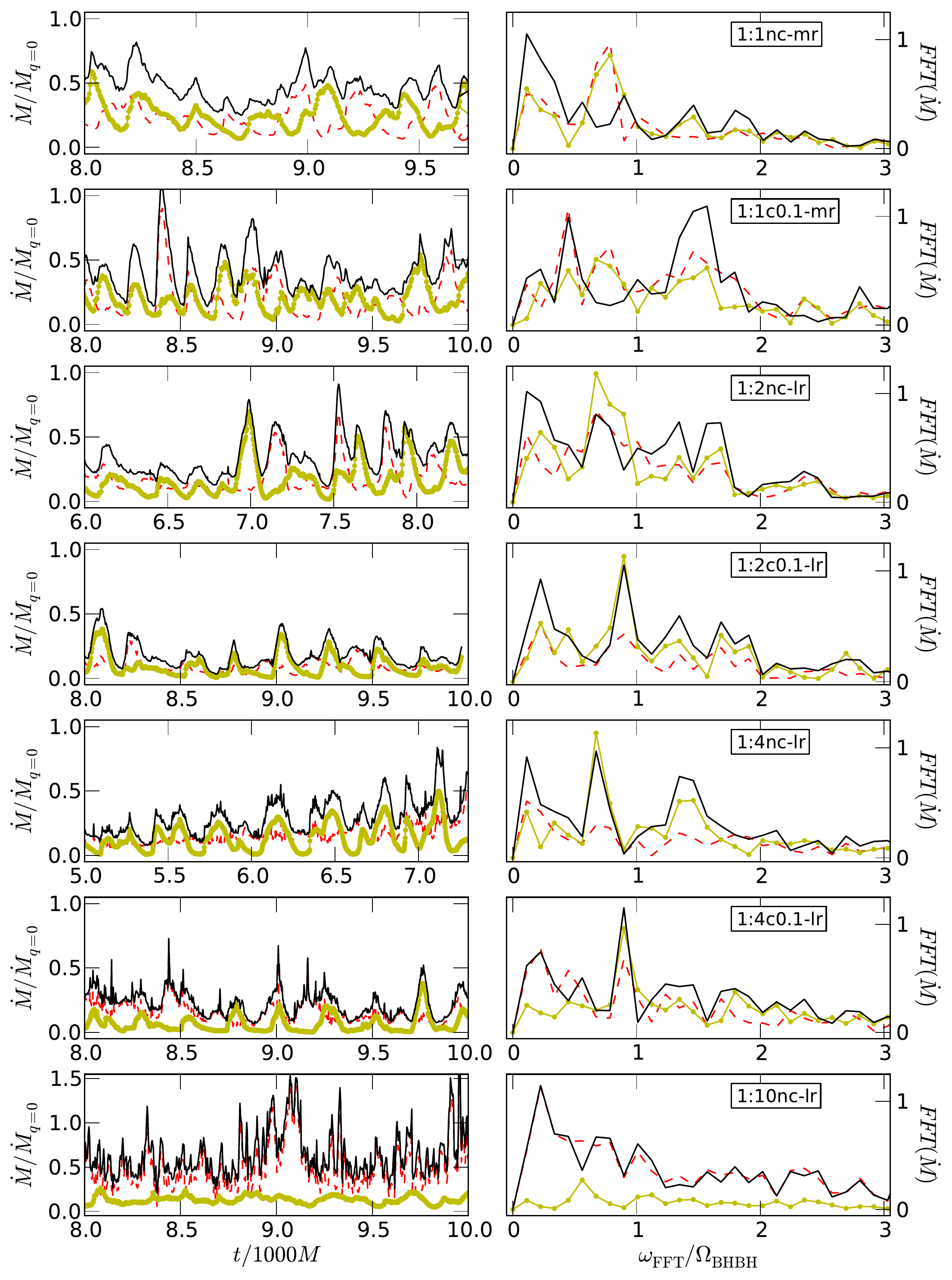}
        \caption{Left panels: Late evolution of mass accretion rate
          onto the binary (solid, black lines), primary (red, dashed
          lines), and secondary (solid-dotted, yellow lines). Right
          panels: Fourier transform of the accretion rates shown on
          the left.  From top to bottom we show the $q = 1$ cooling
          and no-cooling, $q=0.5$ cooling and no-cooling, $q=0.25$
          cooling and no-cooling, and $q=0.1$ no-cooling cases.
          \label{fig:q01c0.1-hr-fft}
        }
  \end{figure*}

\subsubsection{Accretion rates}
\label{sec:mdot}

We compute the accretion rates through the individual BHs as well as
the total accretion rate onto the binary, and normalize these 
by the (time-averaged) single, non-spinning BH accretion rate.

For a given maximum rest-mass density and total BH mass, we find the
highest accretion rate in the single BH case. This is consistent with
the expectation that the absence of a tidal-torque barrier will allow
matter to flow more easily toward the BH(s).

By contrast, the tidal torque is maximized for $q=1$ (all else being
equal), so the expectation is that the accretion rate will be
minimum for $q=1$.  In agreement with this expectation we find the
accretion rate in the $q=1$ cases to be smaller than all other mass
ratios we consider here. 

In Table~\ref{tab:results} we list the average accretion rate vs mass
ratio, and plot the results in Fig.~\ref{fig:diag-vs-q}. The general
trend is that lower mass ratios have higher accretion
rates. In the $q=0.1$ case the average accretion rate is about $50\%$
that of the single BH case with the same initial maximum rest-mass
density and total BHBH mass, while the average accretion rate in the
equal-mass case is roughly $33\%$ of that in the $q=0$ case.

For $q=1$ and $0.5$ both black holes accrete at comparable rates
whether cooling is applied or not (see four upper rows, left panels
Fig.~\ref{fig:q01c0.1-hr-fft}).
However, we observe that often the accretion rates on the individual
BHs are anti-phased, i.e., accretion occurs for half an orbit primarily
on one BH and then for the second half of the orbit on the other BH. This
behavior in the relaxed state is due to an ``alternating''
pattern in which denser material primarily plunges first through one
stream and then through the other. 

In Fig.~\ref{fig:q01c0.1-hr-fft} we also plot the accretion rates for
$q=0.25$ and $0.1$, with and without cooling (three lower rows, left
panels).  It is apparent that for $q=0.25$ the accretion rate onto the
primary is comparable to that onto the secondary when $\Lambda =
0$. However, when $\Lambda \neq 0$ the dominant contribution to the
total accretion rate comes from the primary. In the $q=0.1$ case we
observe that the dominant contribution to the total accretion rate
comes from the primary whether cooling is applied or not, and the same
holds true for the $q=0.125$ case.

This result in the $q=0.1$ case can be qualitatively understood
through a rough analogy to Bondi-Hoyle-Lyttleton accretion: The secondary
has a smaller mass and moves faster on its orbit reducing its
effective capture cross section as suggested by Bondi-Hoyle-Lyttleton
accretion (see e.g.  \cite{Farris:2009mt}). Also the surface area of
the secondary is roughly a factor $M_{\rm BH}^2/m_{\rm bh}^2 \sim 100$
smaller than that of the primary. Note however, that the secondary
plays a role in stripping matter off the inner disk edge effectively
as it orbits closest to (or even through) the disk, so the accretion
rate onto the secondary is not generally expected to be 100 times
smaller than the accretion rate onto the primary. 

In particular, in the $q=0.1$ no-cooling case a dense persistent structure
co-orbits with the secondary (see Fig.~\ref{fig:q10nc-lr_snapshots_rho_xy}). The density in this
structure exceeds the density of matter near the primary by more than a factor
of two.

\subsubsection{Variability}

We now report results from the Fourier analysis of the accretion rate.
These can be seen in the right panels of Fig.~\ref{fig:q01c0.1-hr-fft}.
A summary of the primary Fourier modes for the different cases is
presented in Table~\ref{tab:results}.

In the $q=1$ case a Fourier analysis of $\dot{M}_{\rm BH}$ (accretion
rate onto the primary) and $\dot{m}_{\rm bh}$ (accretion rate onto the
secondary) reveals a characteristic frequency near $(2/3) M\Omega_{\rm
  BHBH}$, in agreement with \cite{Farris:2012ux}. The analysis of
$\dot{M}_{\rm BHBH}$ (the total {\it binary} accretion rate) gives a
dominant Fourier mode with a frequency twice as high. We observe a
peak at the binary period only for $q=0.5$ and $0.25$, and only for the
cooling cases (see Fig.~\ref{fig:q01c0.1-hr-fft}). These results
indicate that if the variability in the accretion rate is directly translated
into a variability of EM signatures, inferring the binary frequency
from EM observations may not be straightforward.
We observe variability for other mass ratios as well. The frequencies for
the equal mass case also appear in other cases, in addition to other
weaker contributions, but a clear trend is not evident. The most
prominent and clean periodic signature occurs in the $q=0.5$
no-cooling case (see Fig.~\ref{fig:q01c0.1-hr-fft} and discussion in
\cite{D'Orazio:2012nz}). Other strong Fourier modes are observed in
$q=0.5$ cooling and the $q=0.25$ cases. For $q=0.125$ and $q=0.1$ no
significant periodicities are observed.

In the $q=0.1$ case the variability is dominated by variations in the
accretion flow onto the primary. The Fourier analysis yields rather
irregular accretion, i.e.~not very pronounced frequencies. The secondary
accretes at several pronounced frequencies, but the amplitude of the
variations is much smaller. 

In \cite{D'Orazio:2012nz,Farris:2013uqa} the dependence on $q$ of the variability was
studied for geometrically thin (``locally isothermal'' disks) and proposed as
a key feature to observationally distinguish accreting BHBHs from
standard, single BH AGNs. In our Fourier analysis the individual
peaks are less significant than in \cite{D'Orazio:2012nz,Farris:2013uqa} and the
Fourier spectrum yields a more complex structure. This discrepancy is
likely due to a combination of additional effects including
differences in the viscosity prescription (i.e.~MHD turbulence
vs.~$\alpha$-viscosity), cooling prescriptions, thin vs.~thick disks,
2D vs 3D, and the EOS. These differences result in
geometrically thin vs thick disks, which are seen to have different
variability.

\subsubsection{Luminosities}
We compute the cooling ($L_{\rm cool}$) and Poynting ($L_{\rm EM}$)
luminosities, as well as the energy loss rate due to
outflowing matter ($L_{\rm gas}$) for all cases. We typically find $L_{\rm gas}$ is
comparable to $L_{\rm EM}$ and quite smaller than $L_{\rm cool}$. We
highlight representative cases and summarize all values for
the different $q$ characterizing the relaxed state in
Table~\ref{tab:results}.

The large variability seen in the accretion rate is only partially
reflected in the cooling luminosities and not reflected in the
Poynting luminosities. However, these conclusions need to be confirmed
with a self-consistent treatment involving radiative transfer.

In all cooling cases we find that the cooling luminosity is
significantly larger than the Poynting luminosity and the energy loss 
rate due to outflowing matter by almost a factor of 10 (see Table~\ref{tab:results}).

\begin{figure}[h]
      \includegraphics[width=0.485\textwidth]{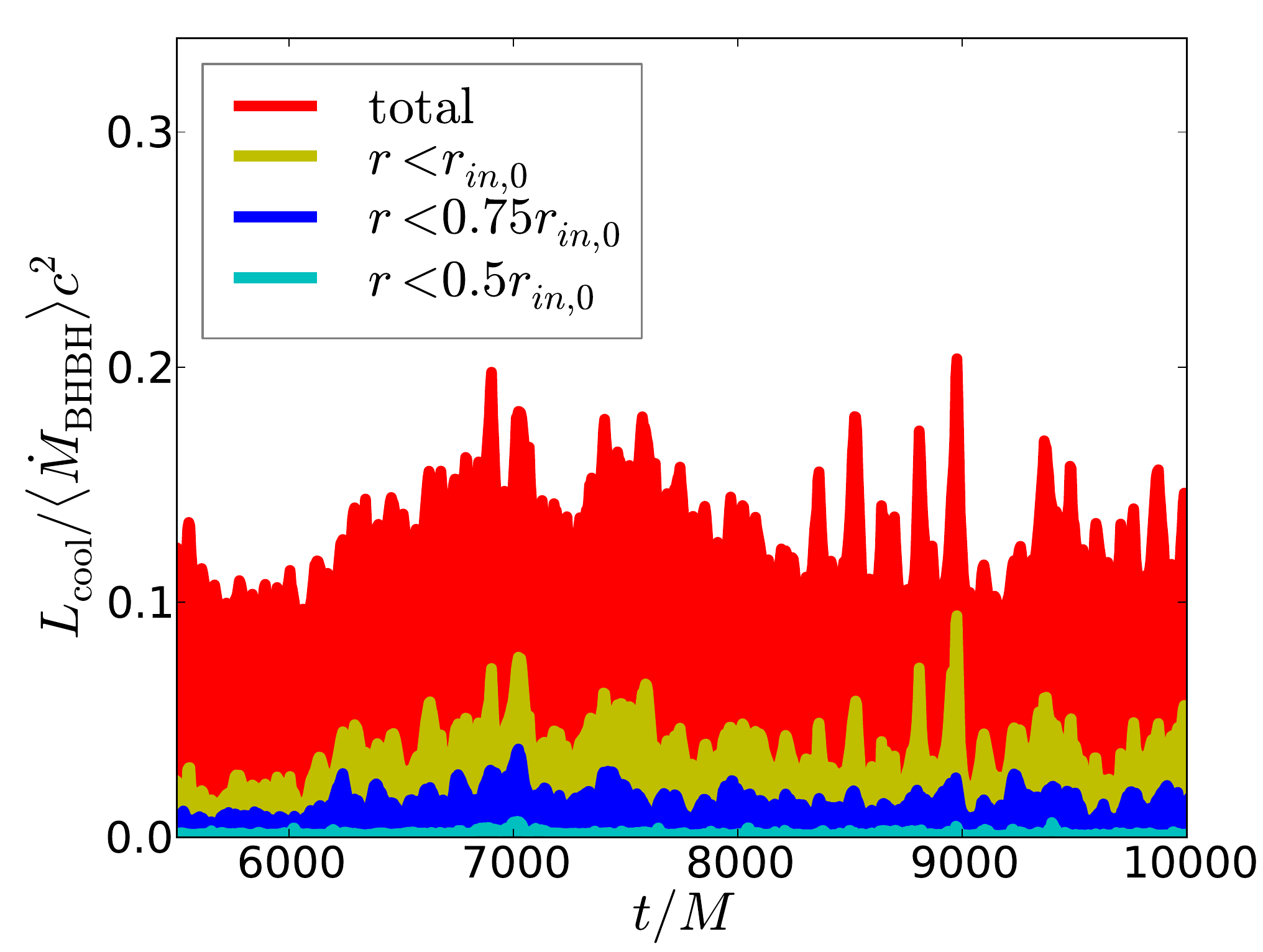}
      \caption{Contribution to $L_{\rm cool}$ from the inner cavity and
        outer disk for the $q=1$ cooling case. The notation $r <
        R_0$ means the cooling luminosity integral $\int_V \Lambda u_0
        \sqrt{-g} d^3x$, where $V$ is the volume within a coordinate
        sphere of radius $R_0$.  Note that about $30\%$ of the total
        cooling luminosity arises within $r_{\rm in,0}$. \label{fig:Lcool}}
\end{figure}

We estimate and compare the contributions to the cooling luminosity
from various regions in the disk: the outer disk, the inner edge, and
the cavity (see Fig.~\ref{fig:Lcool}). We find that although the outer
disk gives the largest contribution, the inner edge and cavity
interior are a substantial portion ($\sim 30\%$) of the total cooling
luminosity. Therefore, the activity in the cavity cannot be ignored, as
has been done in earlier studies.

\subsubsection{Opacities}
We estimate the Thompson scattering ($\tau_{\rm es}$) and
free-free absorption ($\tau_{\rm ff}$) optical depths in all cases. We do not find a strong dependence of $\tau^*=(3\tau_{\rm
  es}\tau_{\rm ff})^{1/2}$, $\tau_{\rm es}$ and $\tau_{\rm ff}$ on the
mass ratio. Our crude analysis shows $\tau^*\sim {\cal O}(1) (L_b/L_{\rm Edd})^{9/16}(M/10^8M_\odot)^{-1/16}$ throughout the
bulk of the disk. In conjunction with the radiation pressure dominance
found for near Eddington accretion rates, this result justifies our
choice of $\Gamma = 4/3$ for the bulk of the disk. Some cases are
marginal, however given the crudeness of our estimate and scaling
arguments the choice of adiabatic index is adequate.  The dominant source
of opacity in all cases is electron scattering. Within the cavity we
find that outside the accretion streams the matter is optically
thin. This means that radiation from the cavity can freely stream out,
and it is likely that (depending on the local temperature) cyclotron
lines may give rise to a nonthermal component to the emergent EM
spectrum. 

We find that $\tau^*$ is affected by cooling. The runs with cooling
have larger $\tau^*$ than those without cooling.

\subsubsection{Characteristic EM radiation frequencies}

The characteristic effective temperatures [see Eq. \eqref{Teffective}] are 
\labeq{}{
T_{\rm eff}
\sim 10^5\bra{\frac{L_{\rm b}}{L_{\rm Edd}}}^{1/4}\bra{\frac{M}{10^8M_\odot}}^{-1/4}\rm K.
} 
The corresponding characteristic thermal radiation frequencies ($\nu_{\rm bb}
\sim k_B T_{\rm eff}/h$) are reduced by a redshift factor
$1/(1+z)$, and are 
\labeq{}{
\begin{split}
\nu_{\rm bb} \sim &\ 10^{15}\bra{\frac{\varepsilon}{0.08}}^{1/4} \bra{\frac{M}{10^8M_\odot}}^{-1/2} \\
 &\ \qquad \qquad\quad  \bra{\frac{\dot M}{2.25M_\odot/\rm yr}}^{1/4}(1+z)^{-1} {\rm Hz} \\
 \sim &\ 10^{15} \bra{\frac{M}{10^8M_\odot}}^{-1/4} \bra{\frac{L_{\rm b}}{L_{\rm Edd}}}^{1/4}(1+z)^{-1} \rm Hz.
\end{split}
}

The lower limit to the equatorial temperature in the cavity for all cases 
(assuming $\rho_0\epsilon = aT^4$) is 
\labeq{}{
T \sim 10^5\bra{\frac{L_{\rm
      b}}{L_{\rm Edd}}}^{1/4}\bra{\frac{M}{10^8M_\odot}}^{-1/4}\rm K,
}
implying that the electrons are nonrelativistic, and hence they emit
cyclotron and not synchrotron radiation.  This result, too, should be
confirmed with radiative transfer calculations. Typical cyclotron
frequencies in the cavity then are 
\labeq{}{
\nu_{\rm cy} \sim 10^{6}\bra{\frac{M}{10^8M_\odot}}^{-1/2}\bra{\frac{L_{\rm b}}{L_{\rm
    Edd}}}^{1/2}(1+z)^{-1}\rm Hz.
}

Note that due to the large radiation pressure near the Eddington
limit, one expects that any dust will be blown away from the disk
and may accumulate at much larger radii than our computational domain. This
dust is likely to absorb the optical/UV radiation and re-emit it in the IR
\cite{Novikov73,Lynden-Bell:1971}.

  \subsubsection{Outflows and jets}

\begin{figure*}[t]
  \begin{center}
      \includegraphics[width=0.49\textwidth]{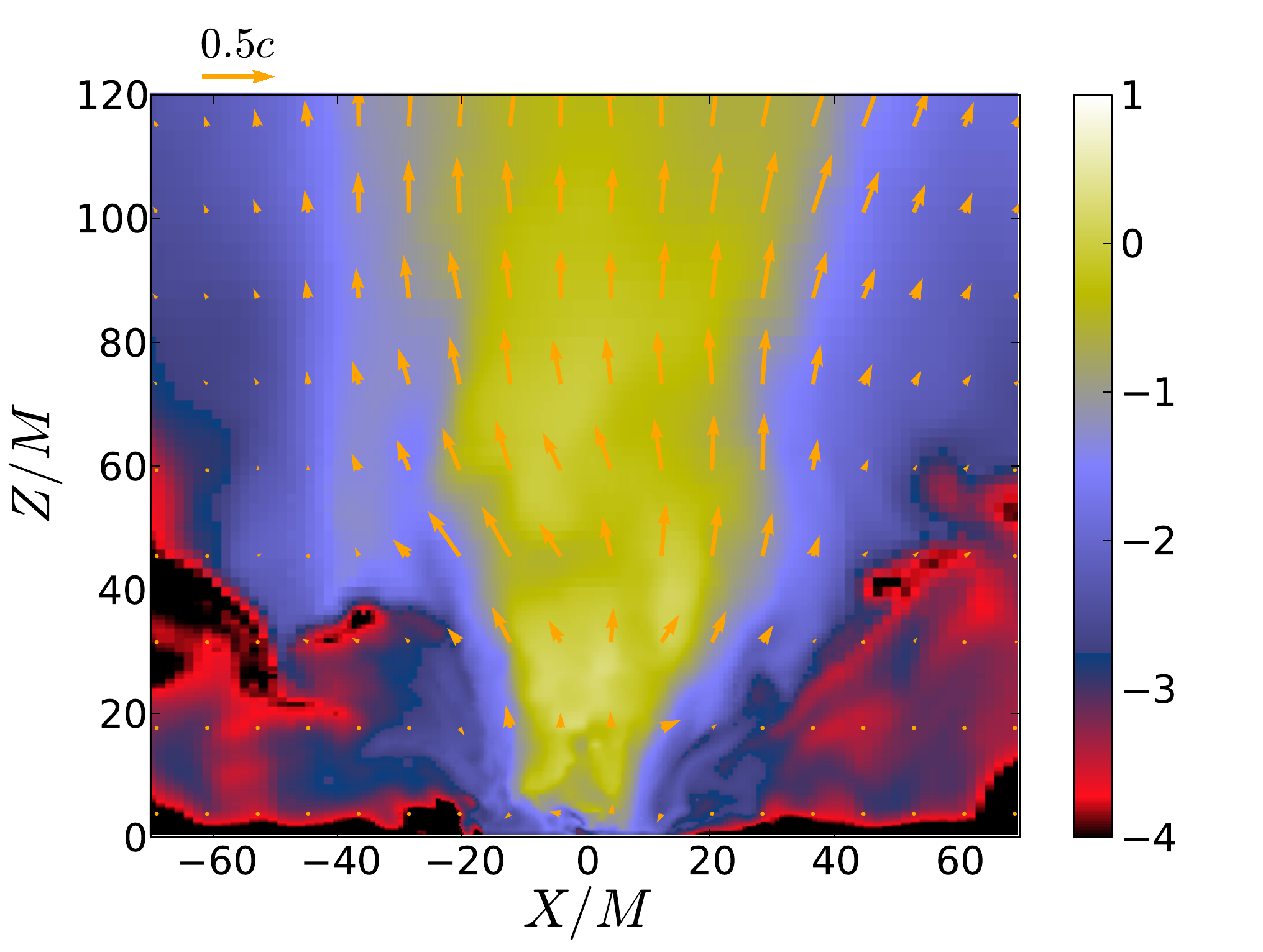}
      \includegraphics[width=0.49\textwidth]{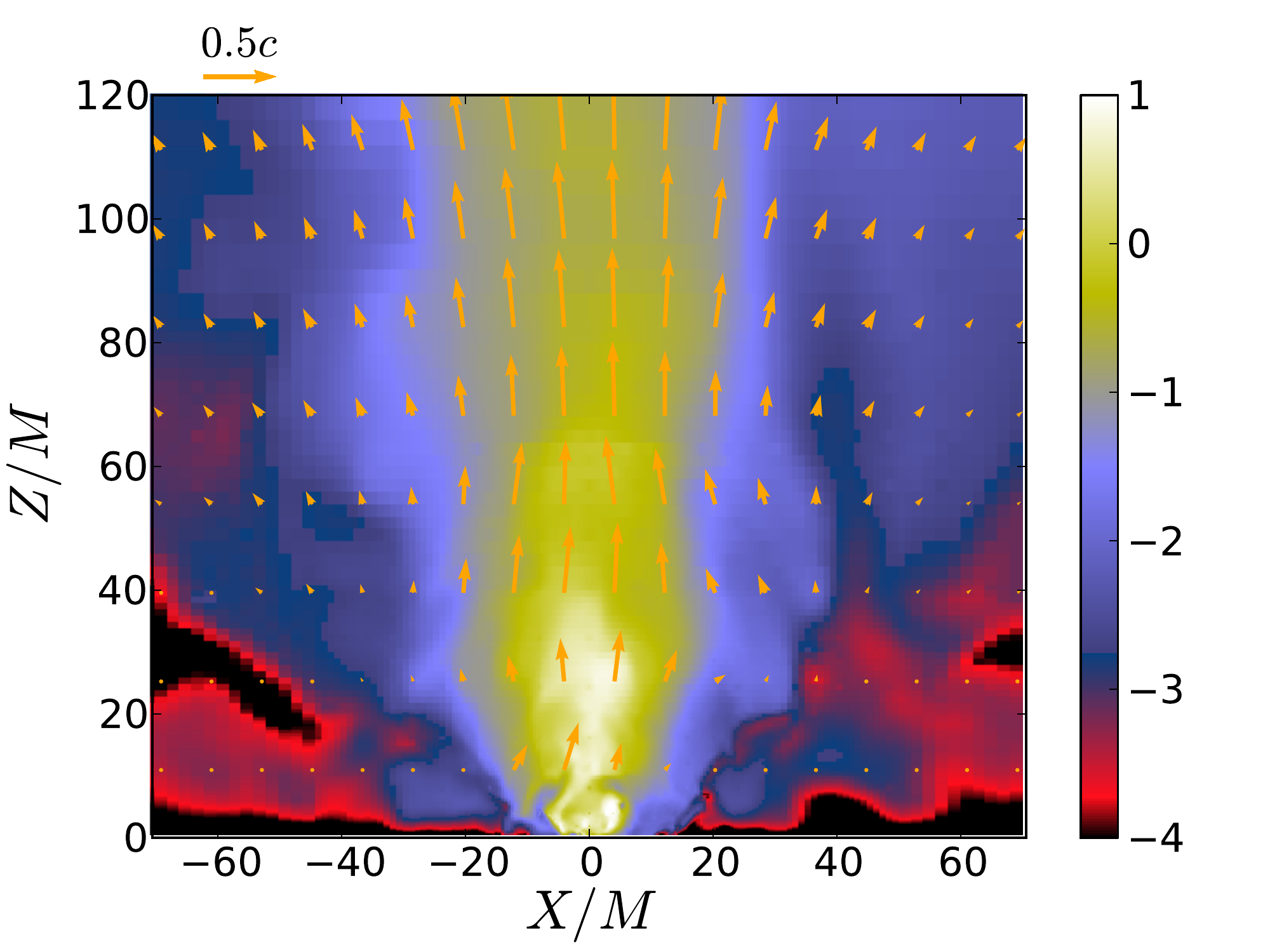}
        \caption{Contours of
          $b^2/2\rho_0$, (log color scale) in a meridional slice, and
          fluid velocity (arrows) at $t\sim 10000M$.  Left panel:
          $q=1$ cooling case. Right panel: $q=0.1$ cooling case.
          \label{fig:q10nc-lr-b2-over-rho}}
  \end{center}
\end{figure*}

\begin{figure*}[h]
  \begin{center}
      \includegraphics[trim =0cm 0cm 1.1cm 0cm,clip=True,width=0.95\textwidth]{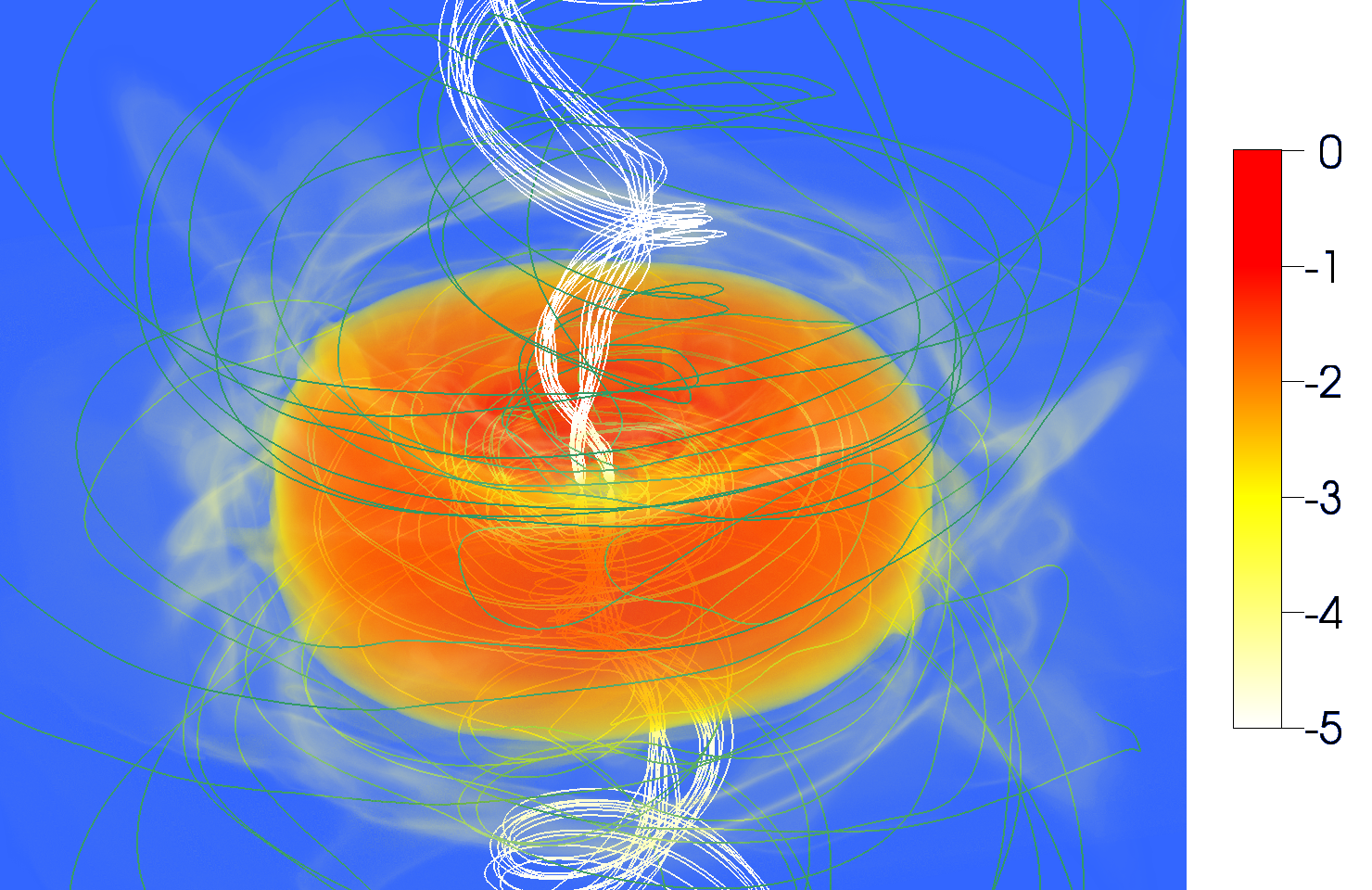}
      \includegraphics[trim =0cm 0cm 1.1cm 0cm,clip=True,width=0.95\textwidth]{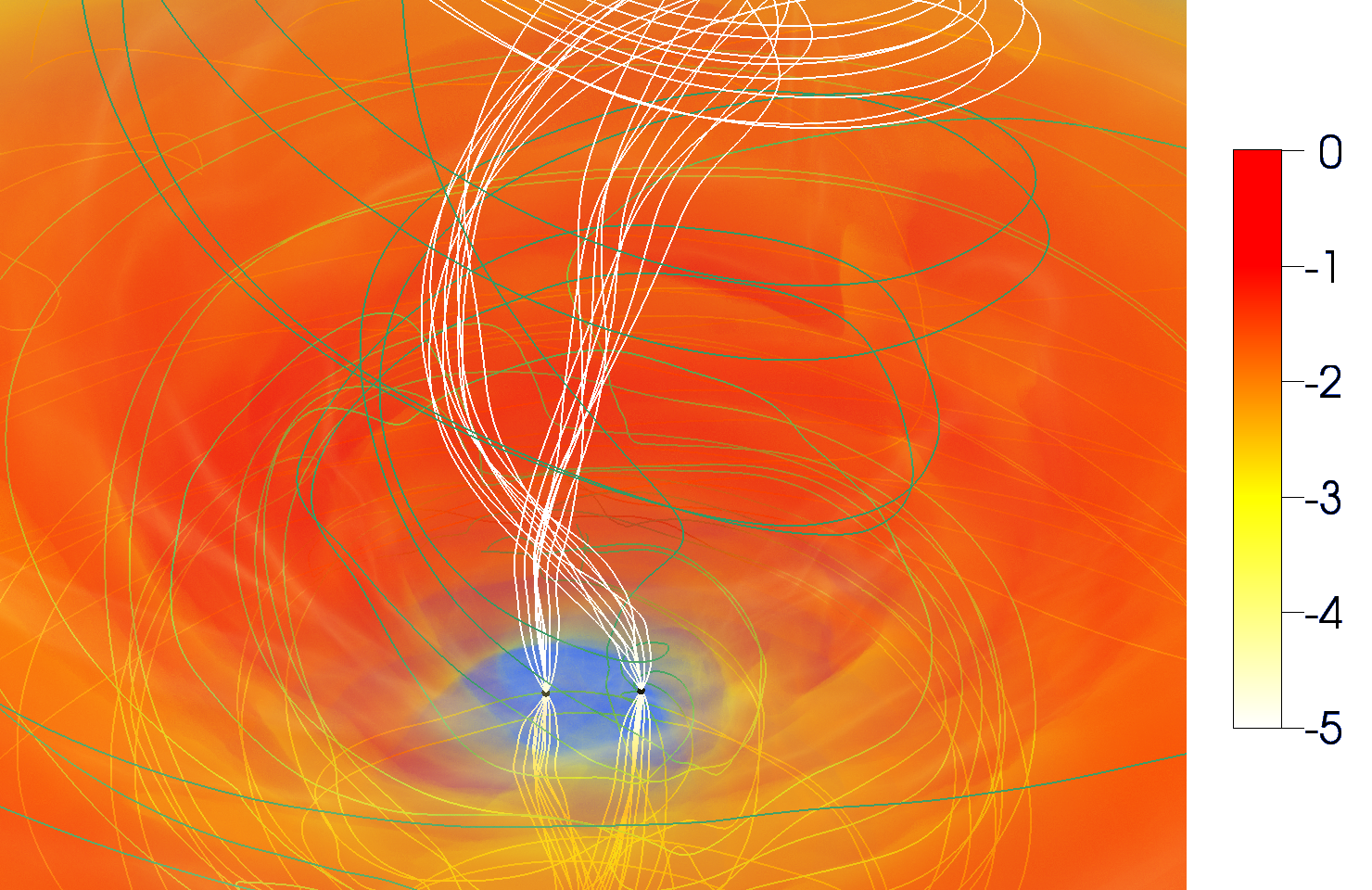}
        \caption{Volume rendering of rest-mass density normalized to
          its initial maximum value (color coding) and magnetic field
          lines for the $q=1$ cooling (medium resolution) case.  White
          field lines emanate from the BH apparent horizons.
          Incipient jets are launched from these systems.
          \label{fig:q01c-hr_Bfield}}
  \end{center}
\end{figure*}

In Fig.~\ref{fig:q10nc-lr-b2-over-rho} ($q=0.1$ and $1$ cases) we plot the
ratio $b^2/(2\rho_0)$, which equals the terminal Lorentz factor in
axisymmetric steady-state jet flows. Close to the BH, values
approaching $b^2/(2\rho_0)\sim 20$ are common, dropping to
$b^2/(2\rho_0) \sim 10$ at larger heights $Z$ (in the funnel).
We observe mildly relativistic outflows in all cases. Our 3D
visualizations of the B-field lines (see
Fig.~\ref{fig:q01c-hr_Bfield}) unambiguously show that there are field
lines emanating from each BH horizon and extending into the polar
regions. Near the BHs [$r\sim \mathcal{O}(10M)$] this leads to a dual
jet structure that at larger radii [$r\sim \mathcal{O}(100M)$] merges
into one common helical structure. Due to this effect the dual jets
may not be detectable individually. In the context of force-free
simulations around binary black holes, the existence of individually
detectable dual jets was proposed in
\cite{Palenzuela:2010nf}. However, in \cite{Moesta:2011bn} it was
shown that while a dual-jet component is present, it is subdominant
with respect to the predominantly quadrupolar EM emission, thereby
casting tight constraints on the detectability of such dual
jets. Regardless, the ``cavity'' contains a lot of dense matter so
that the assumption of the force-free limit of ideal MHD may not be
applicable. An MHD calculation can resolve this question, and our
results suggest that for twin jets to be detectable as individual
jets, it may require either BH spins misaligned with the orbital
angular momentum or tilted accretion disks \cite{Hayasaki:2012wu}.
Although for geometrically thin disks binary-disk misalignment may be
unlikely \cite{Miller:2013gya}, it may be possible for geometrically
thick disks.

\section{Conclusions}
\label{sec:conclusions}

We presented general relativistic magnetohydrodynamic simulations of
magnetized circumbinary accretion disks onto binary black holes with
mass ratios ranging from 1:1 to 1:10.  We model the disks using a
$\Gamma$-law equation of state with $\Gamma = 4/3$ -- appropriate for
optically thick, thermal radiation pressure-dominated fluids. We focus on a
disk near decoupling and perform our computations for $\sim 10000M$ (45
binary orbits).  We compute the disk structure after the disk has
reached a quasi-relaxed state. This dynamically quasi-steady state is a result of
binary tidal torques balancing viscous torques arising from MHD
turbulence triggered by the magnetorotational instability (MRI). The
tidal and viscous torques heat the gas, which is expected to radiate
and cool. To bracket this possibility we perform runs without
cooling, and runs with extremely rapid cooling, adopting an effective
cooling prescription that resembles a leakage scheme.

We study how the structure of the accretion flow is altered for the five
different mass ratios. We employ several diagnostics, including the
accretion rate and its Fourier transform,
estimates of the electromagnetic luminosity and the expected
characteristic frequencies of the emergent electromagnetic radiation, to compare
the cases. 

In the equal mass case we find that simulations without magnetic fields underestimate the 
accretion rates, and adopting our cooling prescription, the total luminosity of the 
source by two orders of magnitude. This is due to a substantial increase of the amount of 
matter in the "cavity" when magnetic fields are present. We also conclude 
that magnetic fields alter the quasi-steady surface density profile. 
Turbulent B-fields lead to more shock heating than the binary tidal
torques alone, thereby boosting cooling luminosities above the values
found in unmagnetized disks. 

The surface density profile of the disk is sensitive to the
mass ratio mainly in the innermost regions. Cooling leads to a density
enhancement near the inner disk edge (a pile-up) which is strongest in
the equal mass case. 

We find that for all mass ratios a two-stream accretion pattern is
present. These streams are attached to the horizons, with the density
close to the horizons being among the densest part of the accretion
flow. In particular, for the $q=0.1$ case the material overflows into
the inner cavity and (partially) refills the hollow present in the
initial data. During this process a persistent dense structure forms
around the secondary. This behavior suggests that it would be
inadequate to ignore the flow inside the cavity, which only
simulations in full GR can treat correctly.

The average binary accretion rates relative to the single,
non-spinning BH case range from $50\%$ ($q=0.1$) to $24\%$ ($q=0.5$)
with a general trend that the accretion rate becomes smaller as $q
\rightarrow 1$. For $q=0.5$ and $q=1$, both BHs accrete at a comparable
rate (on average), but for roughly half a binary period one of the two
accretion streams is significantly stronger than the other, boosting
the accretion rate onto one BH and diminishing that onto the other. For
smaller $q$ the accretion and variability is
increasingly dominated by the primary BH, especially in the cases with
cooling. For the single BH ($q=0$) significant variability
ceases. In general we do not observe evidence for variability exactly at the
binary period. Two exceptional cases in which the binary
frequency is detected are the $q=0.5$ and $0.25$ cooling models.

We find that the variability in the accretion rate is not reflected in
the variability of either the Poynting or the cooling
luminosity. We also find that our cooling luminosity is always larger than
the Poynting luminosity, though careful GRMHD simulations with radiative
transfer are necessary to confirm these findings.

The cavity radiates an amount comparable to the outer disk.  Only the
innermost regions reflect the strong variability seen in $\dot{M}$,
but in general the radiation from the outer disk smears these
variabilities out. We tentatively conclude that it will be challenging
to distinguish between single BH and binary BH AGN sources, unless
other effects such as binary disk misalignment are present. However,
radiative transfer calculations are required to confirm this result.

We observe nonaxisymmetric structure in the relaxed disk. A ``lump''
forms near the inner disk edge and is strongest when we allow for cooling. 

All of our evolutions reveal magnetic field lines emanating from each
of the two horizons, forming dual jets which merge into one helical
structure above the polar regions.

We estimate that the effective optical depth is $\tau^*\sim {\cal
  O}(1) (L_b/L_{\rm Edd})^{9/16}(M/10^8M_\odot)^{-1/16}$ throughout
the bulk of the disk, hence the disks we consider are optically thick
to absorption.  The characteristic effective temperature of our disk
models is $T_{{\rm eff}} \sim 10^5 \bra{L_{\rm b}/L_{\rm
    Edd}}^{1/4}\bra{M/10^8M_\odot}^{-1/4}\rm K$. Expected frequencies
of the thermal radiation are $\nu_{bb} \sim 10^{15} \bra{L_{\rm
    b}/L_{\rm Edd}}^{1/4}\bra{M/10^8M_\odot}^{-1/4}/(1+z)\rm Hz$
(optical/near UV). Therefore, instruments such as LSST, WFIRST, and
PanSTARRS will be ideally suited to study these sources. In the
cavity, we find that $\tau^* < 1$, hence cyclotron lines may be
directly observable as a nonthermal component of the spectrum.
Typical cyclotron frequencies in the cavity then are $\nu_{\rm cy}
\sim 10^{6}\bra{M/10^8M_\odot}^{-1/2}\bra{L_{\rm b}/L_{\rm
    Edd}}^{1/2}(1+z)^{-1}\rm Hz$, which fall in the radio spectrum.
Although considerable caution must be applied to these estimates, as
we do not account for mircophysics and radiative transfer in our
simulations, they may serve as useful guides to subsequent, more
detailed investigations and to astronomical instruments that may be
able to observe the scenarios we are simulating.

In upcoming work we plan to evolve the models presented here through
inspiral, merger, and the post-merger phases. We expect afterglow
emission \cite{Milosavljevic:2004cg,Tanaka:2009iy} similar to
\cite{Farris:2012ux}. The latter phase will model the process by which
the disk material viscously diffuses into the cavity following merger,
leading to a brightening of the disk \cite{Tanaka:2009iy,Shapiro:2009uy}.
In the future we plan to include more realistic cooling and radiative
transfer and use higher resolution for a more accurate modeling of these
sources. It will be crucial to determine in this context the level of
variability in the luminosity, and the EM spectrum of the source.

\acknowledgements

It is a pleasure to thank the Illinois Relativity group REU team
(Albert Kim, Lingyi Kong, Brian R.~Taylor, and Francis J.~Walsh) for
assistance in creating Figs.~\ref{fig:visit-bfield-id} and
\ref{fig:q01c-hr_Bfield}.  We thank Brian Farris for useful
discusions.  This paper was supported in part by NSF Grants
AST-1002667, PHY-0963136 and PHY-1300903 as well as NASA Grants
NNX11AE11G and NN13AH44G at the University of Illinois at
Urbana-Champaign. VP gratefully acknowledges support from a Fortner
Fellowship at UIUC. HPP acknowledges support by NSERC of Canada, the
Canada Chairs Program, and the Canadian Institute for Advanced
Research. The metric initial data was computed on the GPC
supercomputer at the SciNet HPC Consortium~\cite{Scinet}. SciNet is
funded by: the Canada Foundation for Innovation under the auspices of
Compute Canada; the Government of Ontario; Ontario Research
Fund--Research Excellence; and the University of Toronto. This work
used the Extreme Science and Engineering Discovery Environment
(XSEDE), which is supported by NSF grant number OCI-1053575. This
research is part of the Blue Waters sustained-petascale computing
project, which is supported by the National Science Foundation (award
number OCI 07-25070) and the state of Illinois. Blue Waters is a joint
effort of the University of Illinois at Urbana-Champaign and its
National Center for Supercomputing Applications.

\appendix

\section{Numerical stability of cooling schemes in low density regions}
\label{Appendix:Courant-stability}

The evolution equation for the specific thermal energy
without magnetic fields, but including radiation
with an effective emissivity $\Lambda = dU/d\tau$ as measured
by an observer comoving with the fluid, is given by \cite{Paschalidis:2011ez}
\labeq{radEOM}{
\frac{d\epsilon_{\rm th}}{d\tau} = \frac{P_{\rm th}}{\rho_0^2}\frac{d\rho_0}{d\tau} - \frac{1}{\rho_0}\Lambda,
}
where $\epsilon_{\rm th}\equiv \epsilon-\epsilon_0$, and where
$\epsilon$ is the total specific energy and $\epsilon_0$ the total
specific energy of the fluid element calculated with the EOS at $t=0$,
i.e., $P=K_0 \rho_0^\Gamma$, $\epsilon_0 = P/\rho_0(\Gamma-1)$.  We thus
account for cooling by specifying $\Lambda$.

\subsection{Cooling prescriptions}

Two effective emissivities that have been adopted in the literature
are:
\labeq{cool1}{ \Lambda_1 = \frac{\rho_0 \epsilon_{{\rm th}}}{\tau_c}, }
where $\tau_c$ is the cooling time scale, and

\labeq{cool2}{
\Lambda_2 = \frac{\rho_0\epsilon}{\tau_c}\bigg(\frac{\Delta S}{S_0}+\bigg| \frac{\Delta S}{S_0}\bigg|\bigg),
}
\\
where $S \equiv K = P/\rho_0^\Gamma$ and $S_0$ is both the initial (unshocked) and target value. We
introduced \eqref{cool1} in \cite{Paschalidis:2011ez}, and
\eqref{cool2} was introduced in \cite{Shafee:2008mm,Noble:2012xz}.
While these two schemes achieve similar properties, i.e., drive $K$ back to
its initial value, they have fundamentally different numerical
stability properties. Here we explain why this is so.

First, let us recast our cooling emissivity $\Lambda_1$ in a form that is
similar to $\Lambda_2$ to show that our cooling
prescription also drives $K$ to $K_0$. The pressure is given 
by
\labeq{}{
P = \rho_0 \epsilon (\Gamma-1)
}
and similarly for the initial pressure
\labeq{}{
P_0 = \rho_0 \epsilon_0 (\Gamma-1).
}

The difference between the two is
\labeq{Pres}{
\Delta P = \rho_0 (\epsilon-\epsilon_0)(\Gamma-1) = \rho_0 \epsilon_{\rm th}(\Gamma-1) = P_0 \frac{\epsilon_{\rm th}}{\epsilon_0}.
}
But, 
\labeq{dPoP}{
\frac{\Delta P}{P_0} = \frac{(K-K_0)\rho_0^{\Gamma}}{K_0\rho_0^{\Gamma}} = \frac{\Delta K}{K_0} = \frac{\Delta S}{S_0}.
}

Combining Eqs. \eqref{Pres}, \eqref{dPoP}, and \eqref{cool1} yields
\labeq{cool1b}{ 
\Lambda_1 = \frac{\rho_0 \epsilon_{0}}{\tau_c}\frac{\Delta S}{S_0}.
}

The emissivity $\Lambda_2$ for $\Delta S >0$, i.e., the only case when the emissivity 
is not 0, becomes

\labeq{cool2b}{
\Lambda_2 = \frac{2\rho_0\epsilon}{\tau_c}\frac{\Delta S}{S_0}.
}

Thus, the two prescriptions Eqs. \eqref{cool1b} and \eqref{cool2b} are
similar and both drive the gas to constant initial entropy $S_0$. But
the two prescriptions are not the same. Another way to see this is to
write $\Lambda_2$ in a form that resembles $\Lambda_1$.  Using
Eqs. \eqref{Pres}, \eqref{dPoP}, we can write \eqref{cool2b} as

\labeq{cool2c}{
\Lambda_2 = \frac{2\rho_0\epsilon_{\rm th}}{\tau_c}\frac{\epsilon}{\epsilon_0}.
}
\\

As we will see below, it is the factor $\epsilon/\epsilon_0$
by which $\Lambda_2$ differs from $\Lambda_1$ that leads to the
completely different numerical stability properties of these two
cooling schemes.

\subsection{Numerical properties of cooling prescriptions}

Insert our cooling prescription in Eq. \eqref{radEOM} and 
drop the first term on the right-hand-side (RHS) of Eq. \eqref{radEOM}, i.e, assume
that no adiabatic compression or expansion takes place. Then we obtain
\labeq{radEOM2}{ 
\frac{d\epsilon_{\rm th}}{d\tau} = -  \frac{\epsilon_{{\rm th}}}{\tau_c},
} 
i.e., exponential cooling of the excess thermal energy with cooling
time scale $\tau_c$. For an explicit numerical scheme to be Courant
stable we must set the maximum timestep $\Delta t \lesssim \tau_c$. To
see this we can we use a simple Euler explicit integration scheme to
write Eq. \eqref{radEOM2} in finite difference form as
\labeq{radEOM2FD}{ 
\frac{\epsilon_{\rm th}^{n+1}-\epsilon_{\rm th}^n}{\Delta\tau} = -  \frac{\epsilon_{{\rm th}}^n}{\tau_c}.  
} 
Then the amplification factor is given by
\labeq{Ampfac}{ 
f\equiv\frac{\epsilon_{\rm th}^{n+1}}{\epsilon_{\rm th}^n} = 1 -  \frac{\Delta\tau}{\tau_c}.  
} 
For numerical stability the magnitude of the amplification factor must be less than 1
\labeq{}{
|f| < 1 \Rightarrow \bigg|1 -  \frac{\Delta\tau}{\tau_c}\bigg| < 1 \Rightarrow \Delta\tau < 2\tau_c.  
}

By contrast, inserting $\Lambda_2$ [Eq. \eqref{cool2c}] in Eq. \eqref{radEOM} and 
dropping the first term on the RHS of Eq. \eqref{radEOM}
yields
\labeq{radEOM3}{ 
\frac{d\epsilon_{\rm th}}{d\tau} = -
  2\frac{\epsilon_{{\rm th}}}{\tau_c}\frac{\epsilon}{\epsilon_0}, 
}
hence the effective cooling time scale is now
 \labeq{}{ 
\tau_{\rm eff} =  \frac{1}{2}\frac{\epsilon_0}{\epsilon}\tau_c.
} 
We now require $\Delta t < \tau_{\rm eff}$ for numerical
stability. However, low density regions can become strongly shock
heated, in which case $\epsilon_0\ll \epsilon$, implying $\tau_{\rm
  eff} \ll \tau_c$. As a result, a violation of the Courant stability
condition can quickly occur that will terminate the computations, if
$\Delta t < \tau_{\rm c}$ is used as the stability condition. This is
a disadvantage of the $\Lambda_2$ prescription, as it means that one
must impose a density cutoff for the simulations to be stable and
retain relatively large timesteps. This is something we do not need to
require with our cooling emissivity $\Lambda_1$, as the fixed cooling
time scale $\tau_c$ is, at the same time, the effective cooling
time scale. The difference is important as we want to probe low
density regimes in the disk without artificial cutoffs.






\bibliography{bhbh+disk-paper2}


\end{document}